\documentclass[12pt,twocolumn,apj]{aastex631}

\usepackage{changepage}
\usepackage{booktabs}
\usepackage{aas_macros}
\usepackage{epsfig}
\usepackage{amsmath}
\usepackage{amssymb}
\usepackage{comment}
\usepackage{leftidx}
\definecolor{MyGreen}{rgb}{0.0,0.6,0.3}
\definecolor{MyPurple}{rgb}{0.6,0,0.3}
\usepackage{fancyref}
\usepackage{hyperref}
\hypersetup{colorlinks=true,linkcolor=MyPurple,urlcolor=MyPurple,citecolor=MyGreen}
\usepackage{color,colortbl}
\usepackage{tensind}
\tensordelimiter{?}
\DeclareGraphicsExtensions{.bmp,.png,.jpg,.pdf}
\usepackage{verbatim}
\usepackage[normalem]{ulem}
\usepackage{orcidlink}
\usepackage{soul}
\usepackage{upgreek}
\usepackage{bm}

\def\beq{\begin{equation}}
\def\eeq{\end{equation}}
\def\ba{\begin{eqnarray}}
\def\ea{\end{eqnarray}}
\def\bal{\begin{align}}
\def\eal{\end{align}}

\begin{document}

\title[Planetary Mode Excitation] {Excitation and Damping of Oscillation Modes in Gaseous Planets}

\author{Jim Fuller\orcidlink{0000-0002-4544-0750}}
\email{jfuller@caltech.edu}
\affiliation{TAPIR, Mailcode 350-17, California Institute of Technology, Pasadena, CA 91125, USA}

\author{Marzia Parisi\orcidlink{0000-0003-4064-6634}}
\affiliation{Jet Propulsion Laboratory, California Institute of Technology, Pasadena, CA 91109, USA}

\author{Steve Markham}
\affiliation{New Mexico State University, Department of Astronomy, 1320 Frenger Mall, Las Cruces, NM 88003, USA}

\author{A. James Friedson}
\affiliation{IEP, Jet Propulsion Laboratory, California Institute of Technology, Pasadena, CA 91109, USA}

\author{J. R. Fuentes}
\affiliation{TAPIR, Mailcode 350-17, California Institute of Technology, Pasadena, CA 91125, USA}

\begin{abstract}

The excitation and damping mechanisms for oscillation modes of gas giant planets are undetermined. We show that differential rotation may greatly enhance convective  viscosity in giant planets, resulting in damping times of $t_{\rm damp} \sim 10^5-10^6 \, {\rm years}$ for f~modes and low-order p~modes. Radiative diffusion damps p~modes on time scales of $t_{\rm damp} \sim 10^3-10^7 \, {\rm years}$. While the lethargic convective motions cannot effectively excite f~mode or p~modes, storms driven by condensation of water and/or silicates may play a role. High-order p~modes are most effectively excited by cometary/asteroid impacts. Applying these calculations to solar system planets, water storms, rock storms, and impacts may all contribute to exciting the observed f~modes amplitudes of Saturn via ring seismology. Similar f~mode amplitudes with fractional gravitational perturbations of $\delta \Phi/\Phi \sim 10^{-10}-10^{-9}$ are expected for Jupiter and Uranus, apart from their lowest $\ell$ f~modes which could have larger gravitational perturbations of $\delta \Phi/\Phi \sim 10^{-7}$. Rock storms may contribute to mode driving in Jupiter, while water storms are more important for Uranus. The highest-amplitude p~modes are predicted to have periods of $\sim$10-30 minutes, with surface velocities of $\sim$10 {\rm cm/s} for Jupiter and Saturn, and $\sim$1 {\rm cm/s} for Uranus. These oscillation modes may be detectable with radial velocity measurements, ring seismology, or spacecraft Doppler tracking. However, both the damping and excitation physics are uncertain by orders of magnitude, so more careful examination of the relevant physics is required for robust estimates.

\end{abstract}


\section{Introduction}

Seismology is our most powerful technique for measuring the internal structures of the Earth \citep{romanowicz:10}, the Sun \citep{basu:16}, and various types of stars \citep{chaplin:13,kurtz:22}. There is hope that we can perform similar analyses with seismic data for other planets in the solar system. Indeed, seismic activity has now been measured in Mars by Insight \citep{lognonne:23}, and in Saturn via \text{Cassini} observations of its rings \citep{hedman:13,hedman:14,french:16,french:19,Hedman+2019,french:21,hedman:22}. Analysis of these modes indicates that Saturn has a large and stably stratified core-envelope transition region \citep{fuller:14,Mankovich2021} and has been used to accurately measure Saturn's internal rotation rate \citep{mankovich:19,Dewberry2021,dewberry:22}.

Nearly all of the detected modes in Saturn are fundamental modes (f~modes), which have limited power in constraining its internal structure. More detailed seismic analysis would be possible if we could detect other modes, such as pressure modes (p~modes). Detection of oscillations in Jupiter, Uranus, or Neptune could also constrain their internal structures. The crucial question for new seismology efforts is whether planetary oscillations are excited to amplitudes high enough to be measured. This paper attempts to answer this question by exploring the physics of planetary mode excitation and damping.


Seismic activity in gaseous outer planets is likely to be more analogous to that of the Sun than the Earth, because the outer planets are mostly fluid rather than solid. But many aspects of the physics are very different from those of the Sun. For Jupiter, the radius is smaller by a factor of $\approx \! 10$, mass smaller by a factor of $\approx \! 10^3$, and luminosity smaller by a factor of $\approx 10^9$ compared to the Sun. Disparities are even larger for Saturn, Uranus, and Neptune. The lower luminosity is especially important because there is much less power to drive convective turbulence. Convection is known to stochastically excite the Sun's oscillation modes \citep{goldreich:94,samadi:01a,samadi:01b}, producing velocities of order $\sim \! 40$ cm/s at the photosphere \citep{libbrecht:88}. That same turbulence also damps out the Sun's pulsations, limiting their typical lifetimes to $\sim$10 days \citep{Grigahcene:2005,belkacem:15}. 

On gaseous planets, neither the mode driving mechanism(s) nor the mode damping mechanism(s) are well understood, so mode amplitudes are often uncertain by orders of magnitude. Possible excitation mechanisms include convective turbulence, storms \citep{dederick:18,markham:18,wu:19}, cometary impacts \citep{kanamori:93,lognonne:94,dombard:95,wu:19}, baroclinic or Kelvin-Helmholtz instabilities, or even heat-driven oscillation modes \citep{dederick:17}. Possible damping mechanisms include convective viscosity \citep{markham:18}, interaction with rings \citep{wu:19}, radiative losses/thermal diffusion \citep{mosser:95}, and magnetic dissipation. 

Thankfully, Saturn's oscillation modes are determined to produce fractional gravitational perturbations of $\sim10^{-11}-10^{-9}$ \citep{afigbo:25}, corresponding to surface velocities ranging from $\sim10^{-3}-10^{0}$ cm/s. These measurements are for low-degree ($\ell \sim m \lesssim 20$) f~modes, but they do not constrain amplitudes of any other modes, such as p~modes, which do not have resonances in the rings. Our goal is to use Saturn's observed f~modes to benchmark models which can then be extended to predict amplitudes for Saturn's p~modes, and all modes of other planets.

In the same spirit as other recent work \citep{markham:18,wu:19,zanazzi:25}, we systematically investigate various damping mechanisms (Section \ref{sec:damping}) and excitation mechanisms (Section \ref{sec:excitation}). We find excitation/damping models that can approximately reproduce measured amplitudes of modes in the Sun and in Saturn, and then apply them to Jupiter and Uranus to predict their mode amplitudes (Section \ref{sec:amplitudes}). We discuss uncertainties of the physics in Section \ref{sec:discussion} and conclude in Section \ref{sec:conclusion}.

\section{Models and Modes}
\label{sec:modes}

To compute mode excitation and damping rates, we construct models of giant planets and compute corresponding oscillation modes.

Our Saturn and Jupiter models are constructed using the same procedure described in \cite{fuller:14} and \cite{Mankovich2021}. These models include a convective envelope outside of a radiative core, as suggested by ring seismology. Our Saturn model has a radiative region extending to $r/R=0.65$ with a peak Brunt-V\"ais\"al\"a frequency $N/\omega_{\rm dyn} \sim 2$, and a gravity moment $J_2 \simeq 0.016$. Our Jupiter model also has a small stably stratified core extending to $r/R=0.2$, with gravity moment $J_2 \simeq 0.015$ close to the measured value \citep{Guillot2005}. While the size of the radiative core greatly affects g~modes, it has a fairly small effect on the frequencies and eigenfunctions of the f~modes and p~modes that are the focus of this work.

\begin{figure}
\includegraphics[scale=0.36]{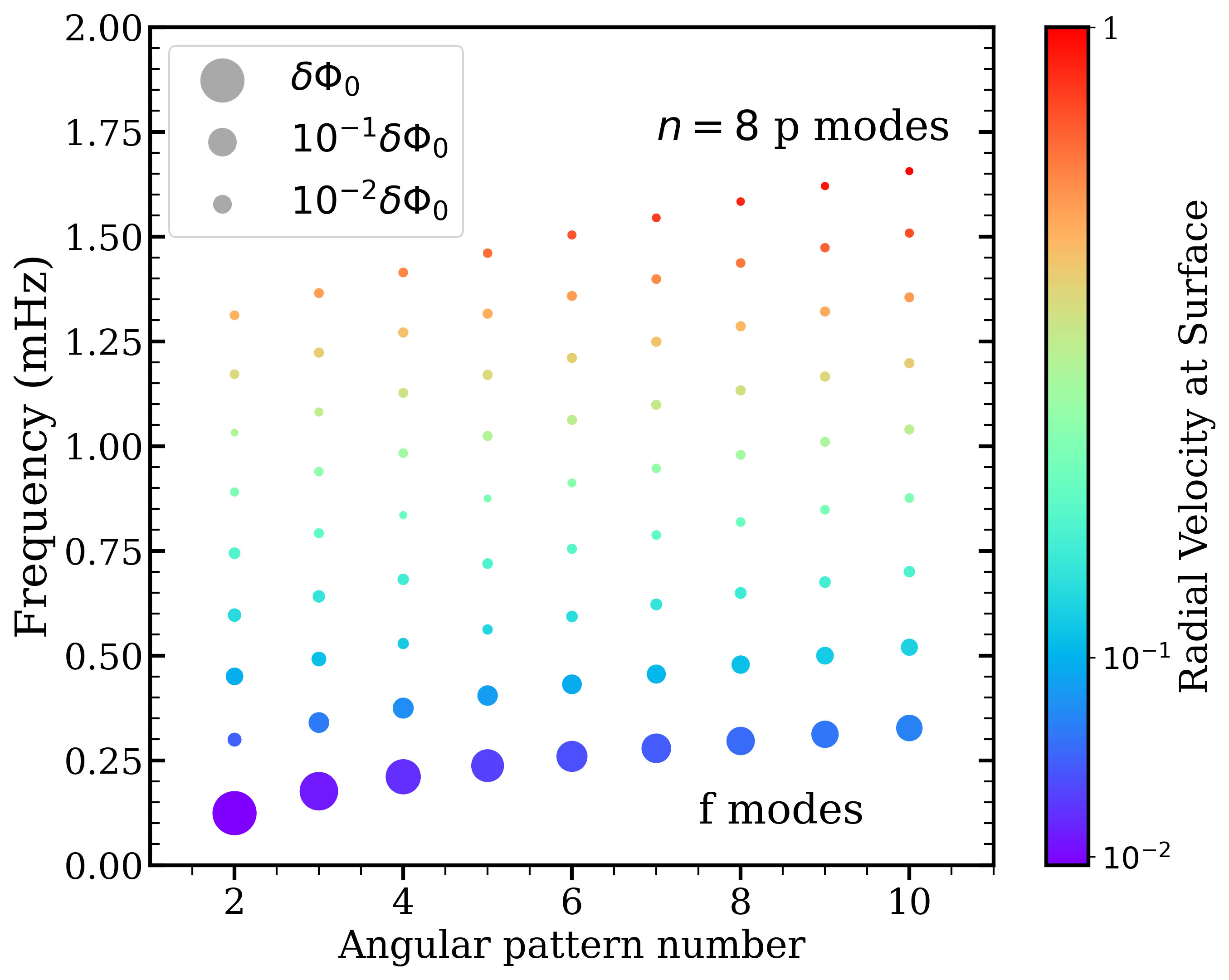}
\caption{\label{fig:Modes-Uranus} Oscillation mode frequencies of our Uranus model as a function of angular number $\ell$. Symbol sizes indicate surface gravitational potential perturbations, while symbol colors indicate surface radial velocities, for modes normalized to have equal energies. The f~modes typically have larger gravitational perturbations than low-order p~modes by a factor of $\sim$100, but smaller radial velocity perturbations by a similar factor. Modes of Jupiter and Saturn models have similar behavior, but with frequencies proportional to $\sqrt{GM/R^3}$ in each case.}
\end{figure}

To calculate oscillation mode frequencies and eigenfunctions, we include rotational effects to second order in $\Omega_{\rm s}$ as described in \cite{fuller:14}. However, for simplicity we neglect coupling between modes of different radial order and angular index $\ell$, i.e., we use the ``pseudo-mode" frequencies and radial eigenfunctions. Since our focus is the highly uncertain mode excitation and damping rates, exact mode frequencies and eigenfunctions are of secondary importance.

For Uranus, we use interior models from \citep{helled:2010} and compute adiabatic oscillation modes using the GYRE pulsation code \citep{townsend:13}, and we neglect rotation effects, which are less important in Uranus. Figure \ref{fig:Modes-Uranus} shows f~modes and p~modes of various angular numbers $\ell$ for our Uranus model (Jupiter and Saturn models have similar characteristics). For modes containing the same energy, the f~modes have much larger gravitational perturbations than p~modes, while they produce much smaller surface velocity fluctuations, with weak dependence on $\ell$.

To define mode amplitudes and eigenfunctions, we adopt an energy normalization condition
\begin{equation}
    \label{eq:norm}
    E_\alpha = \omega_\alpha^2 \int dV \rho |\xi_\alpha|^2 = \frac{G M^2}{R} = E_{\rm bind} \, . 
\end{equation}
Here, $\rho$ is the planetary density, $\omega_\alpha$ is a mode's angular frequency, $\xi_\alpha$ is a mode's displacement eigenfunction, and $\alpha$ indexes different oscillation modes. This energy normalization criterion is similar to that used by \cite{goldreich:94} but different from the inertia criterion used by \cite{wu:19}, so care must be taken when comparing our equations to theirs.

Below, we present results for the $\ell=m$ (prograde) f~modes of our planetary models, extending from $\ell=2$ to $\ell=20$. The f~modes are defined to be the modes with the largest surface gravitational perturbation $|\delta \Phi_\alpha|$ for a given $\ell$ and $m$. Results are expected to be similar for other values of $m$ because the frequencies and eigenfunctions are similar for the f~modes and p~modes of interest. An exception are the low-$\ell=m$ f~modes of Saturn, which are damped by the rings according to our model (Figure \ref{fig:tdamp}).
We also present results for $\ell=m=2$ p~modes, extending from radial order $n=1$ to $n \sim 20$. The p~mode properties depend primarily on mode frequency, so results are expected to be similar for other values of $\ell$ and $m$. 

Once the mode frequencies and eigenfunctions have been calculated, we use them to compute the mode damping times $t_{{\rm damp},\alpha}$ as described in Section \ref{sec:damping}, and the mode excitation rates $\dot{E}_\alpha$ as described in Section \ref{sec:excitation}.

\section{Damping Mechanisms}
\label{sec:damping}

\subsection{Convective Viscosity}
\label{sec:convdamp}

It is well-known that convection can create an effective turbulent viscosity that damps oscillation modes. Following \cite{goldreich:88}, the absorptivity produced by turbulence can be viewed as a three-body interaction between a wave with fluid velocity $\vec{v}_w$, and convective cells with size $h_{\rm con}$ and velocity $\vec{v}_{\rm con}$. They argue that the interaction imparts two sources of velocity perturbations to the cells. The first is $\Delta \vec{v}_w \sim (\vec{k} \cdot \vec{h}_{\rm con}) \vec{v}_w \sim k h_{\rm con} v_w$, where $\vec{k}$ is the wave vector. We believe this description is misleading because it implies a finite effect even if convective velocities are zero, which is non-physical. The convective velocity should enter into this expression.

Instead, the velocity difference imparted to the fluid should be $\Delta v_w \sim k \, x_{\rm con} \, v_w$. Here, $x_{\rm con} = v_{\rm con} \Delta t$ is the distance a fluid element is advected over a time $\Delta t$. The time $\Delta t$ is the duration of the interaction, whose value is $\Delta t \sim {\rm min}(\omega_{\rm con}^{-1},\omega^{-1})$ because it is the shorter between the wave oscillation period and the convection turnover time. We define the convective turnover frequency
\begin{equation}
\label{eq:omcon}
    \omega_{\rm con} \sim \frac{v_{\rm con}}{h_{\rm con}} \, .
\end{equation}
In the limit of fast convection such that $\Delta t \sim \omega_{\rm con}^{-1}$, we obtain $\Delta \vec{v}_w \sim k \, h_{\rm con} v_w$, the same result as \cite{goldreich:88}, but for apparently different reasons. In the limit of slow convection, we obtain $\Delta \vec{v}_w \sim k \, h_{\rm con} (\omega_{\rm con}/\omega) v_w$, which is the secondary effect discussed in \cite{goldreich:88}. 

The amount of energy absorbed by the three-body scattering is $\Delta E \sim \rho h_{\rm con}^3 (\Delta v_w)^2$. The corresponding wave absorptivity is $\alpha \sim \tau^{-1} \Delta E_w/E_w$, where $\tau^{-1}$ is the rate at which the absorptive interactions take place. We argue that $\tau^{-1} \sim \omega_{\rm con}$ in the limit of both fast and slow convection, because it is determined by the rate at which the convective elements can change their velocities so that another absorption event can occur.

In the high-frequency regime of $\omega_{\rm con} > \omega$, this yields an absorptivity $\alpha \sim k^2 h_{\rm con} v_{\rm con}$, and a corresponding convective viscosity $\nu_{\rm con} \sim \alpha/k^2 $ of
\begin{equation}
\label{eq:nucon}
    \nu_{\rm con} \sim h_{\rm con} v_{\rm con} \quad \big({\rm for} \, \, \omega_{\rm con} > \omega \big).
\end{equation}
This is the usual result for high-frequency convective motion. In the opposite limit of $\omega_{\rm con} < \omega$, the effective viscosity is
\begin{equation}
\label{eq:nucon_sup}
    \nu_{\rm con} \sim h_{\rm con} v_{\rm con} \bigg(\frac{\omega_{\rm con}}{\omega}\bigg)^2 \quad \big({\rm for} \, \, \omega_{\rm con} < \omega \big) \, .
\end{equation}

We note that this produces the classic result for low-frequency convective viscosity of \cite{goldreich:77}, but for different reasons. It does not rely on interactions with small-scale resonant eddies in a Kolmogorov cascade, but only on the different interaction times with the largest convective eddies. This appears to be consistent with the results of \cite{goodman:97,ogilvie:12} and the simulations of \cite{Duguidetal2020a,Duguidetal2020b}. The latter find the scaling of equation \ref{eq:nucon_sup} when $\omega_{\rm con} \ll \omega$, but show that the dissipation is produced by the largest convective eddies rather than the resonant eddies. This is consistent with our heuristic argument and will apply when we consider the effects of differential rotation below.

Finally, we note that \cite{Terquem2021,Terquem2023} have proposed a different scaling for convective dissipation, which leads to more efficient dissipation. However, this mechanism has been questioned by \cite{Barker2021}.



\subsection{Effect of Rotation}

Convection is strongly altered by rotation in giant planets where the rotation rate $\Omega$ is typically much larger than convective turnover frequencies. This was demonstrated by \cite{stevenson:79} and confirmed with simulations (e.g., \citealt{barker:14}). In non-rotating convection, the convective velocities are
\begin{equation}
\label{eq:vcon}
v_{\rm con} \sim (F/\rho)^{1/3} \, , 
\end{equation}
where $F \sim L/(4 \pi r^2)$ is the convective energy flux. The associated convective turnover frequency is given by equation \ref{eq:omcon}.

In rotating convection, the convective velocities are reduced to
\begin{equation}
\label{eq:vconr}
    v_{\rm con,rot} \sim [F^2/(\rho^2 \Omega h_{\rm con})]^{1/5} \sim v_{\rm con} \bigg(\frac{\omega_{\rm con}}{\Omega}\bigg)^{1/5} \, ,
\end{equation}
a reduction by a factor of $\sim (\omega_{\rm con}/\Omega)^{1/5}$. More importantly, the convective elements become aligned with the rotation axis, and have small length scales perpendicular to this axis, with
\begin{equation}
\label{eq:Hperpr}
    h_{\perp,{\rm rot}} \sim \bigg(\frac{F h_{\rm con}^2}{\rho \Omega^3}\bigg)^{1/5} \sim h_{\rm con} \bigg(\frac{\omega_{\rm con}}{\Omega}\bigg)^{3/5} \, ,
\end{equation}
a reduction by a factor of $(\omega_{\rm con}/\Omega)^{3/5}$ relative to the non-rotating case. This entails convective turnover frequencies of
\begin{equation}
\label{eq:omconr}
    \omega_{\rm con,rot} \sim \frac{v_{\rm con,rot}}{h_{\perp,{\rm rot}}} \sim \omega_{\rm con} \bigg(\frac{\omega_{\rm con}}{\Omega}\bigg)^{-2/5} \, ,
\end{equation}
an increase in frequency due to the smaller size of the convective eddies.

Using equation \ref{eq:nucon_sup} with the rotating value of convective velocity, length scale, and turnover frequency, we find
\begin{equation}
\label{eq:nuconr}
\nu_{\rm con,rot} \sim \nu_{\rm con} \, .
\end{equation}
Remarkably, the factors of $\Omega$ cancel and we obtain the same convective viscosity for rotating convection as non-rotating convection.

\subsection{Effect of Differential Rotation}
\label{sec:diffrot}

Now we consider what happens for a convective element in the presence of differential rotation. For simplicity, we consider cylindrical differential rotation with associated shear, $\omega_{\rm shear} = d u_{\rm rot}/dR$. The convective element will be sheared out in the $\phi-$direction. A convective eddy with radial size $h_R$ will be stretched at a rate $\sim h_R d u_{\rm rot}/dR \sim h_R \omega_{\rm shear}$. After one convective turnover time, the size of the eddy in the $\phi$-direction is $h_\phi \sim (\omega_{\rm shear}/\omega_{\rm con}) h_R$, increasing the correlation length of the convective elements. In giant planet envelopes where $\omega_{\rm shear} \gg \omega_{\rm con}$, the convective eddies are greatly stretched in the $\phi$-direction.

Repeating the analysis of Section \ref{sec:convdamp} for sheared convection, the wave interacting with a stretched convective element of size $h_\phi$ will obtain a velocity difference of $\Delta v_w \sim k_\phi h_\phi v_w \sim k_\phi h_R (\omega_{\rm shear}/\omega_{\rm con}) v_w$. This corresponds to an absorptivity and viscosity enhanced by a factor of $\sim (\omega_{\rm shear}/\omega_{\rm con})^2$ relative to the case without shear.
In the case of low-frequency convective motions ($\omega_{\rm con} < \omega$) relevant to planets, the corresponding convective viscosity becomes 
\beq
\label{eq:nuconef}
\nu_{\rm con,sh} \sim h_{\rm con} v_{\rm con} \bigg( \frac{k_\phi}{k} \frac{\omega_{\rm shear}}{\omega} \bigg)^2 \, . 
\eeq

One could argue that this calculation is flawed because the convective cells cannot remain coherent for time scales longer than $1/\omega_{\rm shear}$ since they are sheared out on this time scale. In this case, the size of the convective element would still be $h_{\rm con}$, but the effective convective turnover frequency would be $\sim \! \omega_{\rm shear}$. The low-frequency effective convective viscosity would then be
\begin{equation}
\label{eq:nuconef2}
    \nu_{\rm con,sh} \sim h_{\rm con} v_{\rm con} \bigg(\frac{\omega_{\rm shear}}{\omega}\bigg)^2
\end{equation}
which is very similar to equation \ref{eq:nuconef}. In both cases, the convective viscosity is suppressed by a factor ($\omega_{\rm shear}/\omega)^2$ rather than $(\omega_{\rm con}/\omega)^2$. In the outer solar system planets, $\omega_{\rm shear} \gg \omega_{\rm con}$ in much of the planetary interior, so the convective viscosities of equations \ref{eq:nuconef} and \ref{eq:nuconef2} are much larger than previously expected.

To support this heuristic picture, Appendix \ref{sec:shearvisc} presents a more careful analysis of convection in the presence of shear. These calculations reveal terms with the same scaling as equations \ref{eq:nuconef} and \ref{eq:nuconef2}. Appendix \ref{sec:shearvisc} also verifies the increased correlation lengths of sheared convective eddies by analyzing simulation results.

\begin{figure}
\includegraphics[scale=0.36]{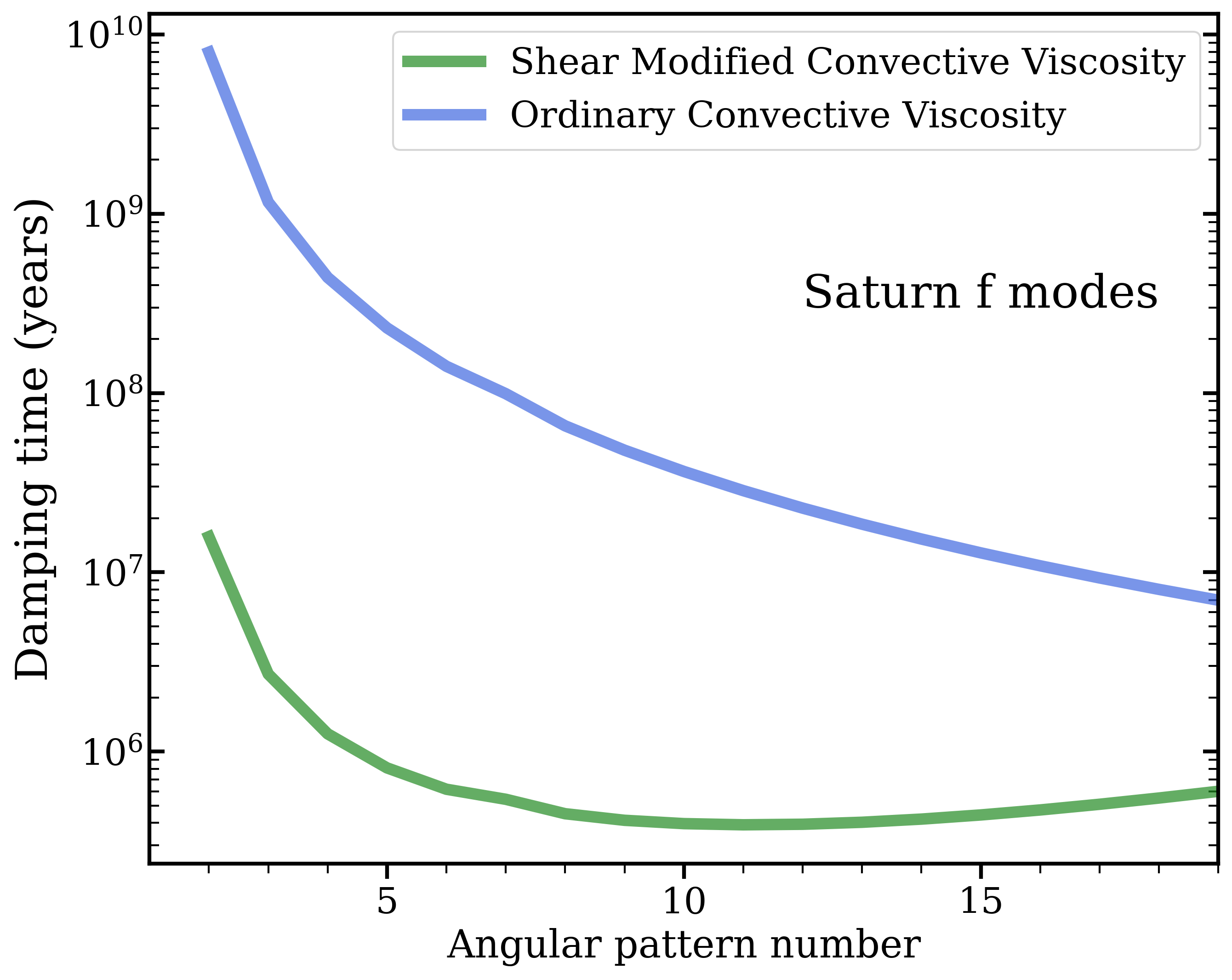}
\caption{\label{fig:tdamprot} Damping time of $\ell=m$ f~modes in Saturn, as a function of angular number $\ell$, due to different models of convective viscosity. The blue line shows the simple estimate of equation \ref{eq:nucon_sup}, while the green line accounts for enhancement due to zonal winds (equation \ref{eq:nuconrotef}).}
\end{figure}

Accounting for both rotation and differential rotation, we combine equations \ref{eq:vconr}, \ref{eq:Hperpr}, and \ref{eq:nuconef2} to estimate the effective convective viscosity 
\beq
\label{eq:nuconrotef}
\nu_{\rm con,ef} \sim h_{\rm con} v_{\rm con} \bigg(\frac{\omega_{\rm shear}}{\omega} \bigg)^2 \bigg( \frac{\omega_{\rm con}}{\Omega} \bigg)^{4/5} \, . 
\eeq
which applies when $\omega_{\rm shear} > \omega_{\rm con,rot}$.

The energy damping rate produced by the convective viscosity is 
\begin{equation}
    \dot{E}_{\rm con} \sim \omega_\alpha^2 |a_\alpha|^2 \int_0^R  4 \pi r^2 \rho \nu_{\rm con,ef} |\vec{\nabla} \vec{\xi}_\alpha|^2 dr \, ,
\end{equation}
where $a_\alpha$ is the dimensionless mode amplitude. This corresponds to a mode damping rate
\begin{equation}
\label{eq:tdampcon}
    t_{\rm damp,con}^{-1} = \frac{\dot{E}_{\rm con}}{E_\alpha} \sim \frac{\omega_\alpha^2}{E_{\rm bind}} \int_0^R 4 \pi r^2 \rho \nu_{\rm con,ef} |\vec{\nabla} \vec{\xi}_\alpha|^2 dr \, ,
\end{equation}
and we have used the mode normalization condition of equation \ref{eq:norm}. 

In our planetary models, we calculate $\omega_{\rm shear}$ due to latitudinal differential rotation via $\omega_{\rm shear} = d u_{\rm wind}/(r d\theta) \sim \Delta u_{\rm wind}/(R\Delta \theta)$, where $u_{\rm wind}$ is the measured rotation speed at the surface of the planet. By examining the measured surface wind speeds (\citealt{guillot:23} and references therein), we estimate $\Delta u_{\rm wind} \sim 350$ m/s, 200 m/s, and 300 m/s for Saturn, Jupiter, and Uranus respectively, with corresponding angular widths of $\Delta \theta \sim 0.7$, 0.35, and 0.96 rad. These crude order-of-magnitude estimates ignore the more intricate zonal wind pattern observed for Jupiter and Saturn but are sufficient for our purposes. 

The differential rotation is believed to be confined to the outer non-electrically conductive layers of the planets \citep{liu:08}, giving way to rigid rotation in the interior. This entails a shear layer at this transition, with $\omega_{\rm shear} \sim d u_{\rm wind}/dr \sim \Delta u_{\rm wind}/\Delta R$, where $\Delta R$ is the radial width of the transition layer. The base of the inferred transition layers occur at $r/R \approx$ 0.86, 0.965, and 0.97 in Saturn, Jupiter, and Uranus, respectively (\citealt{Dewberry2021,galanti:21,guillot:23} and references therein). We assume the thickness of this layer is 1/4 of its depth, such that $\Delta R/R \approx 0.035$, 0.00875, 0.0075 in Saturn, Jupiter, and Uranus. We assume a linear shear profile so that the value of $\omega_{\rm shear} = \Delta u_{\rm wind}/\Delta R$ is constant in this layer.

With these estimates, the value of $\omega_{\rm shear}$ is zero in the rigidly rotating interior, and it obtains its largest values due to the radial shear in the transition layers, giving $f_{\rm shear} = \omega_{\rm shear}/(2 \pi) \sim 0.03 \, {\rm mHz}$, $0.05 \, {\rm mHz}$, and $0.16 \, {\rm mHz}$ in Saturn, Jupiter, and Uranus, respectively. Hence, in our models, most of the convective viscous damping is produced in the transitional layers between the rigidly rotating interior and differentially rotating envelope.

Figure \ref{fig:tdamprot} compares different estimates of damping times for the f modes of our Saturn model for different types of convective viscosity. We use $h_{\rm con} = H$, where $H$ is the pressure scale height. Damping times are very long for ordinary convective viscosity due to the low-frequency suppression (equation \ref{eq:nucon_sup}). However, the effect of shear is to greatly enhance the convective viscosity, decreasing damping times by a factor of $\sim 10-10^3$ relative to normal convection.


\subsection{Rings}
\label{sec:rings}

\begin{figure}
\includegraphics[scale=0.36]{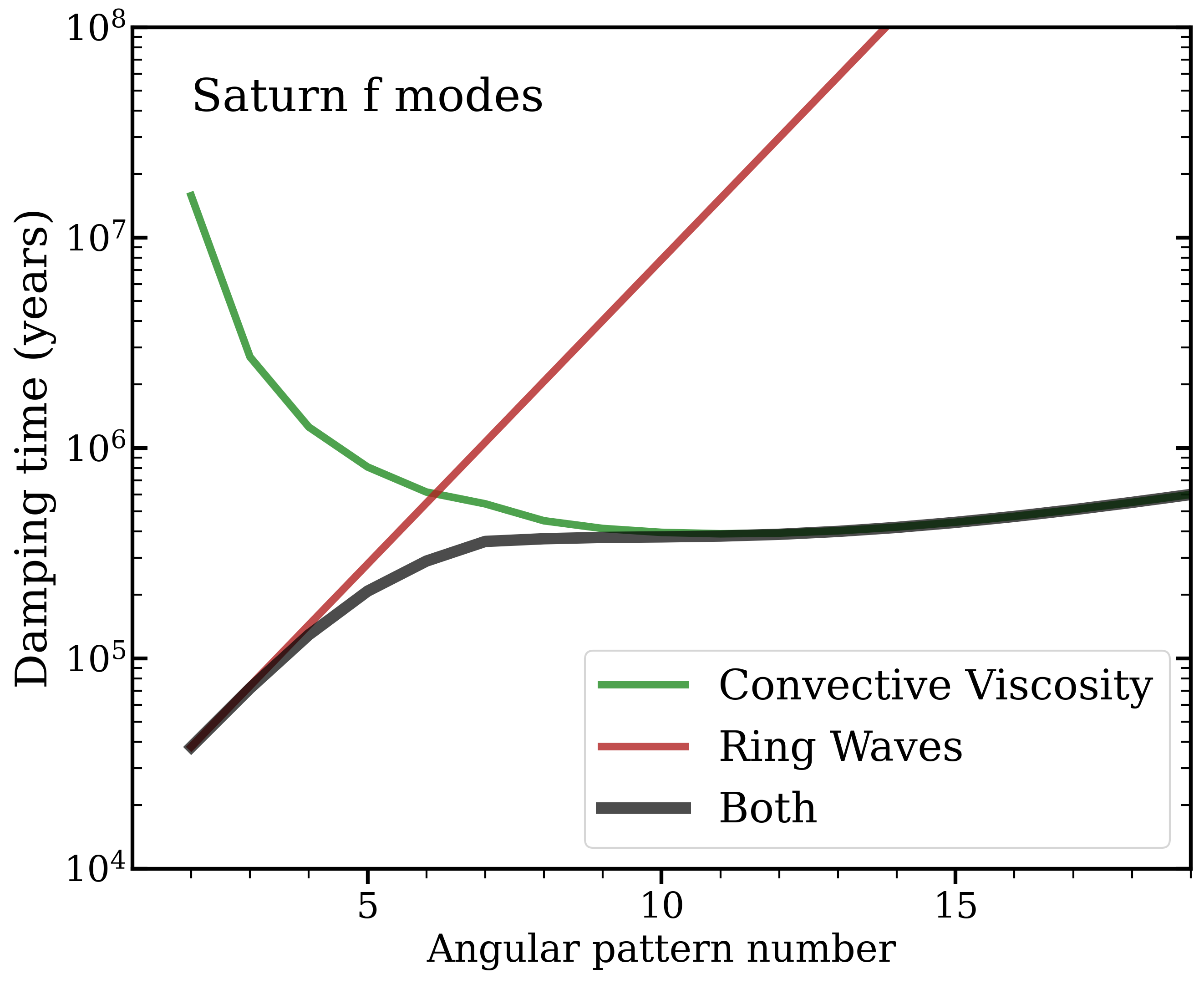}
\caption{\label{fig:tdamp} Damping time of $\ell=m$ f~modes of Saturn as a function of $\ell$. The green line is damping due to convective viscosity (equation \ref{eq:tdampcon}), the maroon line accounts for damping via interaction with the rings (equation \ref{eq:tdampring}), and the black line is the combined effect of both.}
\end{figure}

For f~modes containing a Lindblad resonance in the rings, density wave excitation within the rings may contribute a significant amount of damping, as found by \cite{wu:19}. We approximate the damping times in their table with the formula 
\begin{equation}
\label{eq:tdampring}
    t_{\rm damp,ring} = 10^4 \, e^{2 \ell/3} \, {\rm yr} \, .
\end{equation}
The exponential dependence arises from the fact that the mode gravitational perturbation at a given semi-major axis falls off exponentially with $\ell$. It is important to remember that most modes do not suffer this damping because they do not have Lindblad resonances in the ring \citep{marley:93}. Equation \ref{eq:tdampring} applies only to f~modes with $\ell \simeq m$. 

Figure \ref{fig:tdamp} compares damping produced by convective viscosity and ring wave excitation for the f~modes of our Saturn model. We find that the rings dominate the damping for modes with $\ell \lesssim 6$, while convective viscosity dominates for $\ell \gtrsim 6$. According to these estimates, the modes always have lifetimes longer than $\sim 3 \times 10^4 \, {\rm yr}$, corresponding to very large quality factors $Q \sim t_{\rm damp}/P \sim 10^{10}$.

\subsection{Radiative Diffusion}

Radiative diffusion is very important in stars, but it is less effective in planets due to their long thermal times. Nevertheless, it is likely to be important for p~modes. \cite{markham:18}, \cite{wu:19}, and \cite{zanazzi:25} have analytically estimated wave damping via radiative diffusion. 

We estimate damping rates due to radiative diffusion by computing non-adiabatic oscillation modes of planetary models using the GYRE pulsation code \citep{townsend:13}. We first build planetary models of Jupiter, Saturn, and Uranus using the MESA stellar evolution code (see Appendix \ref{sec:mesa}). We then compute non-adiabatic f~modes and p~modes using GYRE. We find the damping times are very long for f~modes ($t_{\rm damp} \sim {\rm Gyr}$), larger than those produced by convective viscosity, in agreement with \cite{wu:19}.

However, for p~modes, radiative diffusion can be the most important source of damping, as shown in Figure \ref{fig:tdamp-p}. For Saturn, we find that radiative diffusion is the dominant damping mechanism for p~modes with frequencies larger than $f \sim 0.6 \, {\rm mHz}$, corresponding to radial orders larger than $n_p \gtrsim 3$. The damping rate scales approximately as $\gamma_{\rm diff} \propto \omega^{4}$, corresponding to a quality factor $Q = \omega/\gamma \sim 10^{14} [\omega/(10^{-3} \, {\rm rad/s})]^{-3}$. We find a similar scaling with $\omega$ (but slightly weaker) compared to \cite{zanazzi:25}. As discussed in that work, the steep scaling with $\omega$ arises largely due to larger wavenumbers that scale approximately as $k \approx \omega^2/g$ near the surface \citep{goldreich:94}. A local radiative diffusion scaling as $\gamma \sim K k^2$, where $K$ is the thermal conductivity, leads to greater damping rates for higher frequency modes. The different scaling with $\omega$ than the analytic results of \cite{markham:18} arises because they focused on radiative damping in the optically thin atmosphere, while \cite{wu:19} focused on f~modes rather than p~modes.

\subsection{Wave leakage into atmosphere}
\label{sec:tunneling}

Another source of damping is wave propagation into the atmosphere. This occurs for waves with $\omega \gtrsim \omega_{\rm ac}$, where $\omega_{\rm ac} \approx c_s/(2 H)$ is the acoustic cutoff frequency near the surface of the planet. To estimate the acoustic cutoff frequency, we use $H = P/(\rho g) = c_s^2/\gamma g$. We assume a diatomic ideal gas with $c_s^2 = \gamma k_B T/\mu$ with $\gamma=7/5$ and mean molecular weight $\mu = 2.5 m_p$ appropriate for a helium and molecular hydrogen mixture. We use $T = 100 \, {\rm K}$, 120 K, and 76 K for Saturn, Jupiter, and Uranus yielding $f_{\rm ac} = \omega_{\rm ac}/(2 \pi) \approx 1.8 \, {\rm mHz}$, 3.9 mHz, and 1.7 mHz. Above the acoustic cutoff frequency, waves leak into the atmosphere, carrying energy at a rate \begin{equation}
\label{eq:edampac}
    \dot{E}_{\rm leak} \sim  R^4 \rho_s \omega^2 |a_\alpha|^2 \xi_{r,s}^2 c_{s,s} \, ,
\end{equation}
where the $s$ subscript should be evaluated near Saturn's surface. We multiply this rate by an adhoc factor of $\big[ 1/2 + 1/2 \tanh\big(40 (\omega-\omega_{\rm ac})/\omega_{\rm ac}\big) \big]$ to slightly smooth the transition between modes below and above the acoustic cutoff frequency.

\subsection{Storms}
\label{sec:stormdamp}

Storms have been suggested \citep{markham:18} as an important excitation mechanism, which we discuss in Section \ref{sec:storms}. They could also be important for mode damping.
We treat damping via storms the same way as damping via convection, but with a different convective velocity $v_{\rm st}$ (see Section \ref{sec:storms}). The mode energy dissipation rate produced while the storm is active is thus
\begin{equation}
    \dot{E}_{\rm st} \sim \nu_{\rm st} \rho_{\rm st} H_{\rm st}^3 \omega_\alpha^2 |a_\alpha|^2 |\vec{\nabla} \vec{\xi}_\alpha(r_{\rm st})|^2 \, ,
\end{equation}
where quantities with st subscript are evaluated at the depth of the storm. Over the lifetime of the storm, $t_{\rm st} \sim \pi/\omega_{\rm st}$, the energy damped from the mode can be written
\begin{equation}
    E_{\rm damp,st} \sim (\omega_\alpha t_{\rm st})^2 E_{\rm st} |a_\alpha|^2 |\vec{\nabla} \vec{\xi}_\alpha(r_{\rm st})|^2 \, ,
\end{equation}
where $E_{\rm st} \sim \rho_{\rm st} H_{\rm st}^3 v_{\rm st}^2$ is the total kinetic energy in the storm.

\begin{figure}
\includegraphics[scale=0.36]{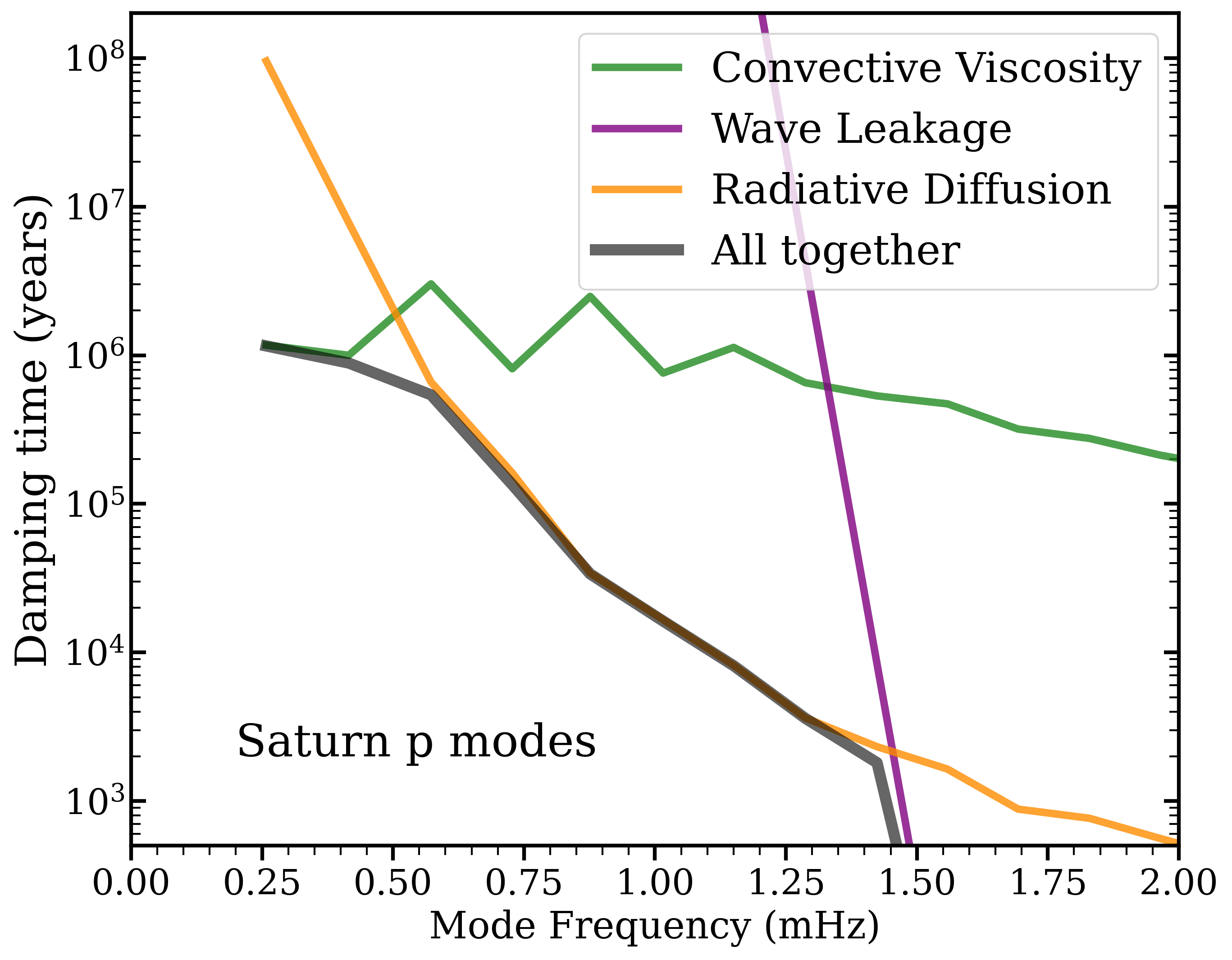}
\caption{\label{fig:tdamp-p} The damping time of $\ell=2$ p~modes of Saturn, as a function of mode frequency. The green line shows damping due to convective viscosity (equation \ref{eq:tdampcon}), the orange line accounts for damping due to radiative diffusion, and the purple line accounts for wave damping above the acoustic cutoff frequency (equation \ref{eq:edampac}). Modes with frequencies larger than $f \gtrsim f_{\rm ac} \sim 1.5 \, {\rm mHz}$ are strongly damped due to the latter effect.}
\end{figure}

Given a storm recurrence rate of $t_{\rm recur,st}$, the long-term average energy dissipation rate is thus
\begin{equation}
\label{eq:edampstorm}
    \dot{E}_{\rm st} \sim (\omega_\alpha t_{\rm st})^2 \frac{E_{\rm st}}{t_{\rm recur,st}} |a_\alpha|^2 |\vec{\nabla} \vec{\xi}_\alpha(r_{\rm st})|^2 \, .
\end{equation}

We find that the damping produced by storms is always sub-dominant to the mechanisms discussed above, for Jupiter, Saturn, and Uranus. However, it is only smaller by $\sim$1 order of magnitude for p~modes, so storm damping could be important if we have underestimated its strength, or overestimated the damping produced by other effects.

\subsection{Results}

\begin{figure}
\includegraphics[scale=0.36]{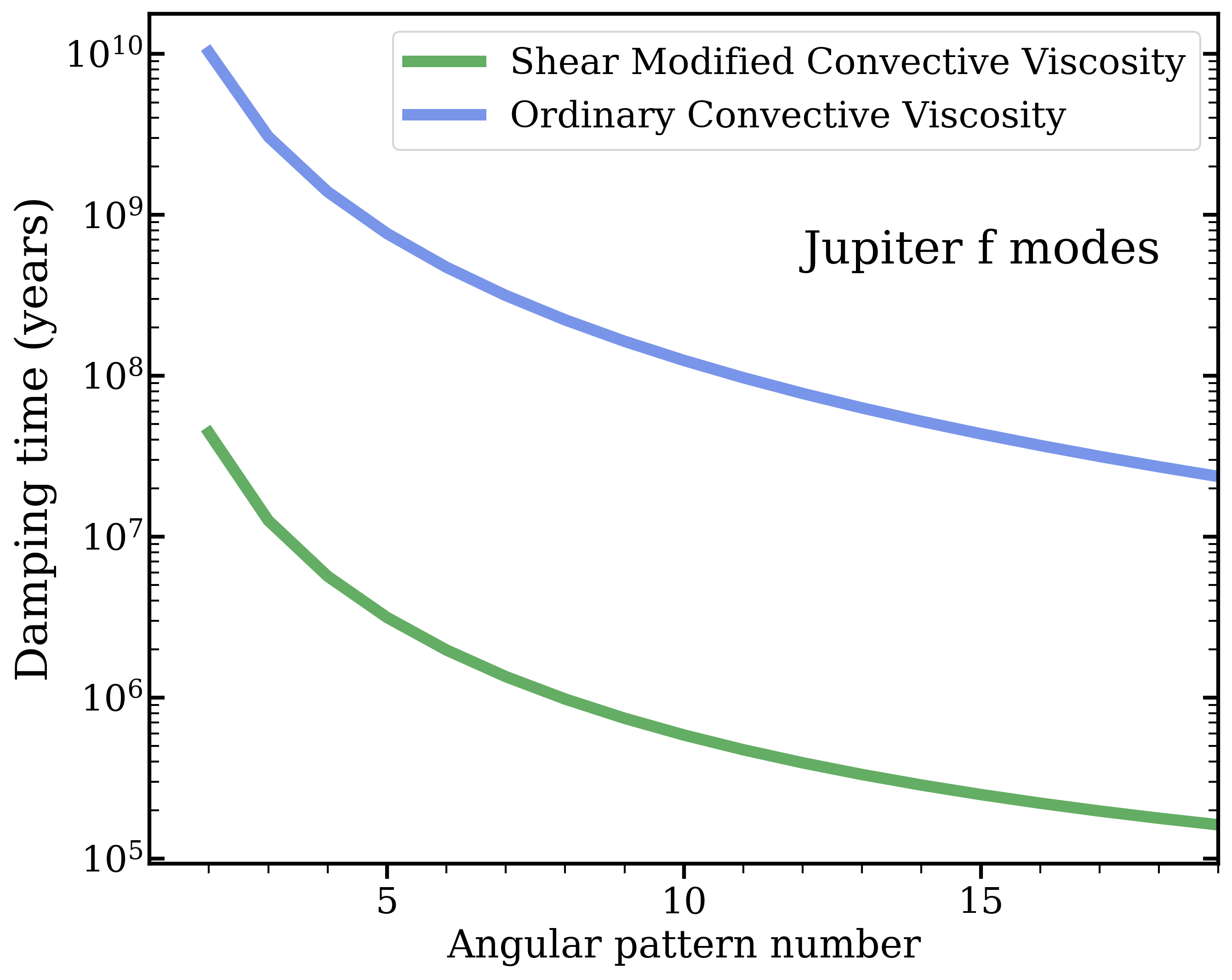}
\includegraphics[scale=0.36]{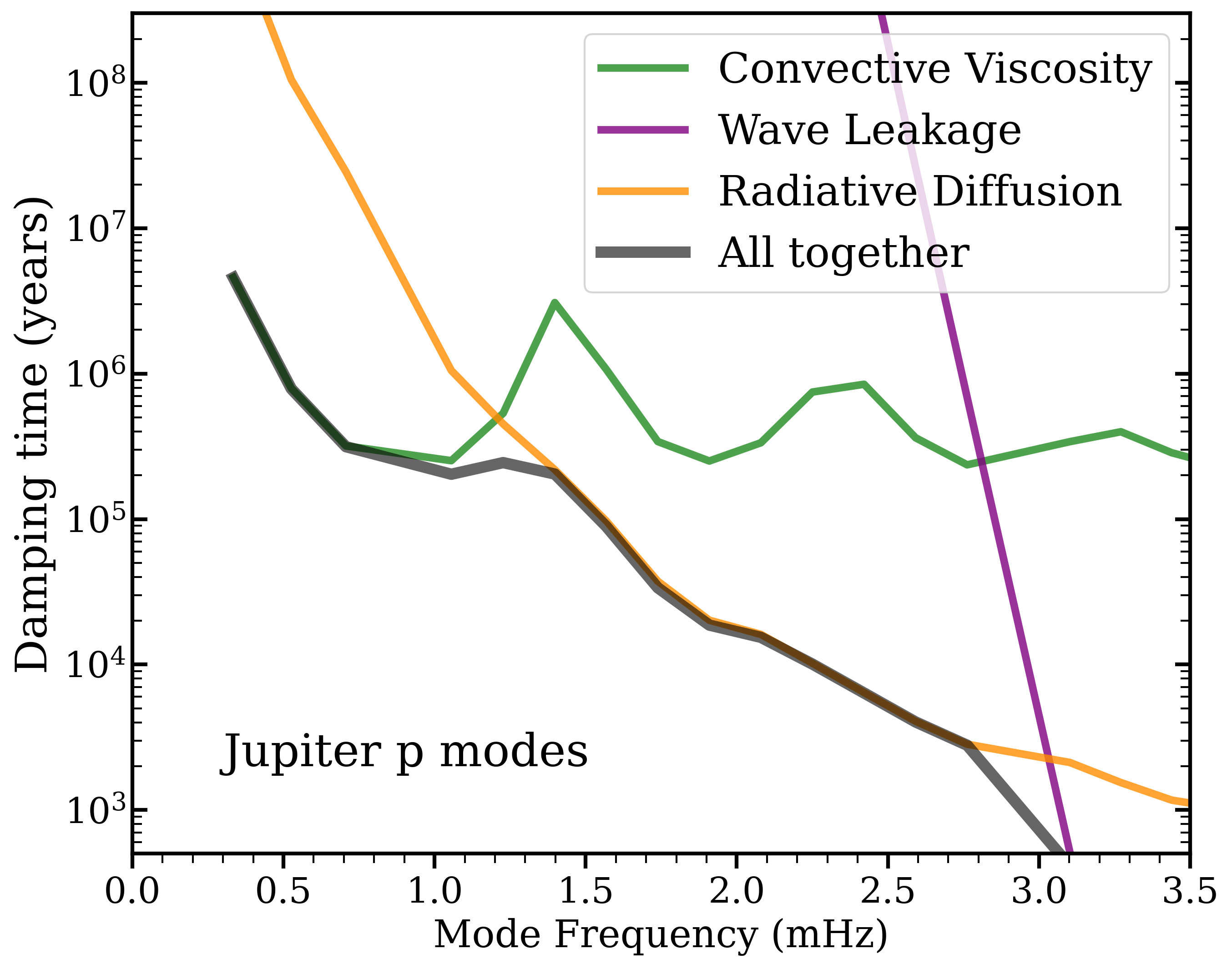}
\caption{\label{fig:tdampJupiter} Similar to Figures \ref{fig:tdamprot} and \ref{fig:tdamp-p}, now showing the damping time of modes in Jupiter. {\bf Top:} Damping time of $\ell=m$ f~modes as a function of $\ell$ due to different models of convective viscosity. {\bf Bottom:} Damping time of $\ell=2$ p~modes as a function of angular frequency.}
\end{figure}

Figures \ref{fig:tdamp} and \ref{fig:tdamp-p} show predicted damping times for Saturn, finding that the rings damp the lowest-$\ell$ f~modes, convective viscosity damps high-$\ell$ f~modes and low-order p~modes, while radiative diffusion damps high-order p~modes. 

Figure \ref{fig:tdampJupiter} shows predicted mode damping times in Jupiter. They are similar to those for Saturn in most respects. An important difference is that the low-$\ell$ f~modes cannot be damped by rings for Jupiter, greatly increasing the predicted value of $t_{\rm damp}$. Another difference is that Jupiter's shear layer lies closer to the surface, at $r/R \simeq 0.97$ rather than $r/R \simeq 0.88$, changing where the convective viscosity peaks. The wavy structure in the p~mode damping times via convection are due to this localized damping: modes with minima or maxima in their velocity eigenfunctions at the shear layer are more weakly damped via viscosity because the value of $|\vec{\nabla} \vec{\xi}_\alpha|$ is small in that layer. 

\begin{figure}
\includegraphics[scale=0.36]{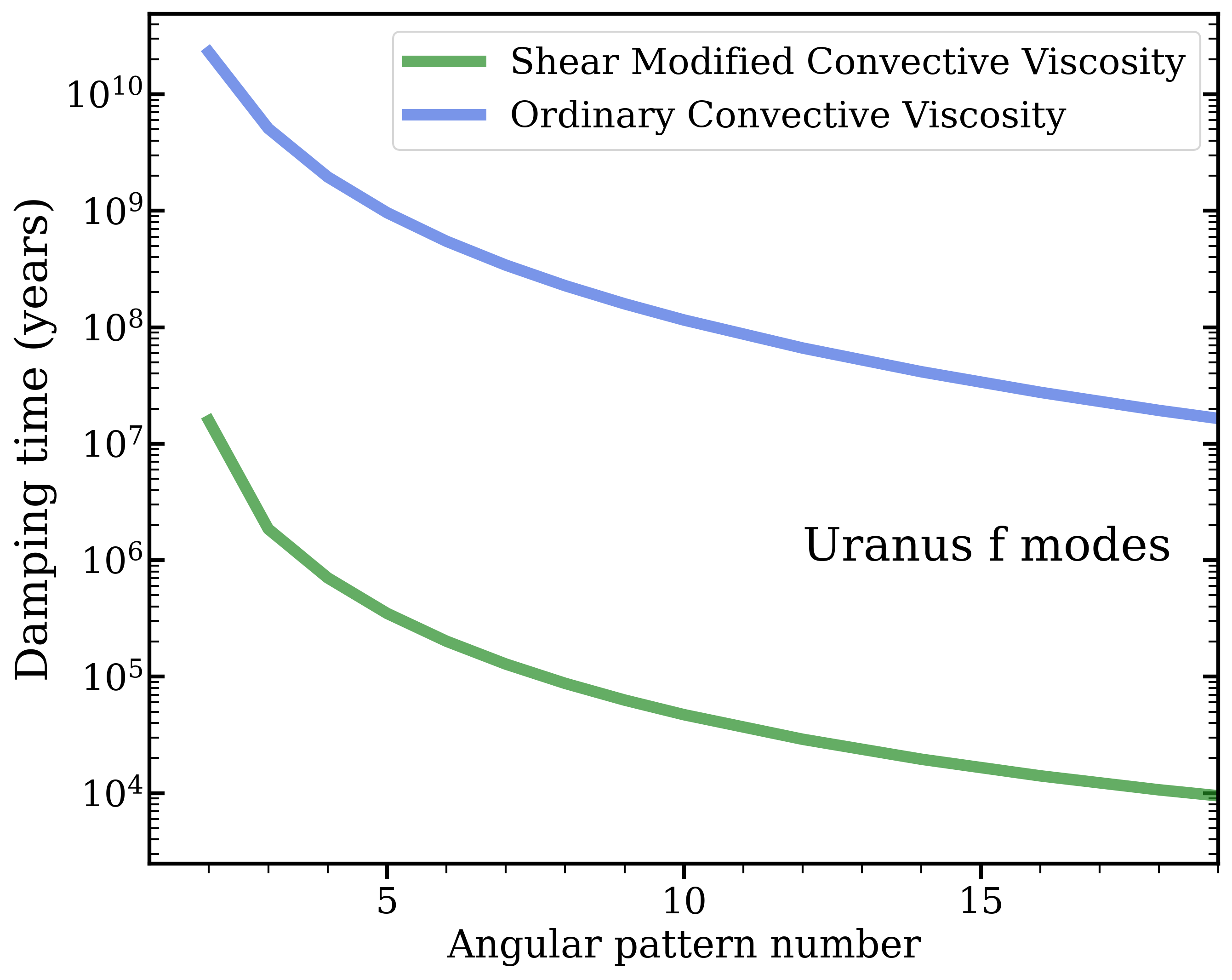}
\includegraphics[scale=0.36]{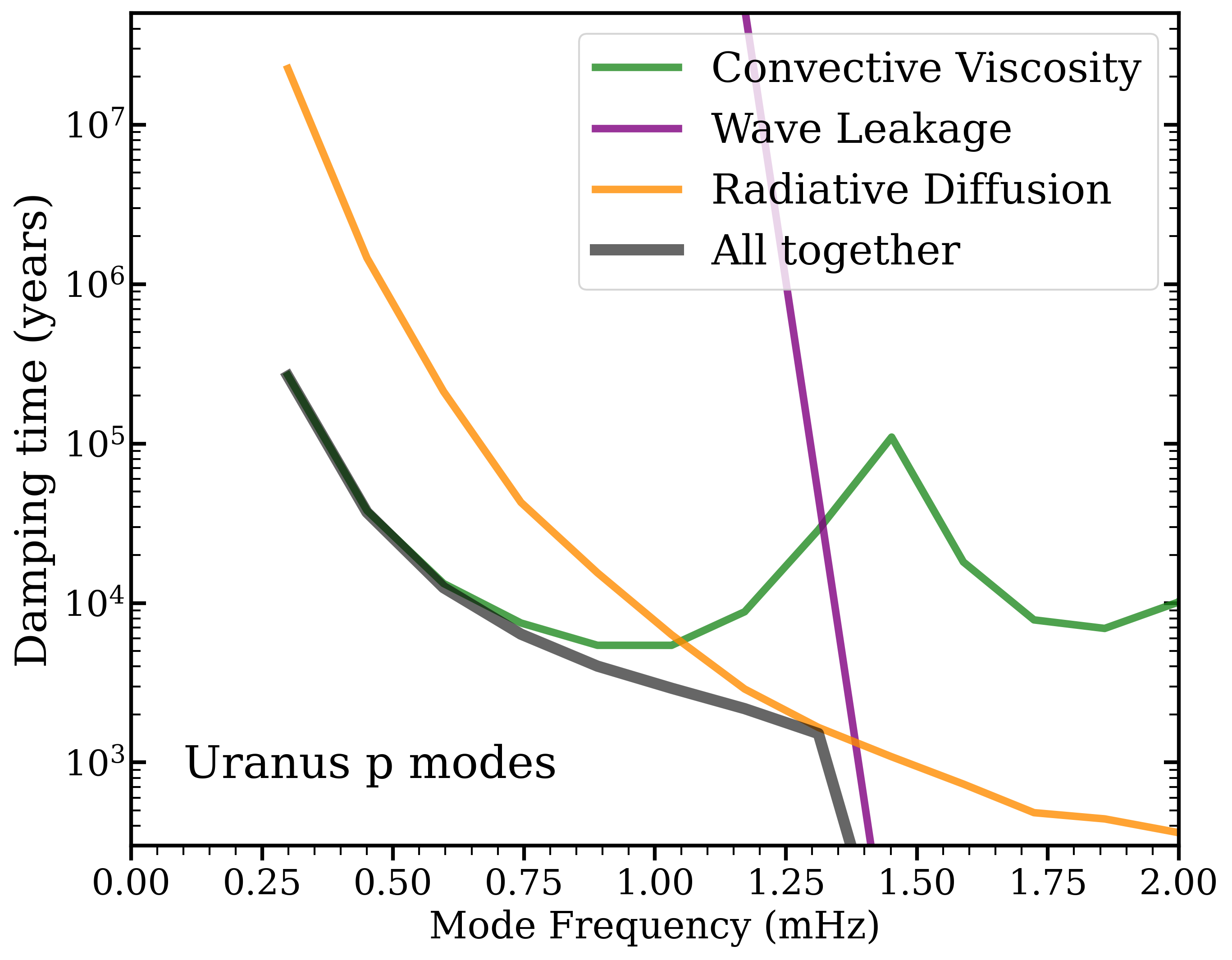}
\caption{\label{fig:tdampUranus} Similar to Figure \ref{fig:tdampJupiter} for modes in Uranus.}
\end{figure}

We apply the same calculations to Uranus, with mode damping times shown in Figure \ref{fig:tdampUranus}. The f~mode damping times are predicted to be shorter in Uranus, especially for the high-$\ell$ f~modes with damping times of $\sim \! 10^4$ years. The higher damping results from the larger shear in our Uranus model due to its high-speed winds, shallow shear layer, and smaller radius. Similarly, the p~modes of our model have shorter damping times of $\sim \! 10^4$ years, with less contribution from radiative diffusion. This will result in lower predicted mode amplitudes in Uranus compared to Jupiter and Saturn.

\section{Excitation Mechanisms}
\label{sec:excitation}

\subsection{Convection}
\label{sec:convection}

The heat flux from giant planets drives turbulent convection that can stochastically excite planetary oscillation modes. We adapt the classic calculations of \cite{goldreich:90,goldreich:94} for solar p~modes to planetary f~modes and p~modes. A crucial difference for planets is that the convective turnover frequencies in planets are much smaller, typically below any p~mode frequencies. Additionally, the low luminosities of planets means the convective eddy energies are small, and we shall see that these two effects greatly diminish mode excitation via convective turbulence. 

Following \cite{goldreich:90} and \cite{goldreich:94}, the rate at which the amplitude of a mode is excited, per unit time and volume, is
\begin{equation}
    \label{eq:adot}
    \frac{d\dot{a}_\alpha}{dV} \sim \frac{\omega_\alpha \rho v_h^2}{E_{\rm bind}} |\vec{\nabla} \vec{\xi}_\alpha | \, ,
\end{equation}
This is ``quadrupole" forcing from Reynold's stresses as opposed to the monopole and dipole forcing discussed below. Convection is assumed to remain coherent over one convective turnover time $\sim 1/\omega_h$, with a convective eddy size $\sim \! h$.

The change in mode amplitude due to one convective eddy is found by integrating equation \ref{eq:adot} over one convective turnover time and one convective eddy volume. If the mode frequency $\omega_\alpha$ is lower than the convective turnover frequency $\omega_h$, this yields
\begin{equation}
    \label{eq:da}
    \Delta a_\alpha \sim \frac{\omega_\alpha}{\omega_h} \frac{\rho h^3 v_h^2}{E_{\rm bind}} |\vec{\nabla} \vec{\xi}_\alpha | \, ,
\end{equation}
This corresponds to a change in energy $\Delta E_\alpha = |\Delta a_\alpha|^2 E_{\rm bind}$. The corresponding power input to the mode is 
\begin{align}
    \label{eq:power}
    \dot{E}_\alpha &\sim \frac{\omega_\alpha^2}{\omega_h} \frac{\rho^2 h^6 v_h^4}{E_{\rm bind}} |\vec{\nabla} \vec{\xi}_\alpha |^2 \nonumber \\
    & \sim \frac{\omega_\alpha^2}{\omega_h} \frac{E_h^2}{E_{\rm bind}} |\vec{\nabla} \vec{\xi}_\alpha |^2 \, ,
\end{align}
where $E_{\rm h} \sim \rho h^3 v_h^2$ is the kinetic energy of a convective element with size $h$. 

Since each convective element is assumed to be uncorrelated, the amplitude change of equation \ref{eq:da} due to each element is uncorrelated. Stochastic excitation due to $N$ convective elements thus produces an average amplitude change of $\sim \sqrt{N} \Delta a_\alpha$, or an energy change of $\sim N \Delta E_\alpha$, so the power input due to many convective elements scales linearly with the number of elements.

The total energy input per unit radius is obtained by summing over the convective elements in a shell of thickness $dr$, $dN = 4 \pi r^2 dr/h^3$, giving
\begin{equation}
    \label{eq:power_int}
    \frac{d \dot{E}_\alpha}{dr} \sim \frac{\omega_\alpha^2}{\omega_h} \frac{4 \pi r^2 \rho^2 h^3 v_h^4}{E_{\rm bind}} |\vec{\nabla} \vec{\xi}_\alpha |^2 \, ,
\end{equation}
The total power input into a mode is then given by integrating equation \ref{eq:power_int} over radius.

\subsubsection{Low-frequency convection}
\label{sec:lowfreq}

Mode excitation is highly suppressed when the convective turnover frequency $\omega_h$ is below the mode frequency $\omega_\alpha$, because integration of equation \ref{eq:adot} over one convective turnover time nearly vanishes since the mode oscillates many times during this period. The form of this suppression is unclear and depends on the spatial/frequency spectrum of convective turbulence (see disscussion in \citealt{samadi:11}). For example, in a simple model where the convection is modeled as a gaussian pulse of duration $1/\omega_h$, i.e., $v_h(t) \sim v_h e^{-(\omega_h t)^2}$, we would obtain a suppression factor of $\Delta a_\alpha$ of $\sim \! e^{-(\omega_\alpha/\omega_h)^2}$ relative to equation \ref{eq:da}. This would produce very little mode excitation when $\omega_h \lesssim \omega_\alpha$. 

In the model of \cite{goldreich:94}, a shallower power-law suppression is obtained by considering smaller convective eddies in a turbulent cascade. Assuming Kolmogorov turbulence, the convective velocity at scale $h$ is $v_h \sim v_{\rm con} (h/H)^{1/3}$, where $v_{\rm con}$ is the convective velocity from equation \ref{eq:vcon}. The associated convective turnover frequencies at scale $h$ is $\omega_h = v_h/h \sim \omega_{\rm con}(h/H)^{-2/3}$. The largest eddies that can contribute to mode excitation are those with $\omega_h \sim \omega_\alpha$, giving $h \sim H (\omega_{\alpha}/\omega_{\rm con})^{-3/2}$, and convective velocities $v_h \sim v_{\rm con} (\omega_\alpha/\omega_{\rm con})^{-1/2}$. Using these relations in equation \ref{eq:power_int}, we obtain 
\begin{equation}
    \label{eq:power_int_om}
    \frac{d \dot{E}_\alpha}{dr} \sim  \frac{\omega_\alpha^2}{\omega_{\rm con}} \bigg(\frac{\omega_\alpha}{\omega_{\rm con}}\bigg)^{-15/2} \frac{4 \pi r^2 \rho^2 H^3 v_{\rm con}^4}{E_{\rm bind}} |\vec{\nabla} \vec{\xi}_\alpha |^2 \, ,
\end{equation}
This estimate is valid in regions where $\omega_\alpha \gtrsim \omega_{\rm con}$, and exhibits the classic $(\omega_{\rm con}/\omega_\alpha)^{15/2}$ suppression of \cite{goldreich:94} when the convective turnover frequency is small. But it may not always be valid because it assumes Kolmogorov turbulence, which may not be appropriate for planetary convection zones, especially when effects of rotation and differential rotation are considered.

\subsubsection{Rotating convection}

Let us now consider how rotation alters the estimate of stochastic excitation above. As discussed around equation \ref{eq:vconr}, the convective velocities $v_h$ and eddy sizes $h$ are thought to be reduced by the effect of Coriolis forces, reducing the stochastic excitation power. However, the corresponding convective turnover frequency, $\omega_{\rm con,rot}$ is higher by a factor of $(\Omega/\omega_{\rm con})^{2/5}$ from equation \ref{eq:omconr}, increasing stochastic excitation power in planets.

To estimate mode excitation rates in this regime, we assume that convective eddies maintain their relative dimensions and that their velocities are Kolmogorov on smaller scales. This means $v_{h,{\rm rot}} = v_{\rm con,rot} (h/H)^{1/3}$ with $v_{\rm con,rot}$ given by equation \ref{eq:vconr}, $h_z = H (h/H)$, and $h_\perp = H_{\perp,{\rm rot}} (h/H)$ with $H_{\perp,{\rm rot}}$ given by equation \ref{eq:Hperpr}. Running through the same arguments leading up to equation \ref{eq:power_int_om}, we obtain
\begin{align}
    \label{eq:power_int_om_r}
    \frac{d \dot{E}_{\alpha,{\rm rot}}}{dr} &\sim \frac{\omega_\alpha^2}{\omega_{\rm con,rot}} \bigg(\frac{\omega_\alpha}{\omega_{\rm con,rot}}\bigg)^{\!\!-15/2} \nonumber \\
    & \times \frac{4 \pi r^2 \rho^2 H H_{\perp,{\rm rot}}^2 v_{\rm con,rot}^4}{E_{\rm bind}} |\vec{\nabla} \vec{\xi}_\alpha |^2 \nonumber \\
    &\sim \frac{d \dot{E}_{\alpha}}{dr} \bigg(\frac{\omega_{\rm con}}{\Omega}\bigg)^{-3/5} \, ,
\end{align}
where the second line follows from substituting our relations for rotating convection, and $d E_\alpha/dr$ is the expression from equation \ref{eq:power_int_om}.
Hence the net effect of rotation is to produce a small enhancement of mode excitation when $\omega_{\rm con} < \Omega < \omega_\alpha$.

\subsubsection{Including shear}

We previously argued that differential rotation substantially enhances convective viscosity by elongating convective eddies in the $\phi$-direction, or by increasing the effective convective turnover frequency to $\omega_{\rm shear}$. The same effect would in principle occur for stochastic mode excitation via Reynold's stresses. However, this enhancement will not exist for resonant eddies with $\omega_h \sim \omega_\alpha > \omega_{\rm shear}$. Therefore we do not expect any alteration to the excitation rate, unless the large-scale (non-resonant) eddies are involved in mode excitation.

\subsubsection{Results}

\begin{figure}
\includegraphics[scale=0.36]{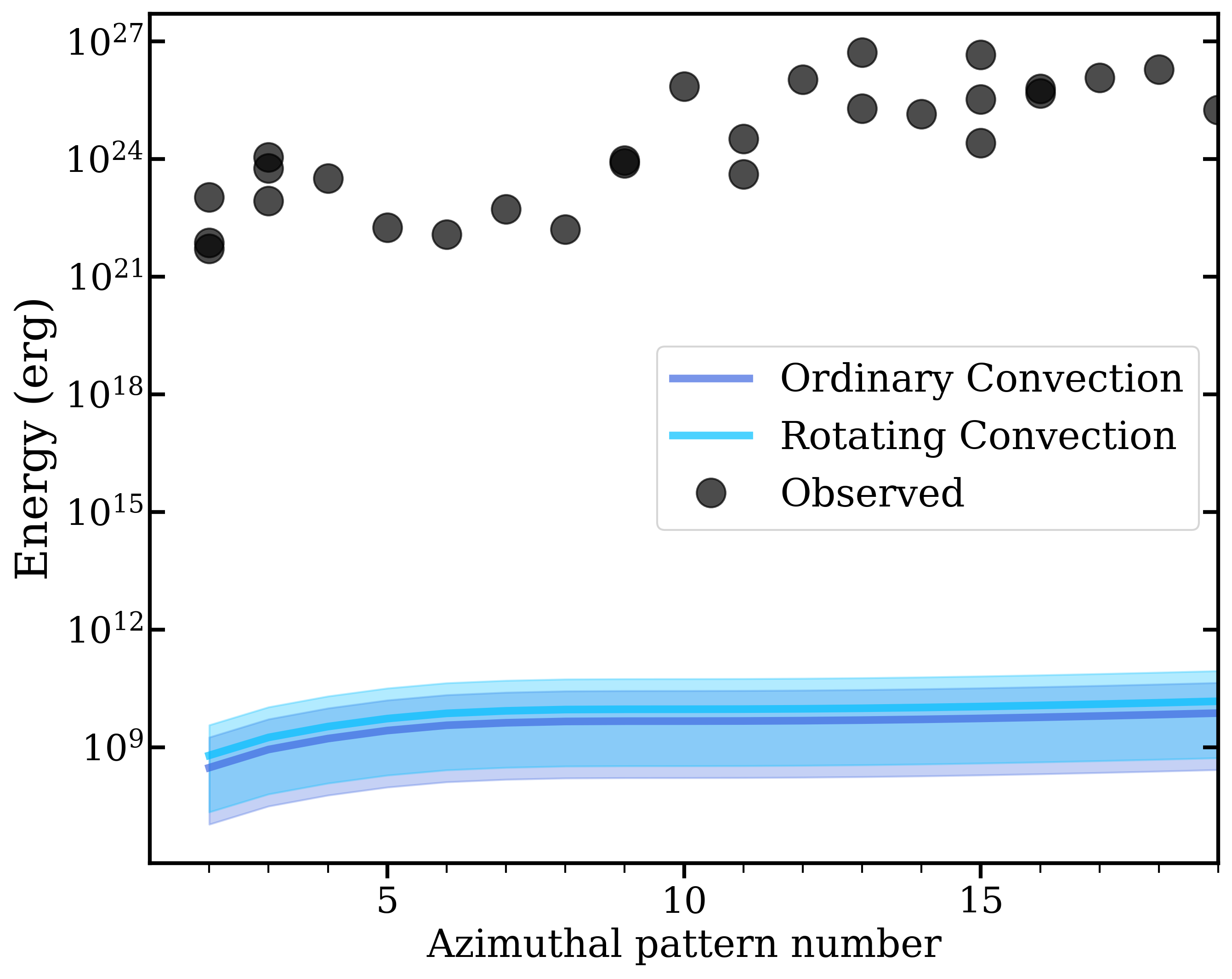}
\caption{\label{fig:Emodescon} {\bf Top:} Energies of Saturn's \bm{$\ell \sim m$} f~modes as a function of angular wavenumber $\ell$. Black dots are observed modes from \citealt{afigbo:25}. The curves show predicted energies for rotating and non-rotating convective excitation, and the damping rates shown in Figure \ref{fig:tdamp}. Convective excitation cannot account for the observed mode energies.}
\end{figure}

We compute mode excitation rates via the non-rotating (equation \ref{eq:power_int_om}) and rotating (equation \ref{eq:power_int_om_r}) models of convection. The corresponding equilibrium mode energy is then
\begin{equation}
\label{eq:ealpha}
    E_\alpha = \dot{E}_\alpha t_{{\rm damp},\alpha} \, ,
\end{equation}
and we consider damping from both shear-modified convection and the rings, equations \ref{eq:tdampcon} and \ref{eq:tdampring}. Figure \ref{fig:Emodescon} shows predicted and observed mode energies for Saturn's f~modes. The convective excitation models fall short of observed mode energies by $\sim$14 orders of magnitude. We conclude that ordinary convection cannot account for excitation of Saturn's f~modes, unless one of our assumptions is terribly wrong. Other mechanisms for mode excitation discussed below are much more promising.

\subsection{Storms}
\label{sec:storms}

Storms may be a very important excitation mechanism in giant planets, due to their ability to release large amounts of energy during short amounts of time. This is enabled by chemistry, because energy can be stored in the latent heat of evaporated species (e.g., water, ammonia, or even silicates). During a storm, this energy is rapidly released into heat that generates kinetic energy (wind), which can then excite oscillation modes via Reynold's stresses. Storms are more effective than convection because the energy of a storm can be much larger than that of an equally sized convective element, enhancing excitation since it scales as $E_h^2$ (equation \ref{eq:power}). The larger energy of storms also produces higher turbulent turnover frequencies, circumventing the suppression that affects low-frequency convective motions. 

Storms may be especially energetic in gaseous planets because of convective inhibition by condensable species such as water vapor \citep{guillot:95}. Since these species have larger mean molecular weight than the surrounding hydrogen/helium, and since more of the vapor can be dissolved at larger depths, it can produce a stabilizing mean molecular weight gradient that counteracts the unstable thermal gradient that drives convection. As heat is lost from above due to radiative cooling, the thermal gradient steepens. \cite{leconte:17} and \cite{friedson:17} show that double-diffusive instabilities can be suppressed by condensation. The temperature gradient will steepen until either the internal heat simply diffuses outwards, or until overturning convection occurs, instigating a storm. As convective elements rise up and cool, the condensates rain out, decreasing the mean molecular weight of the rising elements, accelerating their rise \citep{li:15} and intensifying the storm.

Observed storms on Saturn can have power output temporarily comparable to Saturn's total luminosity. Storm-driven wind speeds are typically $v_{\rm st} \sim 100 \, {\rm m/s}$ \citep{garciamelendo:13}, although peak speeds at the beginning of the storm may be substantially higher, $v_{\rm st} \sim 300 \, {\rm m/s}$ \citep{garciamelendo:17}. Very large storm sizes up to $\sim \! 10^4$ km do exist \citep{fischer:11}, but this typically occurs after an extended phase of geostrophic adjustment lasting for days or weeks. This is likely not relevant for mode excitation because the associated time scales are much longer than f~mode and p~mode periods. Therefore, for the purposes of mode excitation, such large storms may be better modeled as a combination of many smaller storms of size $H_{\rm st} \sim \! H$, which are uncorrelated with each other. 

We estimate the excitation rate produced by storms in a similar to way to that of convection, due to Reynold's stresses produced by the turbulent wind motions present during a storm (see also \citealt{markham:18,wu:19}). We model the storm as a single convective element of height and horizontal size $H_{\rm storm} \sim H$ in the layers they occur. Storms are thought to originate near the bottom of the condensation layer for a given species, so species with higher condensation temperatures (e.g., silicates) drive storms in deeper layers where the scale height and storm size are larger.

Based on the intersection of adiabats with condensation curves, we estimate the base of the water storm layer to occur at temperatures $T_{\rm storm} \approx$ 300 K, 350 K, and 400 K in Jupiter, Saturn, and Uranus, respectively. We estimate that silicate storms occur at temperatures of $T_{\rm storm} \approx$ 2000 K, 2200 K, and 2400 K. To estimate storm energies, we evaluate relevant quantities at the depth corresponding to these temperatures.

There are two energy sources for storms: thermal and chemical energy. The latent heat of condensation converts chemical to thermal energy when an evaporated species condenses. The energy released per unit mass is $f_v L_v$, where $f_v$ is the volatile mass fraction and $L_v$ is the latent heat per unit mass. We use $L_v = 2.3 \times 10^{10} \, {\rm erg}/{\rm g}$ for water and $L_v = 1.2 \times 10^{11} \, {\rm erg}/{\rm g}$ for silicates. Based on the measured abundances from \cite{guillot:23}, we assume a water mass fraction of 4\% in our Jupiter model (about 3x solar metallicity) and a silicate mass fraction of 0.5\%. We double these values in our Saturn model, and quadruple these values in our Uranus model, but their surface compositions are quite uncertain.

The second energy source is thermal energy in a moist parcel due to convective inhibition \citep{guillot:95}, which becomes available as the condensate rains out. The energy available is the thermal energy difference between a moist and dry adiabat. Following \cite{li:15}, the energy per unit mass is $\sim f_v c_v T_{\rm storm}$, where $c_v$ is the specific heat. It turns out that this thermal energy is roughly equal to the latent heat energy for water storms, and is roughly double the latent heat for rock storms. Like \citealt{markham:18} and \citealt{wu:19}, we use the latent heat energy in our estimates below, which is a conservative choice but yields the right order of magnitude.

The velocity of a storm's motions due to latent heat release is
\begin{equation}
    v_{\rm storm}^2 \sim \eta_{\rm storm}^2 f_v L_v \, ,
\end{equation}
where $\eta_{\rm storm}^2$ is the fraction of the latent heat that is converted to kinetic energy. We choose $\eta_{\rm storm} = 0.3$, yielding $v_{\rm storm} \sim 1.3 \times 10^4 \, {\rm cm}/{\rm s}$ for water storms and $v_{\rm storm} \sim 1.0 \times 10^4 \, {\rm cm}/{\rm s}$ for rock storms. These estimates are similar to wind speeds observed \citep{garciamelendo:13,sayanagi:13} in Saturn's storms, as well as those seen in simulations of water-driven storms \citep{garciamelendo:17,sanchezlavega:24}. Corresponding storm energies are $E_{\rm storm} = \pi \rho_{\rm storm} H_{\rm storm}^3 v_{\rm storm}^2 \sim 4 \times 10^{26} \, {\rm erg}$ for water storms and $E_{\rm storm} \sim  2 \times 10^{31} \, {\rm erg}$ for rock storms for our Saturn model. Our water storm energies are much larger than \cite{wu:19} while our rock storm energies are slightly smaller, but both estimates are highly uncertain.


Following the same procedure as Section \ref{sec:convection}, a single storm adds an energy to each mode of
\begin{align}
\label{eq:Ealpha_storm}
    \Delta E_\alpha &\sim \frac{\omega_\alpha^2}{\omega_{\rm storm}^2} \frac{\pi^2 \rho_{\rm storm}^2 H_{\rm storm,r}^6 v_{\rm storm}^4}{E_{\rm bind}} |\vec{\nabla} \vec{\xi}_\alpha|^2 \nonumber \\ 
    & \sim \frac{\omega_\alpha^2}{\omega_{\rm storm}^2} \frac{E_{\rm storm}^2}{E_{\rm bind}} |\vec{\nabla} \vec{\xi}_\alpha|^2\, ,
\end{align}
where we define $\omega_{\rm storm} = \pi v_{\rm storm}/H_{\rm storm}$. We find that water storms typically have $\omega_{\rm storm}$ higher than f~mode frequencies, and comparable to low-order p~modes. However, rock storms typically have $\omega_{\rm storm} \lesssim \omega_{\alpha}$ for high-$\ell$ f~modes and p~modes. When $\omega_{\rm storm} < \omega_\alpha$, we apply the same $(\omega_{\rm storm}/\omega_\alpha)^{15/2}$ suppression used for convection (equation \ref{eq:power_int_om}). This hinders the ability of storms to excite p~modes, especially for rock storms.

We note that equation \ref{eq:Ealpha_storm} is different from the storm excitation rate assumed by \cite{wu:19}. They appeared to apply the dipole forcing (equation \ref{eq:power_imp_dip}) but with $E_{\rm imp}$ replaced by $E_{\rm storm}$. We believe this is not correct for storm excitation, which does not have a source of external momentum, and must therefore rely on monopole and/or quadrupole forcing. We suspect monopole forcing is weak because the storm speeds are subsonic, so little expansion/compression occurs. 

Unlike convective motions, storms are not continuous and have a long recurrence time $t_{\rm storm,recur}$ between events. To estimate this time scale, we follow \cite{li:15}, who equate the recurrence time scale to a thermal time at the condensation level,
\begin{equation}
\label{eq:ttherm}
    t_{\rm therm} \sim \frac{4 \pi R^2 H_{\rm storm} \rho_{\rm storm} f_v L_v}{L} \, ,
\end{equation}
where $L$ is the planet's intrinsic luminosity. \cite{li:15} estimated this time scale to be about 60 years in Saturn, whereas we find a higher value of $t_{\rm therm} \sim 300 \, {\rm yr}$ for water storms and $t_{\rm therm} \sim 2 \times 10^5 \, {\rm yr}$ for rock storms. 

However, equation \ref{eq:ttherm} is the thermal time at the weather layer, which should be much greater than the recurrence time between individual storms. The latter is smaller by the number of storms that can fit across the surface of the planet
\begin{align}
\label{eq:trecur}
    t_{\rm storm,recur} &\sim t_{\rm therm} \frac{\pi H_{\rm storm,\perp}^2}{4 \pi R^2} \nonumber \\
    & \sim \frac{E_{\rm storm}}{\eta_{\rm storm}^2 L} \, .
\end{align}
Note that our recurrence time scale estimate implies a fraction $\eta_{\rm storm}^2$ of the planet's heat flux is used to drive turbulent motion within storms. In Saturn, we estimate $t_{\rm storm,recur} \sim 2 \times 10^{-4} \, {\rm yr}$ for water storms in Saturn, and $t_{\rm storm,recur} \sim 8\, {\rm yr}$ for rock storms. Many water storms should be active at any given time, with occasional larger rock storms occurring on a time scale of $\sim$years-decades.

We speculate that giant storms with initial sizes $H_{\rm storm,\perp} \sim 1000 \, {\rm km}$ like Saturn's giant 2010 storm are actually driven by more energetic rock storms emerging from deeper in the planet, rather than large water storms as assumed in most studies. Our estimated rock storm size and recurrence time scale is similar to those observed for Saturn's giant storms, supporting this hypothesis. The predicted rock storm recurrence time scale in Jupiter is only $\sim$1 month, suggesting we may have seen many such events.

Using this recurrence time scale estimate, the time-averaged energy input rate of storms into a mode is 
\begin{align}
\label{eq:stormpower}
    \dot{E}_\alpha & \sim \frac{\omega_\alpha^2}{\omega_{\rm storm}^2} \frac{E_{\rm storm}^2}{E_{\rm bind}} |\vec{\nabla} \vec{\xi}_\alpha|^2 \frac{1}{t_{\rm storm,recur}} \nonumber \\
    & \sim \eta_{\rm storm}^2 L \frac{\omega_\alpha^2}{\omega_{\rm storm}^2} \frac{E_{\rm storm}}{E_{\rm bind}} |\vec{\nabla} \vec{\xi}_\alpha|^2 \, .
\end{align}
This estimate applies if $\omega_{\rm storm} \gtrsim \omega_\alpha$. If $\omega_{\rm storm} \lesssim \omega_\alpha$, then mode excitation will be suppressed as in Section \ref{sec:lowfreq}.

\subsection{Impacts}
\label{sec:impacts}

Cometary impacts have frequently been suggested to excite planetary oscillations (e.g., \citealt{kanamori:93,zahnle:94,lognonne:94,dombard:95}), and were also closely examined by \cite{wu:19} and \cite{zanazzi:25}. Impacts can excite oscillations via a monopole source due to the compression caused by a cometary impact, or a dipole source due to the momentum delivered by the impact. \cite{wu:19} estimate the amplitude change of a mode caused by each mechanism is
\begin{equation}
\label{eq:imp_mon}
    a_{\alpha,{\rm mon}} \sim \frac{E_{\rm imp}}{E_{\rm bind}} \vec{\nabla} \cdot \vec{\xi}_\alpha(R) \, .
\end{equation}
and 
\begin{equation}
\label{eq:imp_dip}
    a_{\alpha,{\rm dip}} \sim \frac{E_{\rm imp}}{E_{\rm bind}} \frac{\omega_\alpha}{\omega_{\rm dyn}} \frac{\xi_{\alpha,r} (R)}{R} \, .
\end{equation}
Note that we have altered their expressions to account for our normalization, which requires replacing their amplitude with $a_{\alpha,{\rm Wu}} = a_\alpha \omega_{\rm dyn}/\omega_\alpha$, and $\xi_{\alpha,{\rm Wu}} = \xi_\alpha \omega_\alpha/\omega_{\rm dyn}$, where $\omega_{\rm dyn} = \sqrt{GM/R^3}$. These results are also similar to \cite{zanazzi:25} as long as the impactor velocity is $v_{\rm imp} \simeq v_{\rm esc}$.

We disagree with \cite{wu:19} and \cite{zanazzi:25} on the following issue. They assume the impact occurs at a pole such that the value of $\xi_{\alpha,r} (R) \propto Y_{\ell,0}(\theta=0) = \sqrt{(2 \ell+1)/4 \pi}$. However, for an impact occurring at an arbitrary latitude, $\xi_{\alpha,r} (R)$ has (on average) no dependence on $\ell$, so this scaling is overestimated by a factor of $\sqrt{2\ell+1}$. Another way to reach the same conclusion is to realize that a polar impact excites only $m=0$ modes. In reality, for impacts at random latitudes, the same energy would be divided among the $2 \ell+1$ different values of $m$ for a given $\ell$, reducing the amplitude of each mode by a factor of $\sqrt{2\ell+1}$. 

Assuming random locations and phases of the impacts, each mode's amplitude changes by the sum of equations \ref{eq:imp_mon} and \ref{eq:imp_dip} during each impact, producing a random walk in mode amplitude. On average, the energy grows linearly with the number of impacts, i.e., the power input to a mode is
\begin{align}
\label{eq:power_imp_mon}
    \dot{E}_{\alpha,{\rm mon}} &\sim \frac{|a_{\alpha,{\rm mon}}|^2}{t_{\rm imp}} E_{\rm bind} \nonumber \\
    &\sim \frac{E_{\rm imp}^2}{E_{\rm bind} t_{\rm imp}} |\vec{\nabla} \cdot \vec{\xi}_\alpha(R)|^2 \, ,
\end{align}
where $t_{\rm imp}$ is the time between impacts. Additionally, 
\begin{align}
\label{eq:power_imp_dip}
    \dot{E}_{\alpha,{\rm dip}} &\sim \frac{|a_{\alpha,{\rm dip}}|^2}{t_{\rm imp}} E_{\rm bind} \nonumber \\
    &\sim \frac{E_{\rm imp}^2}{E_{\rm bind} t_{\rm imp}} \frac{\omega_\alpha^2}{\omega_{\rm dyn}^2} \frac{|\xi_{\alpha,r}(R)|^2}{R^2} \, ,
\end{align}
For f~modes, $\vec{\nabla} \cdot \vec{\xi}_\alpha \approx 0$ and so the dipole formula of equation \ref{eq:power_imp_dip} dominates. For p~modes, $\vec{\nabla} \cdot \vec{\xi}_{\alpha} \sim (\omega_\alpha^2/g)\xi_r$ near the planetary surface \citep{goldreich:94}, and we have $\dot{E}_{\rm mon} \sim \dot{E}_{\rm dip} (\omega_\alpha/\omega_{\rm dyn})^2$, so the monopole source term of equation \ref{eq:power_imp_mon} dominates. 

The impact rate $\dot{N}$ as a function of impactor size $D$ has been estimated by \cite{zahnle:03,singer:19,nesvorny:23,brasser:25}. They find
\begin{equation}
    \frac{d \dot{N}}{dD} \sim \frac{\dot{N}_0}{D_0} \bigg(\frac{D}{D_0}\bigg)^{-q} \, ,
\end{equation}
where $\dot{N}_0$ is the impact rate at a reference size $D_0$. The rate of impacts for objects above a size $D$ is
\begin{equation}
    \dot{N} (D) \sim \frac{\dot{N}_0}{q-1} \bigg(\frac{D}{D_0}\bigg)^{1-q} \, ,
\end{equation}
and the works above find $q \simeq 3$. We adopt the impact rates of \cite{nesvorny:23}, with $\dot{N}_0 \simeq 6.4 \times 10^{-3} {\rm yr}^{-1}$ (Jupiter), $\dot{N}_0 \simeq 2.6 \times 10^{-3} {\rm yr}^{-1}$ (Saturn), $\dot{N}_0 \simeq 2.4 \times 10^{-3} {\rm yr}^{-1}$ (Uranus), for an impactor size of $D_0 = 1 \, {\rm km}$. The true impact rate may be a factor of a few higher \citep{zahnle:03} or lower \citep{brasser:25}.

From equation \ref{eq:imp_mon} and \ref{eq:imp_dip}, the rate at which energy is transferred to modes is 
\begin{align}
\label{eq:power_imp_tot}
    \dot{E}_{\alpha,{\rm imp}} &\sim \bigg[ \bigg( \vec{\nabla} \cdot \vec{\xi}_\alpha(R) \bigg)^2 + \bigg(\frac{\omega_\alpha}{\omega_{\rm dyn}} \frac{\xi_{\alpha,r} (R)}{R}\bigg)^2 \bigg] \nonumber \\
    & \times \frac{1}{E_{\rm bind}} \int^{D_{\rm max}}_{D_{\rm min}} E_{\rm imp}(D)^2 \frac{d \dot{N}}{dD} dD  \, ,
\end{align}
where $E_{\rm imp}(D)$ is the impactor energy at size $D$. We assume spherical impactors of density $\rho_{\rm imp} = 1 \, {\rm g}/{\rm cm}^3$ and mass $M_{\rm imp} = (4 \pi/3) \rho_{\rm imp} D^3$, infalling at the planetary escape speed such that $E_{\rm imp}(D) = G M M_{\rm imp}/R$. Evaluating the integral of equation \ref{eq:power_imp_tot} yields
\begin{align}
\label{eq:power_imp_tot2}
    \dot{E}_{\alpha,{\rm imp}} &\sim \bigg[ \bigg( \vec{\nabla} \cdot \vec{\xi}_\alpha(R) \bigg)^2 + \bigg(\frac{\omega_\alpha}{\omega_{\rm dyn}} \frac{\xi_{\alpha,r} (R)}{R}\bigg)^2 \bigg] \nonumber \\
    & \times  \frac{\dot{N}_0}{7-q} \frac{E_{\rm imp,0}^2}{E_{\rm bind}} \bigg(\frac{D_{\rm max}}{D_0} \bigg)^{7-q} \, ,
\end{align}
where $E_{\rm imp,0}$ is the impact energy for an impactor of size $D_0$. 

The integral is dominated by the largest impactor sizes $D_{\rm max}$, because the energy imparted to modes scales as $E_{\rm imp}^2 \propto D^6$.
The sensitivity to $D_{\rm max}$ makes it tricky to predict the excitation rate due to impacts. The impactor size distribution is thought to steepen for $D \gtrsim 100 \, {\rm km}$ \citep{brasser:25}, so one could use $D_{\rm max} = 100 \, {\rm km}$. However, the rate of impacts for such large objects is roughly once every $10^7 \, {\rm yr}$, longer than the mode damping times estimated in Section \ref{sec:damping}. Hence it is unlikely for such impacts to contribute to currently observed mode amplitudes.

Instead, the impacts that determine the observed amplitudes are those that recur on time scales of $\sim t_{\rm damp}$, i.e., those for which $\dot{N} (D>D_*) \sim t_{\rm damp}^{-1}$. This implies a relevant impactor size
\begin{equation}
    D_{\rm max} \sim D_0 \bigg(\frac{\dot{N}_0 t_{\rm damp}}{q-1}\bigg)^{1/(q-1)} \, .
\end{equation}
Plugging this into equation \ref{eq:power_imp_tot2} gives 
\begin{align}
\label{eq:power_imp_tot3}
    \dot{E}_{\alpha,{\rm imp}} &\sim \bigg[ \bigg( \vec{\nabla} \cdot \vec{\xi}_\alpha(R) \bigg)^2 + \bigg(\frac{\omega_\alpha}{\omega_{\rm dyn}} \frac{\xi_{\alpha,r} (R)}{R}\bigg)^2 \bigg] \nonumber \\
    & \times \frac{\dot{N}_0}{7-q} \frac{E_{\rm imp,0}^2}{E_{\rm bind}} \bigg(\frac{\dot{N}_0 t_{\rm damp}}{q-1} \bigg)^\frac{7-q}{q-1} \, .
\end{align}
The expected mode energy $E_\alpha = \dot{E}_\alpha t_{\rm damp}$ is thus
\begin{align}
\label{eq:energy_imp_tot3}
    E_{\alpha,{\rm imp}} &\sim \bigg[ \bigg( \vec{\nabla} \cdot \vec{\xi}_\alpha(R) \bigg)^2 + \bigg(\frac{\omega_\alpha}{\omega_{\rm dyn}} \frac{\xi_{\alpha,r} (R)}{R}\bigg)^2 \bigg] \nonumber \\
    & \times \frac{(q-1)^\frac{q-7}{q-1}}{(7-q)} \frac{E_{\rm imp,0}^2}{E_{\rm bind}} \big(\dot{N}_0 t_{\rm damp} \big)^\frac{6}{q-1} \, .
\end{align}
For $q \approx 3$, the mode energy scales approximately as $E_{\alpha,{\rm imp}} \propto t_{\rm damp}^3$, so modes with longer lifetimes are expected to have much larger energies because they have a longer time baseline over which they could have been excited by a large impact. Unfortunately, the steep scaling with the impact rate $\dot{N}_0$ and the mode lifetime $t_{\rm damp}$, both of which are uncertain, makes it difficult to reliably predict mode energies.

\section{Mode Amplitudes}
\label{sec:amplitudes}

Here we compute expected mode energies given the excitation and damping rates discussed above. We use the values of $t_{{\rm damp},\alpha}$ computed in Section \ref{sec:damping} accounting for rotation and shear effects. Accounting for multiple damping methods simultaneously gives
\begin{equation}
    t_{{\rm damp},\alpha}^{-1} = t_{\rm damp, con}^{-1} + t_{\rm damp, ring}^{-1} + t_{\rm damp, diff}^{-1} + t_{\rm damp, leak}^{-1} \, .
\end{equation}
While other damping processes could contribute, one of those above (convective viscosity, interaction with rings, radiative diffusion, wave leakage) is always the dominant damping mechanism for the modes shown in this paper.

Since mode excitation is a stochastic process, the mode amplitude $a$ is expected to have a Gaussian distribution in expected amplitude,
\begin{equation}
    \frac{dP}{d a} = \frac{1}{\sqrt{2 \pi a_{\alpha}^2}} e^{-a^2/2 a_\alpha^2} \, ,
\end{equation}
where $a_\alpha$ is the expected mode amplitude determined by
\begin{equation}
    E_\alpha = a_\alpha^2 \frac{G M^2}{R} \, .
\end{equation}
The corresponding distribution of mode energy is
\begin{equation}
    \frac{dP}{dE} = \frac{1}{\sqrt{2 \pi E_\alpha E}} \, e^{-E/2 E_\alpha} \, .
\end{equation}
By integrating this distribution, one finds that the mode energy lies in the interval $0.023 E_\alpha < E < 3.0 E_\alpha$ 90\% of the time with the mode having lower energy 5\% of the time, and higher energy 5\% of the time. The corresponding range of mode amplitudes is $0.15 a_\alpha < a < 1.7 a_\alpha$. The median mode energy is $E_{\rm ex} = E_\alpha/2$, and median mode amplitude is $a_\alpha/\sqrt{2}$. We plot these median amplitudes/energies and 90\% confidence intervals in the results below.

\begin{figure}
\includegraphics[scale=0.36]{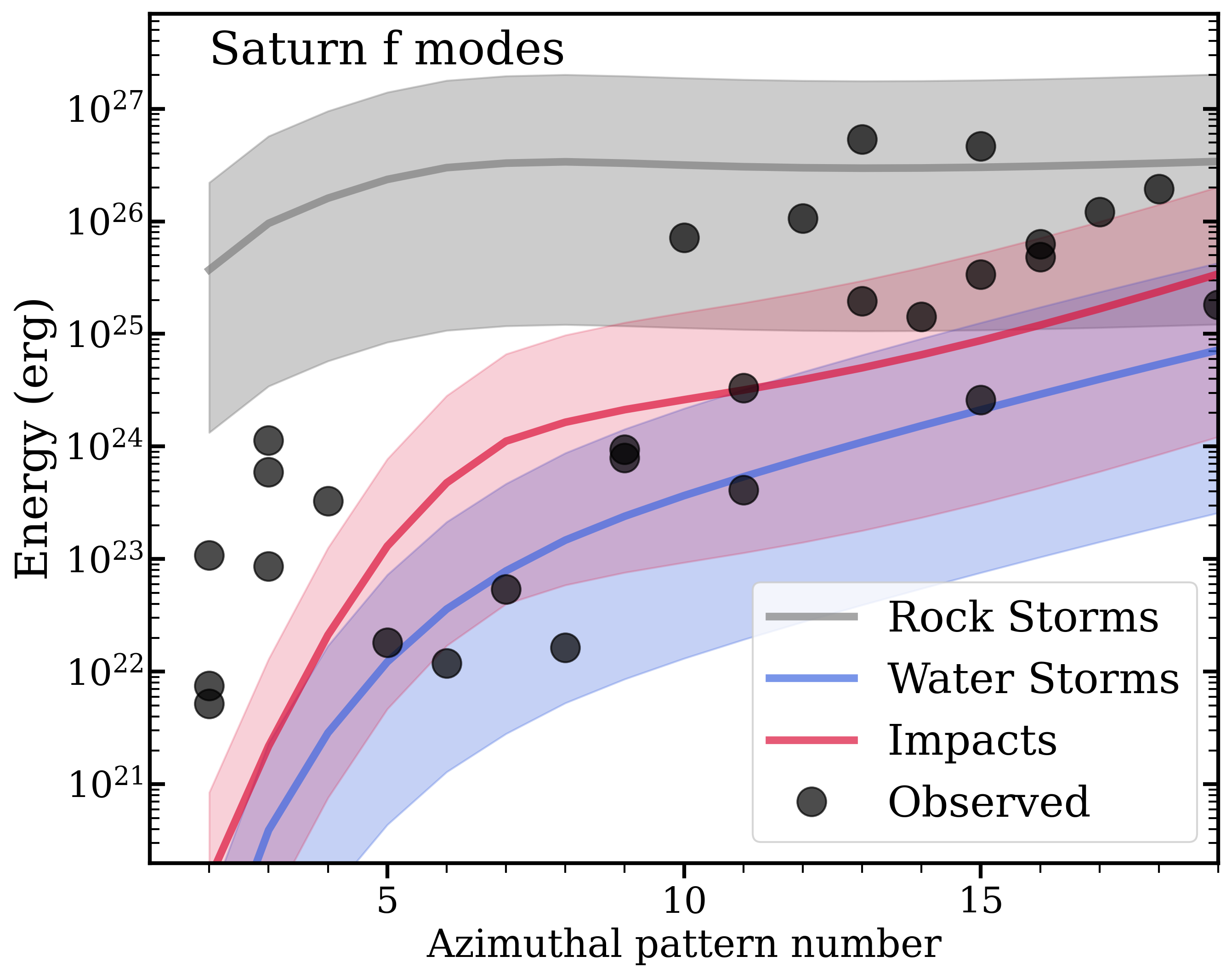}
\includegraphics[scale=0.36]{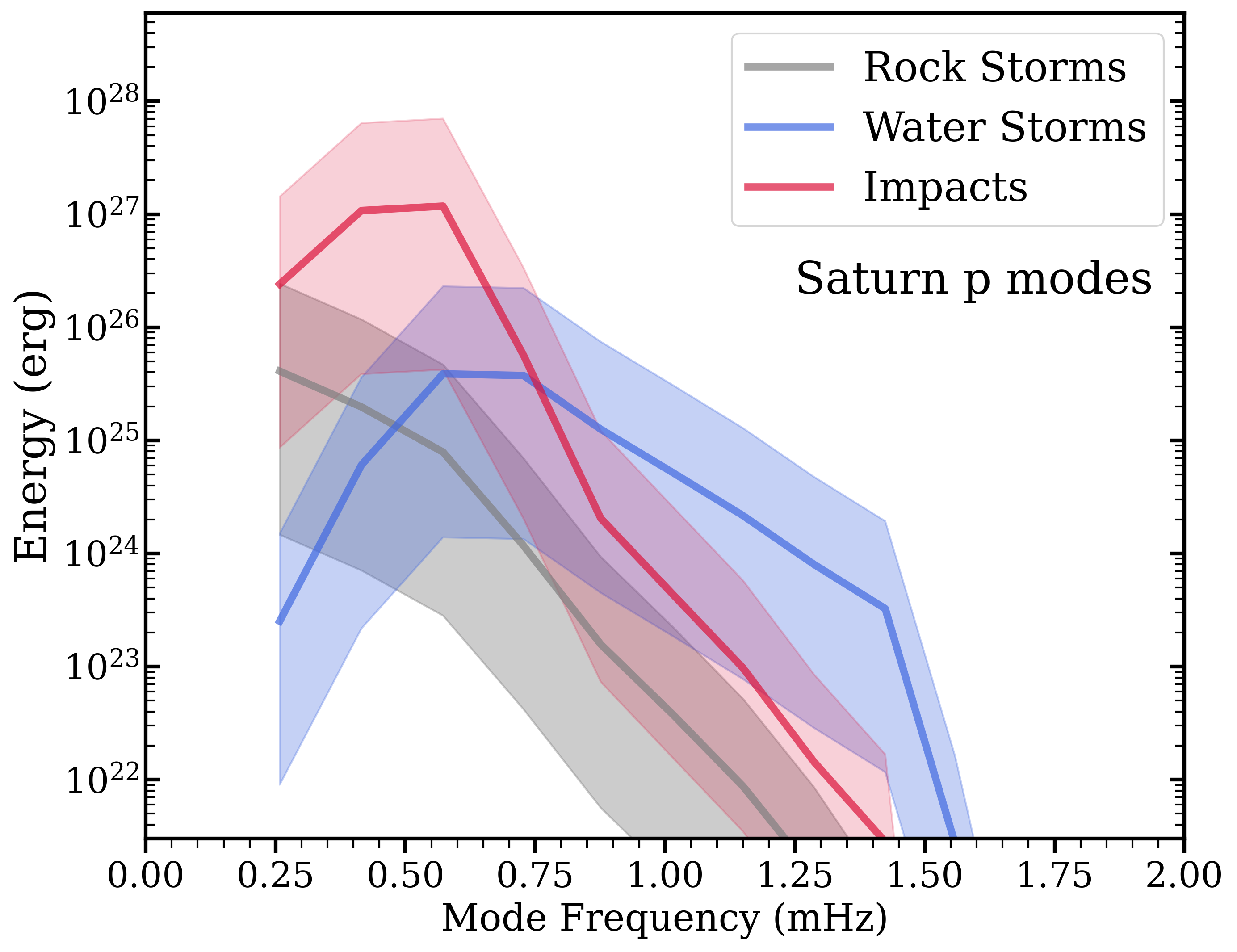}
\caption{\label{fig:Emodes} {\bf Top:} Energies of Saturn's \bm{$\ell \sim m$} f~modes as a function of angular wavenumber $\ell$. Black dots are observed modes from \citealt{afigbo:25}, the gray region is the predicted energy range due to stochastic excitation by rock storms, the blue region is the predicted energy range from water storms, and the red region is the energy range due to cometary impacts. Mode damping times are taken from Figures \ref{fig:tdamp} and \ref{fig:tdamp-p}. {\bf Bottom:} Same as the top panel, but now for Saturn's $\ell=2$ p~modes as a function of mode angular frequency $\omega$.}
\end{figure}

\subsection{Solar Calibration}
\label{sec:sun}

To make sure that our models for convective damping and excitation are reasonable, we compare to observations of the Sun's mode excitation and damping rates \citep{baudin:05}. We build a MESA model of the Sun and compute adiabatic oscillation modes as described in Section \ref{sec:modes}. We then evaluate equation \ref{eq:power_int} for a model of the Sun choosing a length scale of a pressure scale height $H$ for the sizes of the convective elements. Our model matches the measurements of power input to the modes (within a factor of $\sim$2) using convective velocities $v_{\rm con}$ from MESA's mixing length theory. Alternatively, evaluating the convective velocity as $v_{\rm con} = 0.5 [L/(4 \pi \rho r^2)]^{1/3}$ yields almost the same results. This is a satisfactory match given the large uncertainties inherent to our models.

Next, we estimate mode damping rates via equation \ref{eq:nucon} and \ref{eq:nucon_sup}, finding our model overestimates solar p~mode damping rates by a factor of $\sim$10. More sophisticated calculations of mode damping rates (e.g., \citealt{Grigahcene:2005,belkacem:12}) find that both a convective entropy perturbation and a turbulent pressure perturbation contribute to mode damping/driving, with nearly equal magnitudes but opposite signs. Our simple calculation is similar to the magnitude of their terms but would not correctly capture the cancellation between terms.

However, this cancellation may not occur in planets, because rotating convection (e.g., \citealt{stevenson:79}) does not have near-equipartition between kinetic and thermal energy. If the cancellation between terms does not occur in rotating and sheared convection, our rough estimate may not be too inaccurate. More detailed work will be necessary for a robust answer, but we proceed with our estimates of convective viscosity from equations \ref{eq:nucon}, \ref{eq:nucon_sup}, and \ref{eq:nuconef2}.

\subsection{Saturn}
\label{sec:saturn}

Figure \ref{fig:Emodes} shows the predicted mode energies for Saturn's f~modes and p~modes. Corresponding surface gravitational potential perturbations are shown in Figure \ref{fig:phimodes}, and surface radial velocities in Figure \ref{fig:vmodes}. We consider excitation due to water storms, rock storms, and impacts as described in Section {\ref{sec:excitation}. We also estimate observed mode energies, which are computed from the observed gravitational potential perturbations $\delta \Phi_{\rm obs}$ \citep{afigbo:25}. The computed mode energy is
\begin{equation}
    E_{\rm obs,\alpha} = \bigg( \frac{\delta \Phi_{\rm obs}}{\delta \Phi_\alpha(R)} \bigg)^2 E_{\rm bind} \, ,
\end{equation}
where $\delta \Phi_\alpha(R)$ are from our normalized model f~mode eigenfunctions. 

\begin{figure}
\includegraphics[scale=0.36]{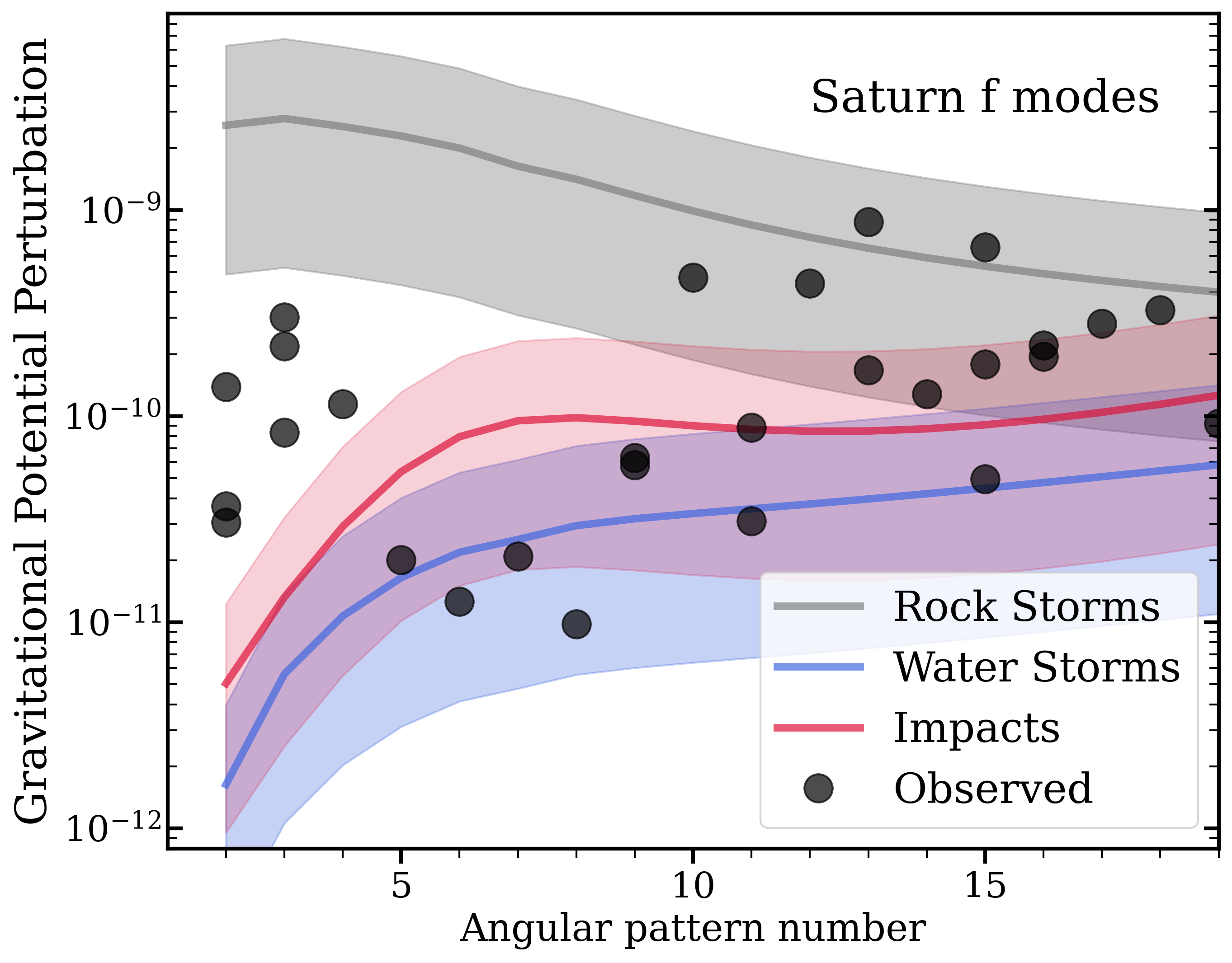}
\includegraphics[scale=0.36]{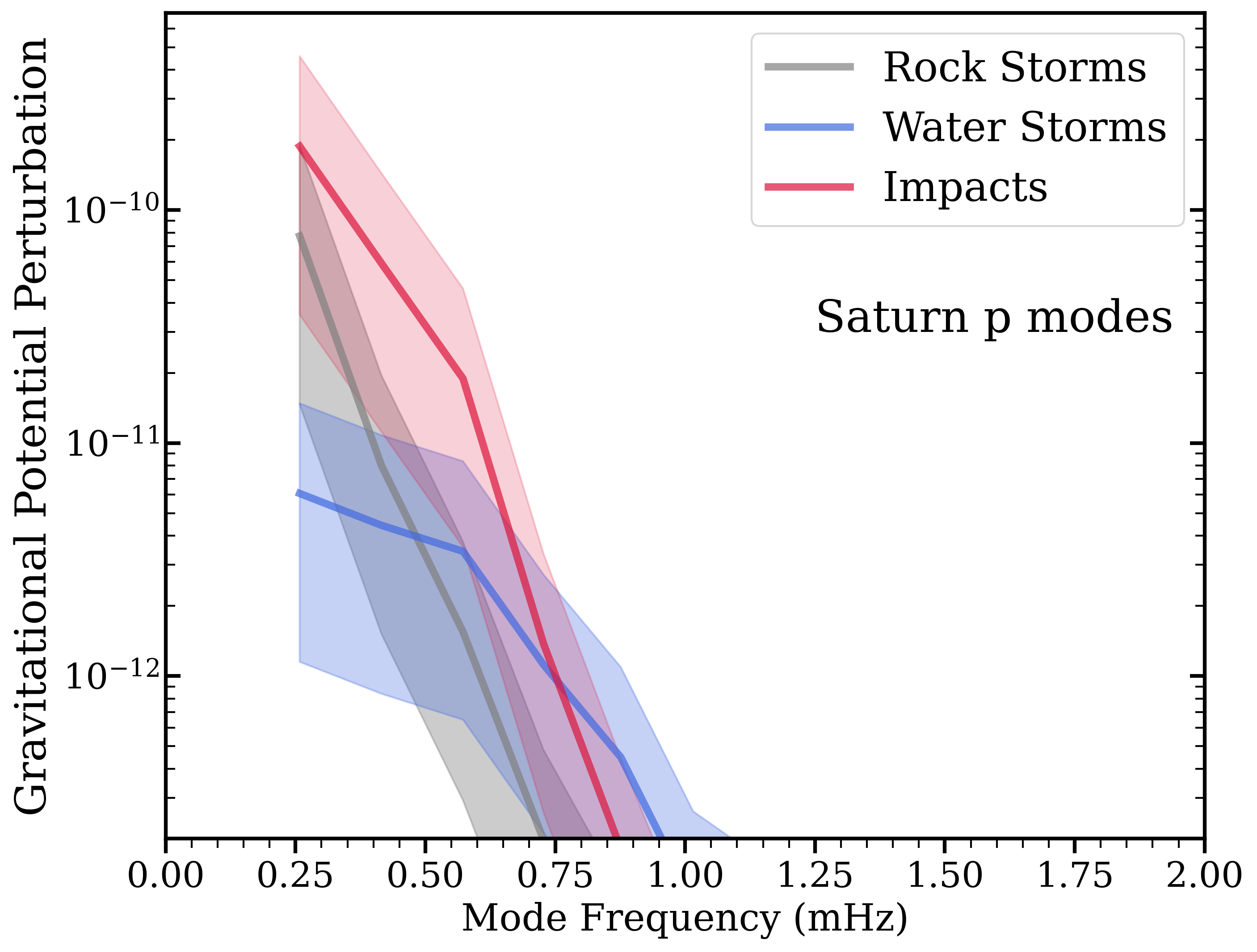}
\caption{\label{fig:phimodes} {\bf Top:} Similar to Figure \ref{fig:Emodes}, but now showing the amplitude of each of Saturn's \bm{$\ell \sim m$} f~mode gravitational potential perturbations, $\delta \Phi/(GM/R)$, measured at the surface of Saturn. {\bf Bottom:} The predicted gravitational perturbations of Saturn's $\ell=2$ p~modes.}
\end{figure}

For Saturn, the rock storm model can account for the observed mode energies for $\ell \geq 10$, but it overestimates the amplitudes of $\ell \lesssim 10$ modes. Note that the largest amplitude $m=2$ f mode on the Maxwell ringlet \citep{french:16} is not included in the \cite{afigbo:25} sample, and this mode has a substantially higher energy and gravitational perturbation than the three $\ell=2$ modes that are shown. The rock storm model may not be too far off for the $\ell=2$ f~mode.

The water storm model underpredicts most of the mode energies by a factor $\sim$10-100. The difference between water and rock storms arises primarily due to their different depths. Water storms occur at a depth of $\sim$14 bars in our Saturn model, where the density and scale height are small. Hence, the storm energy $E_{\rm storm}$ is small, decreasing the excitation rate since $\dot{E}_\alpha \propto E_{\rm storm}^2$ (equation \ref{eq:stormpower}). Rock storms have much larger energies due to larger expected depth of rock storms ($\sim$10 kbar), as pointed out by \cite{markham:18} and \cite{wu:19}. However, they have turnover frequencies $\omega_{\rm st} \sim 0.8 \omega_\alpha$ for our $\ell=2$ f~mode, meaning they have difficulty exciting high-$\ell$ f~modes or p~modes with larger frequencies.

The impact model predicts mode energies that are coincidentally very similar to the water storm models. The model uncertainties are very large because the predicted mode energy (equation \ref{eq:energy_imp_tot3}) scales approximately as the damping time $t_{\rm damp}^{3}$, which itself is quite uncertain.

We remind the reader that low-$\ell$ f~ modes with $m \neq \ell$ do not undergo ring damping and are thus predicted to have larger values of $t_{\rm damp}$ and much larger energies. Though not shown in the plots, the low-$\ell$ f~modes with $m \neq \ell$ are predicted to have larger energies by a factor of $\sim$100 compared to the rock storm model shown in Figure \ref{fig:Emodes}.


The bottom panel of Figure \ref{fig:Emodes} shows predicted energies of Saturn's p~modes. Here, the rock storm model predicts the lower amplitudes because $\omega_{\rm storm} \ll \omega_\alpha$ for p~modes. The water storms are more effective for p~modes because they have $f_{\rm storm} \sim 0.7 \, {\rm mHz}$, comparable to frequencies of low-order p~modes. The impact model predicts the largest mode energies because the impulsive impact can excite high-frequency modes, and because the compressive impact couples well with compressive p~modes (equation \ref{eq:power_imp_mon}). In all models, the predicted mode energy drops above $0.75 \, {\rm mHz}$ where radiative damping shortens mode lifetimes. The mode energy drops very sharply above 1.5 mHz at the acoustic cutoff frequency of Saturn.

\begin{figure}
\includegraphics[scale=0.36]{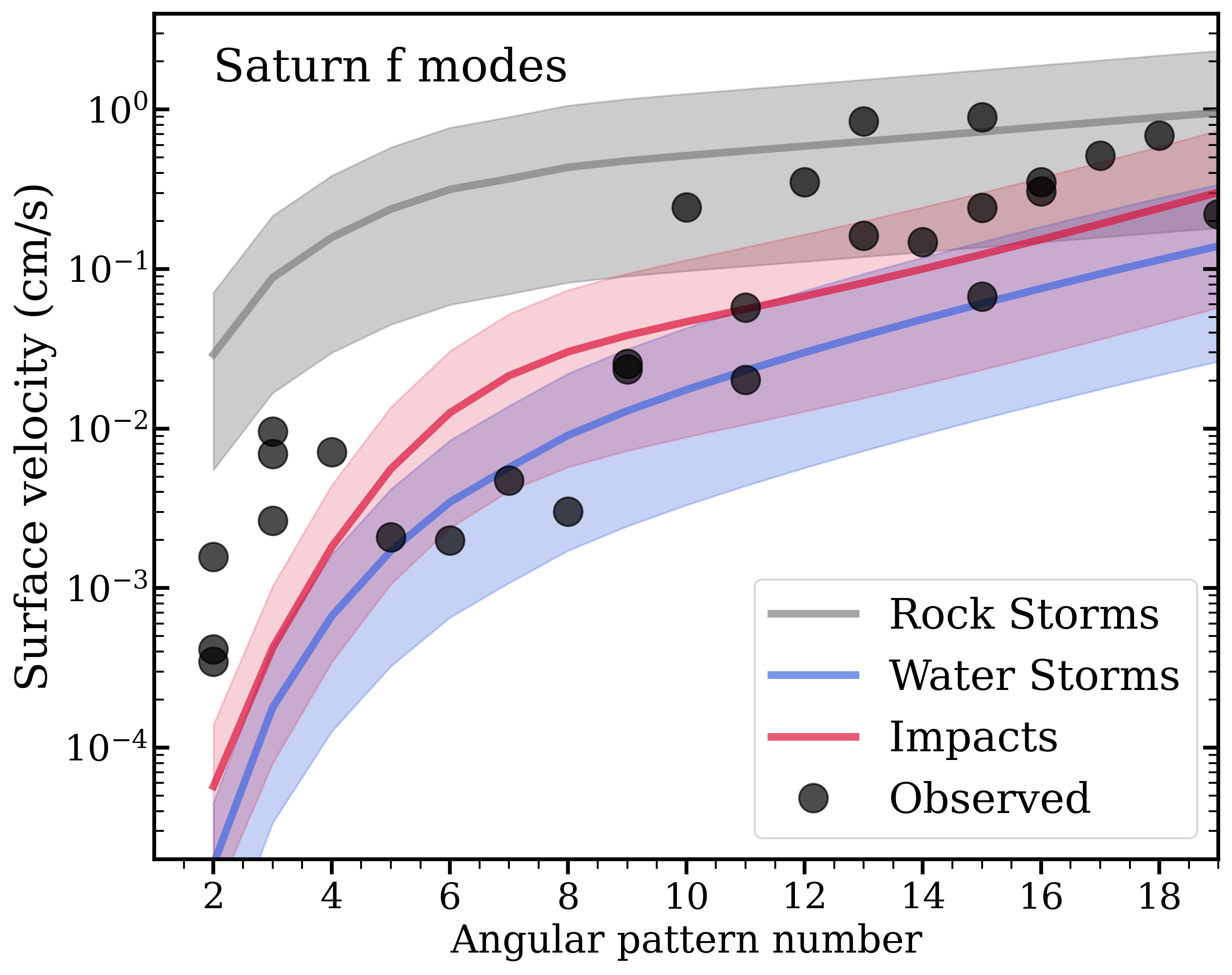}
\includegraphics[scale=0.36]{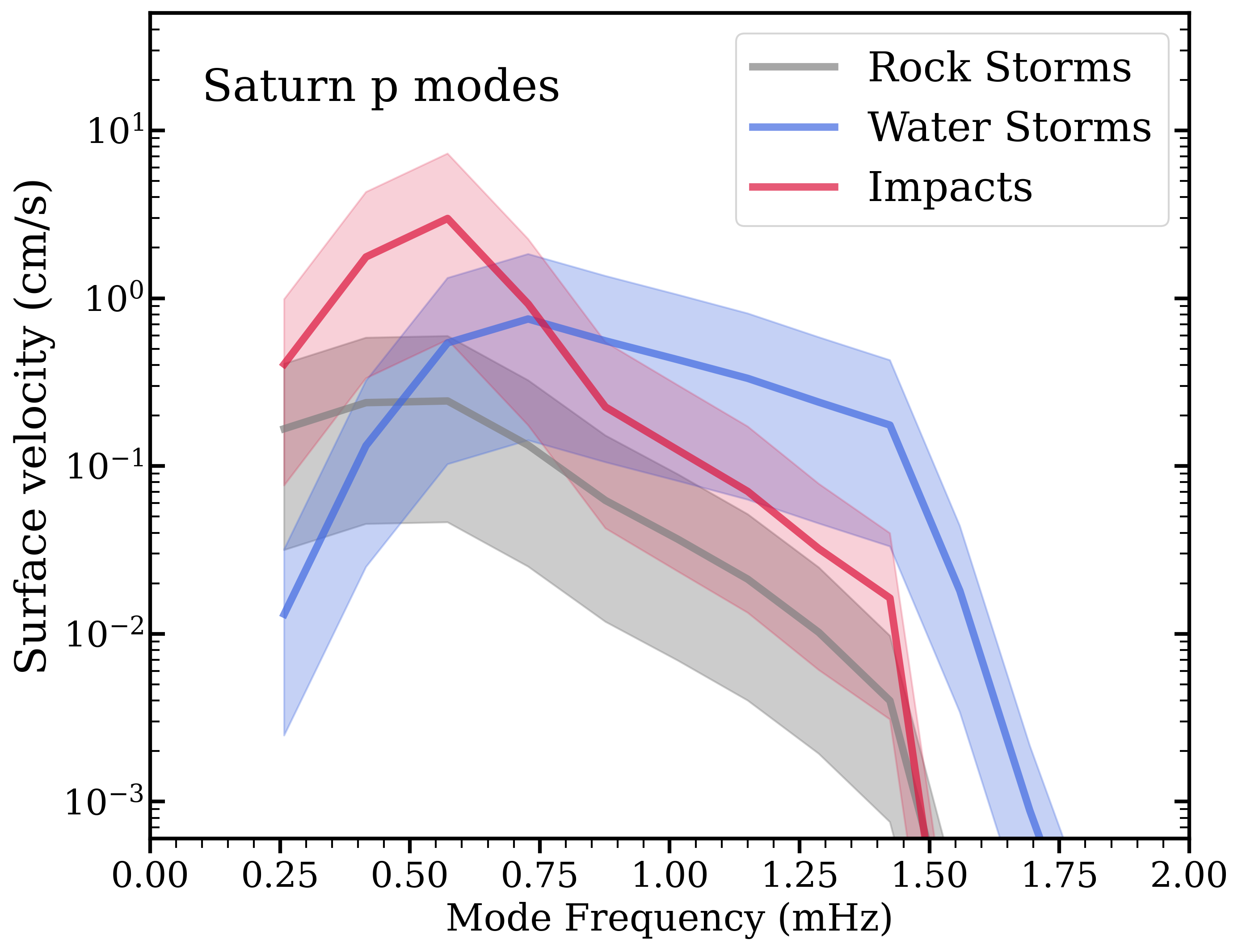}
\caption{\label{fig:vmodes} Similar to Figure \ref{fig:phimodes}, but now showing the surface radial velocity variations produced by Saturn's \bm{$\ell \sim m$} f~modes (top) and $\ell=2$ p~modes (bottom).}
\end{figure}

Figure \ref{fig:phimodes} shows the observed gravitational perturbations of Saturn's f~modes, and predicted perturbations of p~modes. Our impact model predicts the lowest frequency p~modes have gravitational perturbations comparable to f~modes, despite the fact that p~modes eigenfunctions have smaller values of $\delta \Phi_\alpha$. This is a consequence of the larger p~mode energies relative to f~modes. 

Figure \ref{fig:vmodes} shows the predicted surface radial velocity fluctuations for Saturn's f~modes and p~modes. In agreement with previous results, the observed low-$m$ f~modes will produce very small surface velocities less than $\sim \! 1$ cm/s, too small to be detected with existing instrumentation. However, high-$\ell$ f~modes and p~modes may be able to produce detectable surface velocity variations. Our impact model predicts maximum p~mode amplitudes of $\sim$5 cm/s at $f \sim 0.6 \, {\rm mHz}$, potentially detectable with future observations.

\subsection{Jupiter}
\label{sec:jupiter}

\begin{figure}
\includegraphics[scale=0.36]{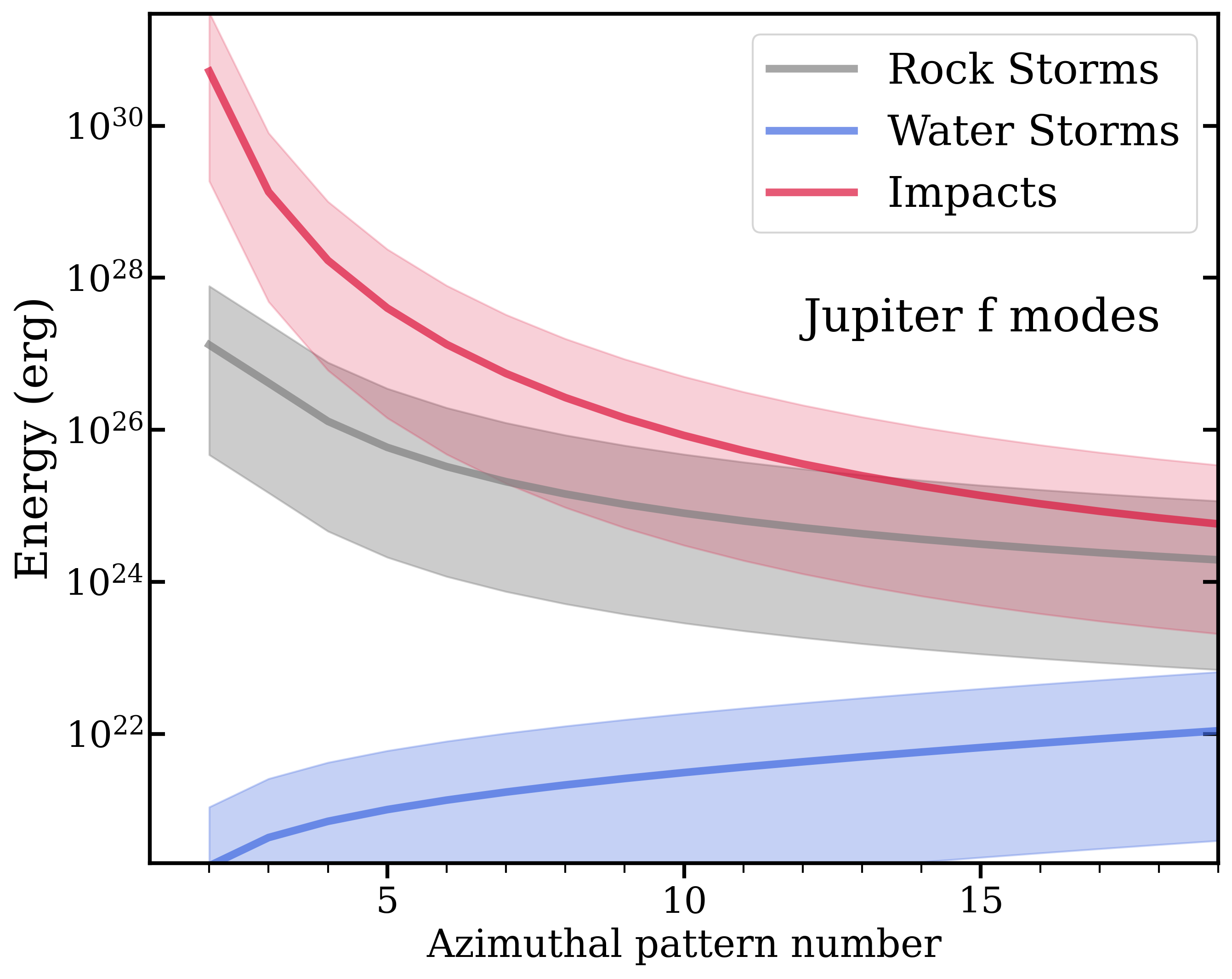}
\includegraphics[scale=0.36]{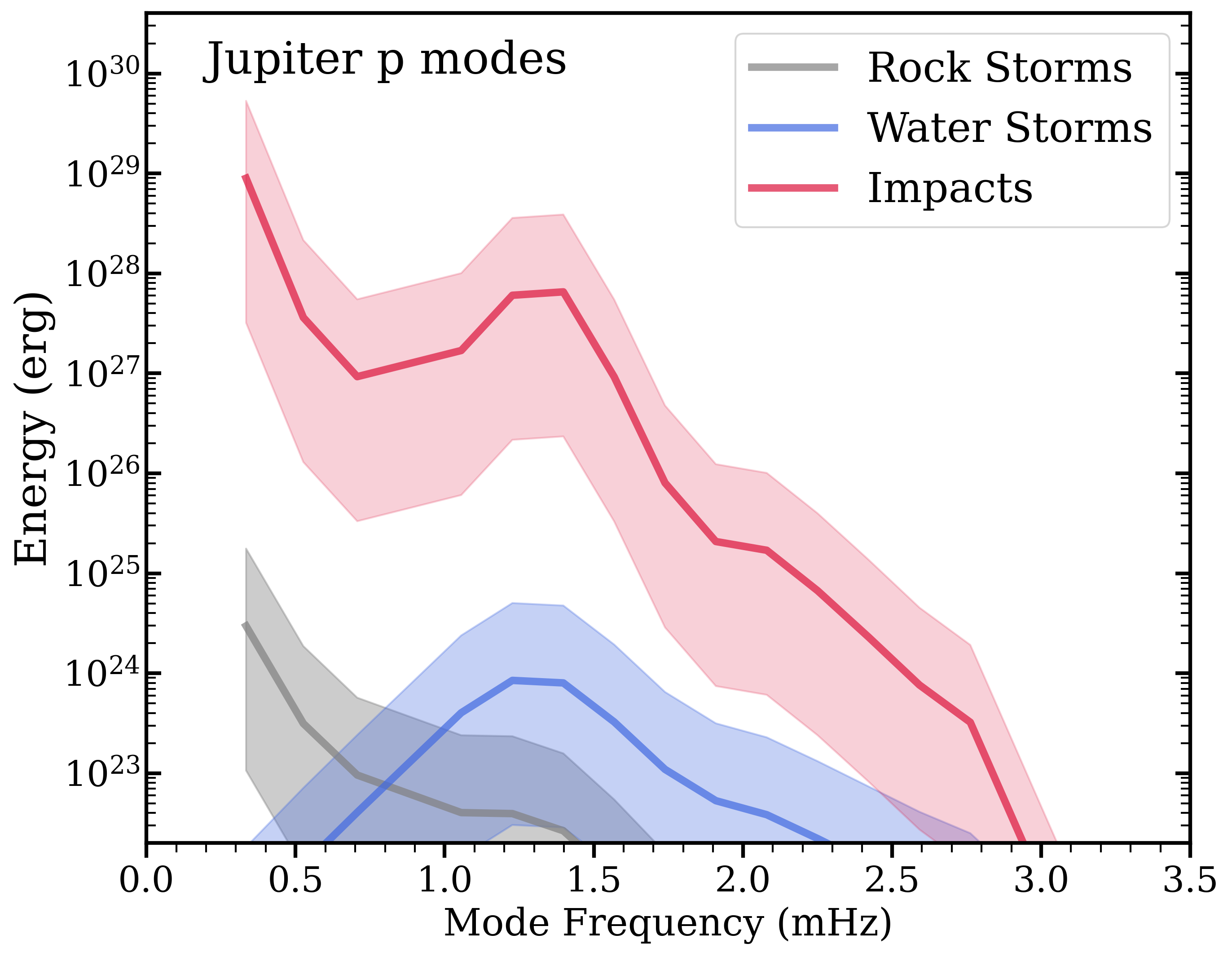}
\caption{\label{fig:EmodesJupiter} Similar to Figure \ref{fig:Emodes}, but now for the $\ell=m$ f~modes of Jupiter (top) and $\ell=2$ p~modes (bottom).}
\end{figure}

We apply the same calculations to our Jupiter model to predict the amplitudes of its putative oscillation modes. Unlike Saturn, Jupiter's modes have not been clearly detected, so our models are not well constrained, but they make testable predictions.

\begin{figure}
\includegraphics[scale=0.36]{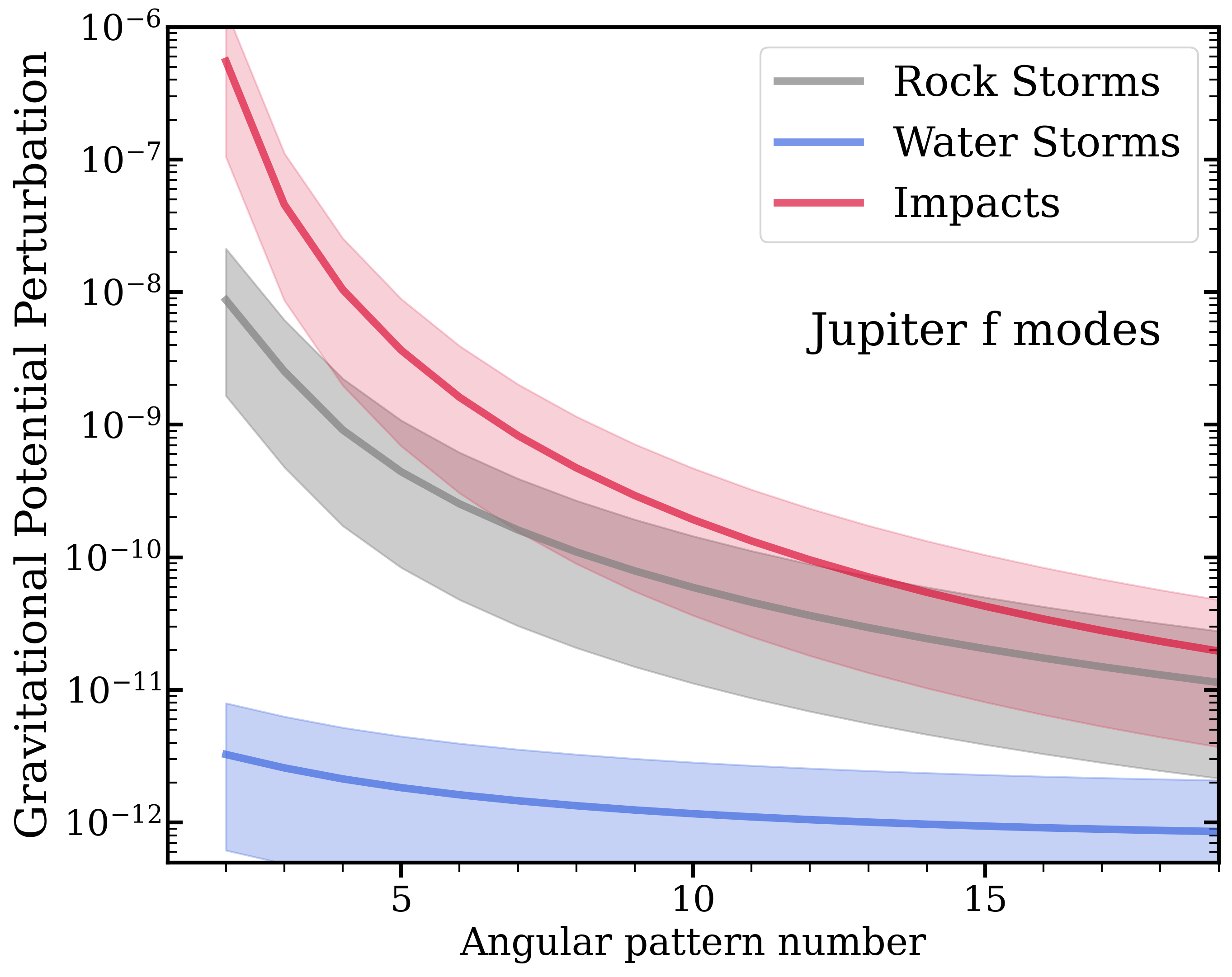}
\includegraphics[scale=0.36]{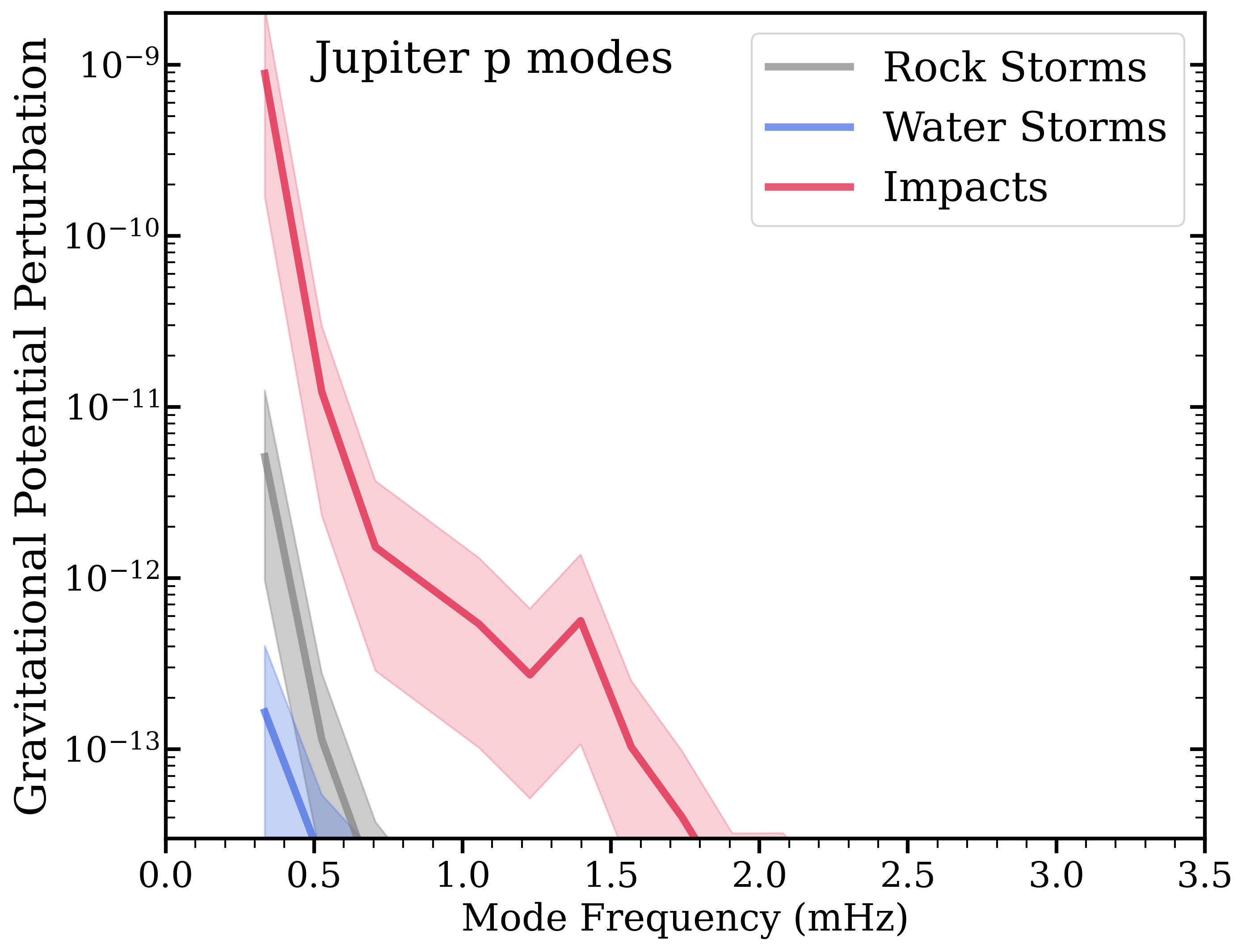}
\caption{\label{fig:phimodesJupiter} Similar to Figure \ref{fig:phimodes}, but now for the $\ell=m$ f~modes of Jupiter (top) and $\ell=2$ p~modes (bottom).}
\end{figure}

Figure \ref{fig:EmodesJupiter} shows predicted mode energies for Jupiter. In this case, rock storms and impacts are predicted to excite f~modes to similar energies at high-$\ell$, though impacts could produce energies as large as $\sim \! 10^{30} \, {\rm erg}$ for low-$\ell$ f~modes. The high predicted amplitudes of low-$\ell$ f~modes for Jupiter arises because there is no ring damping as in Saturn, so they have very long lifetimes. Water storms are less important for Jupiter than for Saturn, because of Jupiter's higher surface temperature. This moves the water weather layer even closer to the surface and decreases the value of $E_{\rm storm}$. Jupiter's p~modes are most efficiently excited by impacts, with predicted energies of $\sim \! 10^{27}-10^{29} \, {\rm erg}$ below frequencies of $\sim \! 2 \, {\rm mHz}$. Rock storms have turnover frequencies too low to efficiently excite p~modes, while water storms have too little energy.

\begin{figure}
\includegraphics[scale=0.36]{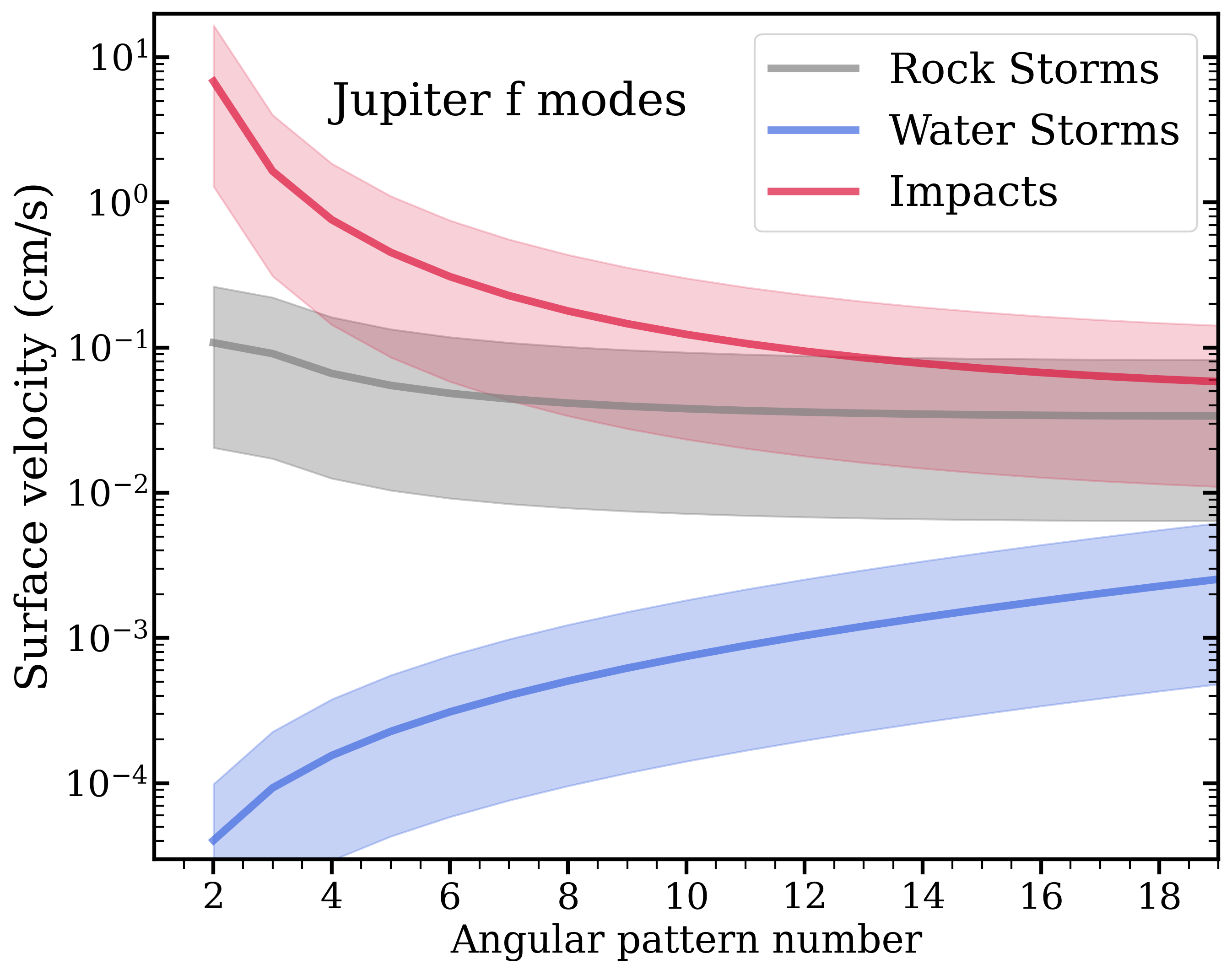}
\includegraphics[scale=0.36]{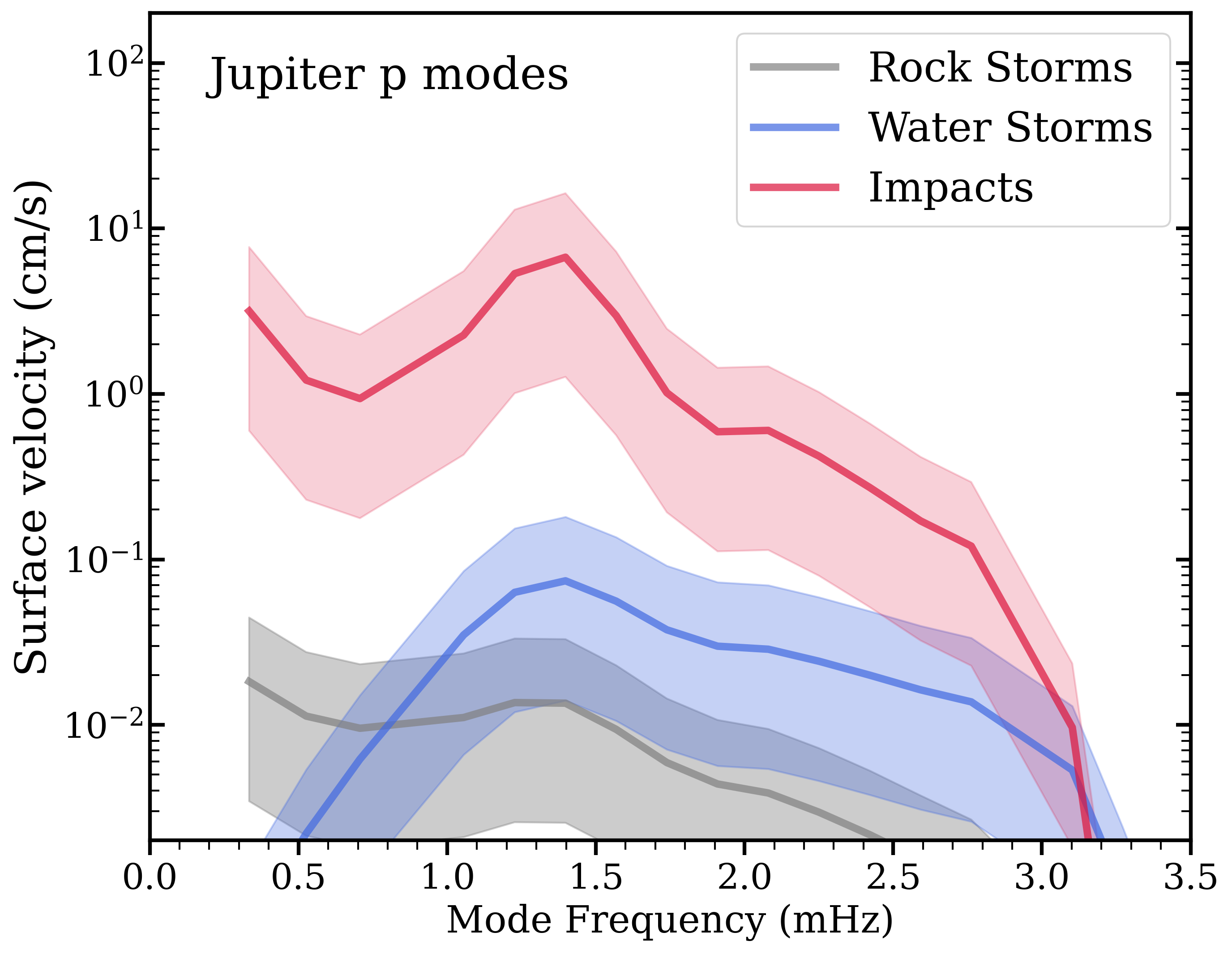}
\caption{\label{fig:vmodesJupiter} Similar to Figure \ref{fig:vmodes}, but now for the $\ell=m$ f~modes of Jupiter (top) and $\ell=2$ p~modes (bottom).}
\end{figure}

Figure \ref{fig:phimodesJupiter} shows predicted gravitational perturbations of modes in our Jupiter model. They are similar to Saturn, except for the larger predicted amplitudes for the lowest-$\ell$ f~modes, for the same reason mentioned above. These low-$\ell$ f~modes could have $\delta \Phi \sim 10^{-7} \, GM/R$ at frequencies of $\omega \sim 0.1 \, {\rm mHz}$. Another difference is that Jupiter's f~mode amplitudes are predicted to decrease at high-$\ell$ due to increasing damping, whereas Saturn's f~modes amplitudes are predicted to increase with $\ell$. Jupiter's lowest frequency p~modes may also obtain gravitational perturbations of $\delta \Phi \sim 10^{-10} \, GM/R$ at frequencies of $\sim 0.5 \, {\rm mHz}$, according to our impact model. 

Figure \ref{fig:vmodesJupiter} shows the predicted surface radial velocity fluctuations of Jupiter's oscillation modes. Interestingly, the large predicted amplitude of the low-$\ell$ f~modes entails surface velocities of $\sim$5 cm/s, potentially detectable in the future. For p~modes, the water and rock storm models predict velocities under 1 cm/s, likely too small to be observed. However, impacts could excite Jupiter's p~modes to amplitudes of $\sim \! 10$ cm/s for frequencies of $f \sim 1.5 \, {\rm mHz}$. These estimates are near current detections and/or upper limits \citep{Gaulme2011,shaw:22}, suggesting that p~modes could soon be detected in Jupiter by ongoing radial velocity measurements.

\subsection{Uranus}
\label{sec:uranus}

\begin{figure}
\includegraphics[scale=0.36]{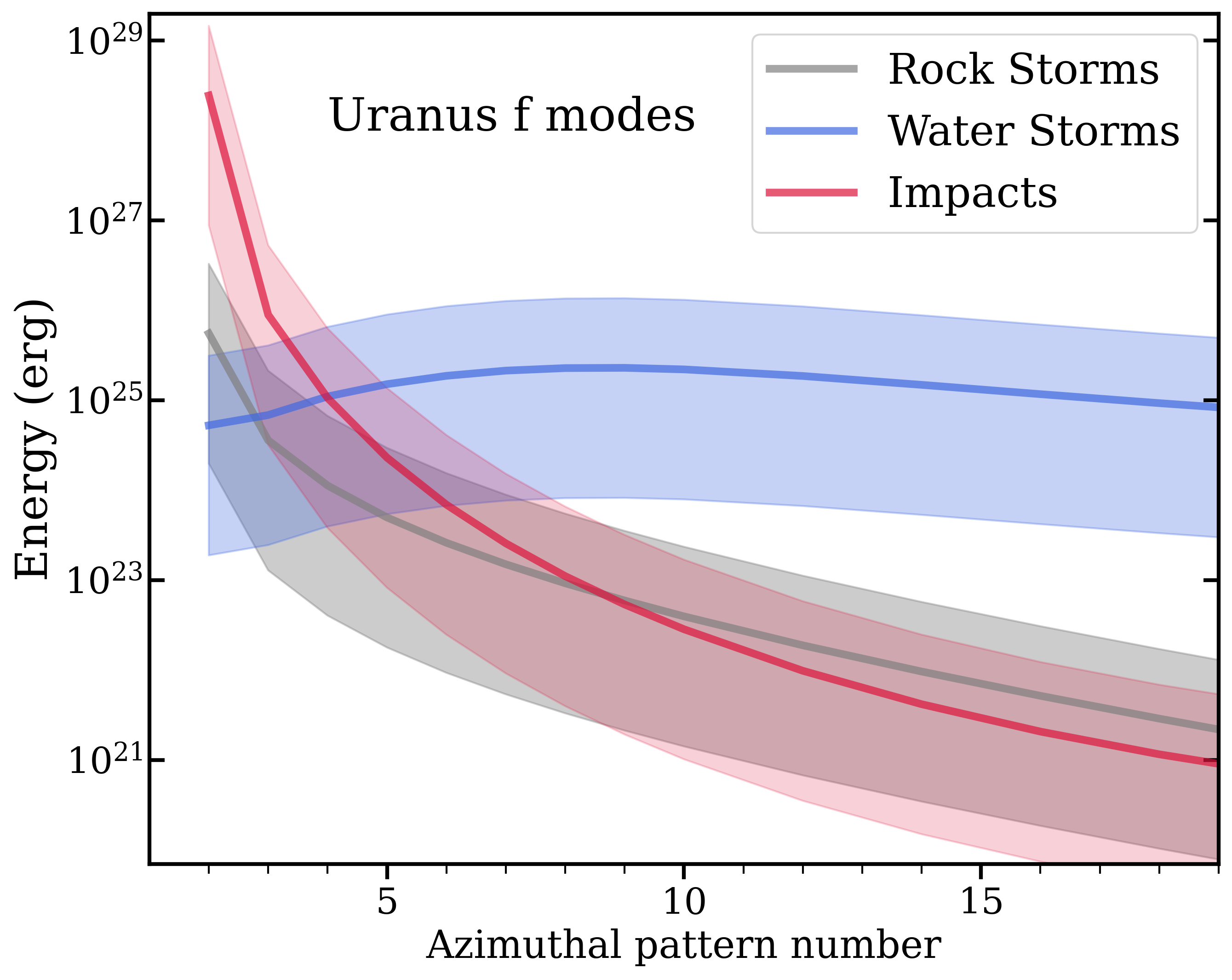}
\includegraphics[scale=0.36]{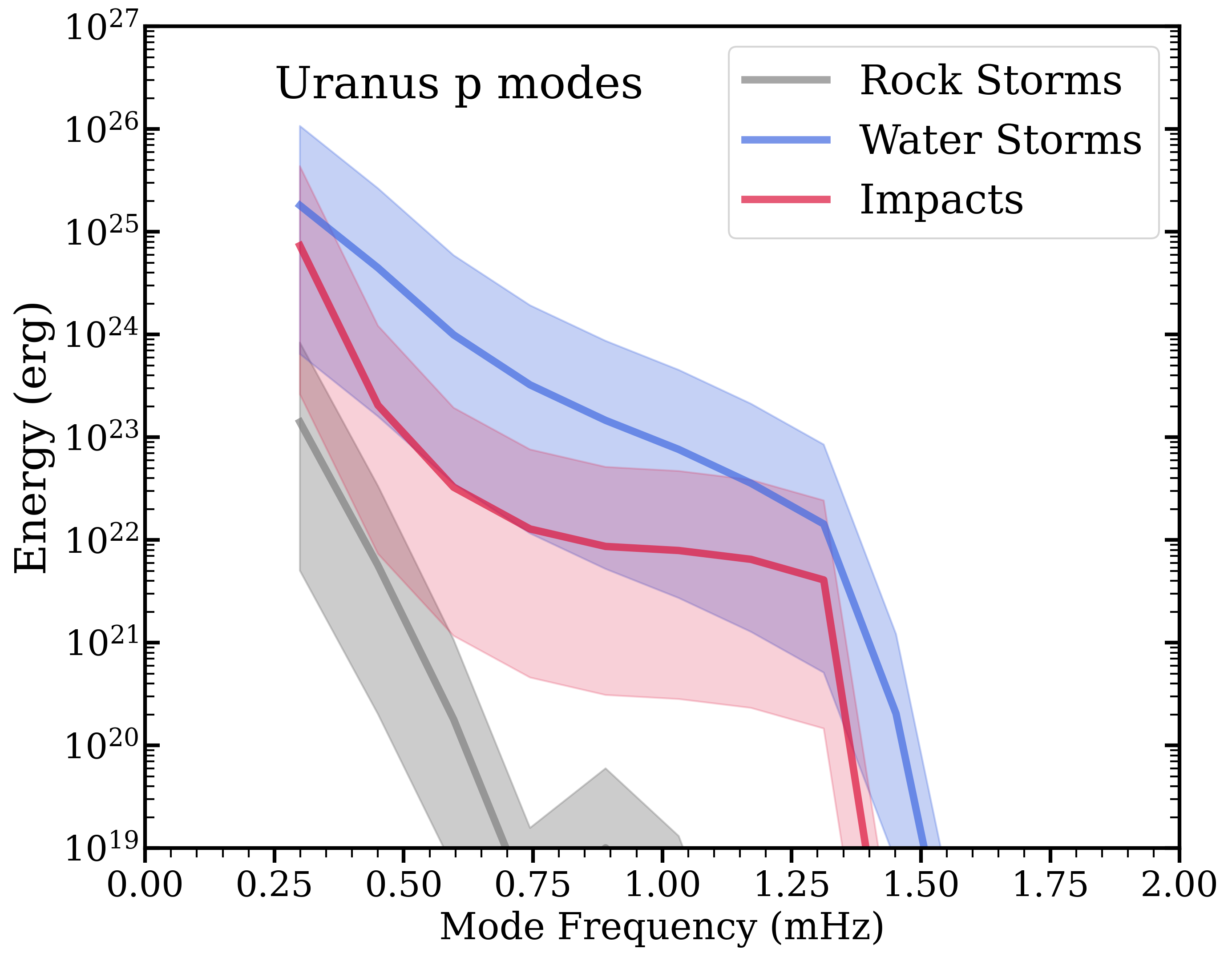}
\caption{\label{fig:EmodesUranus} Similar to Figure \ref{fig:Emodes}, but now for the $\ell=m$ f~modes of Uranus (top) and $\ell=2$ p~modes (bottom).}
\end{figure}

\begin{figure}
\includegraphics[scale=0.36]{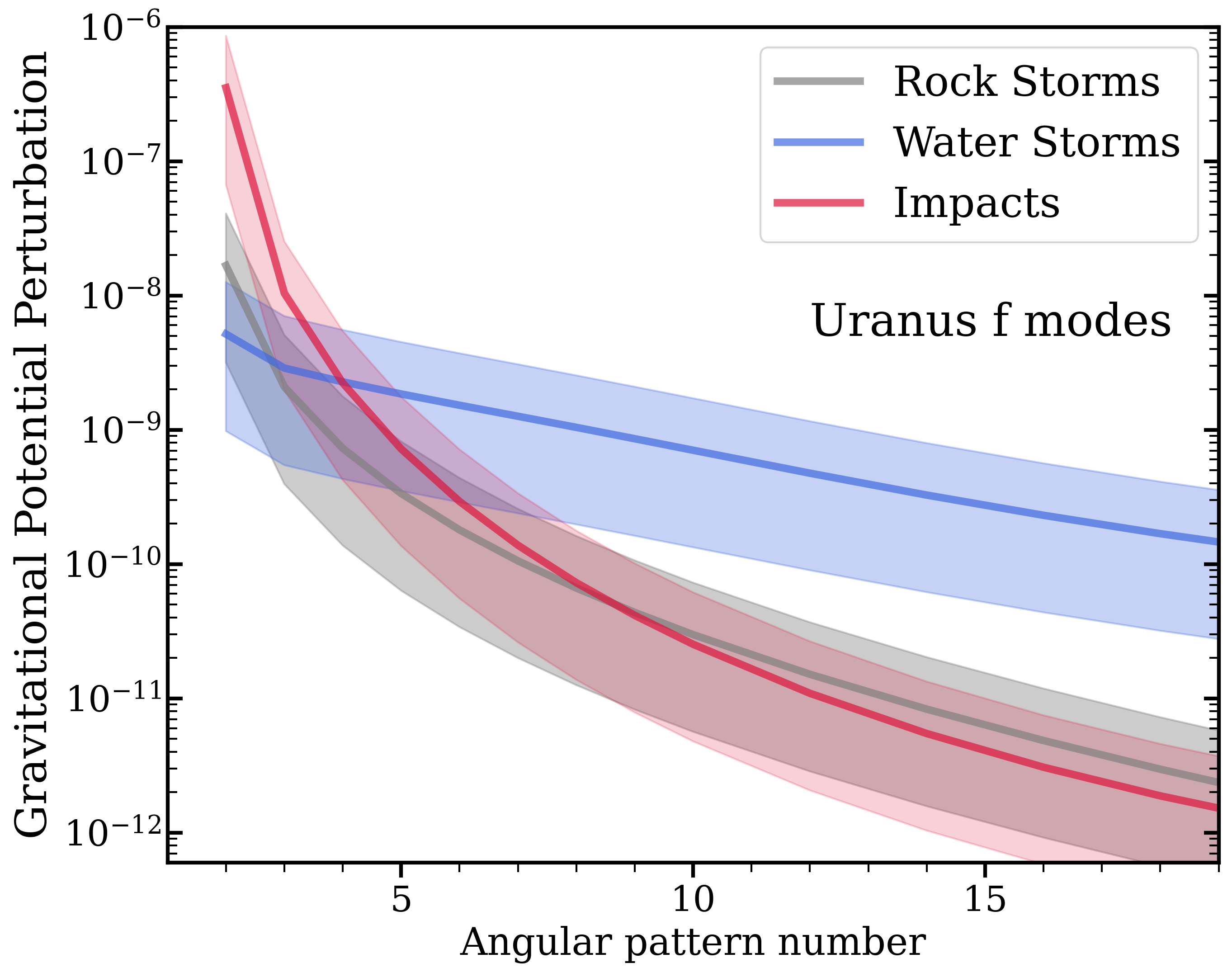}
\includegraphics[scale=0.36]{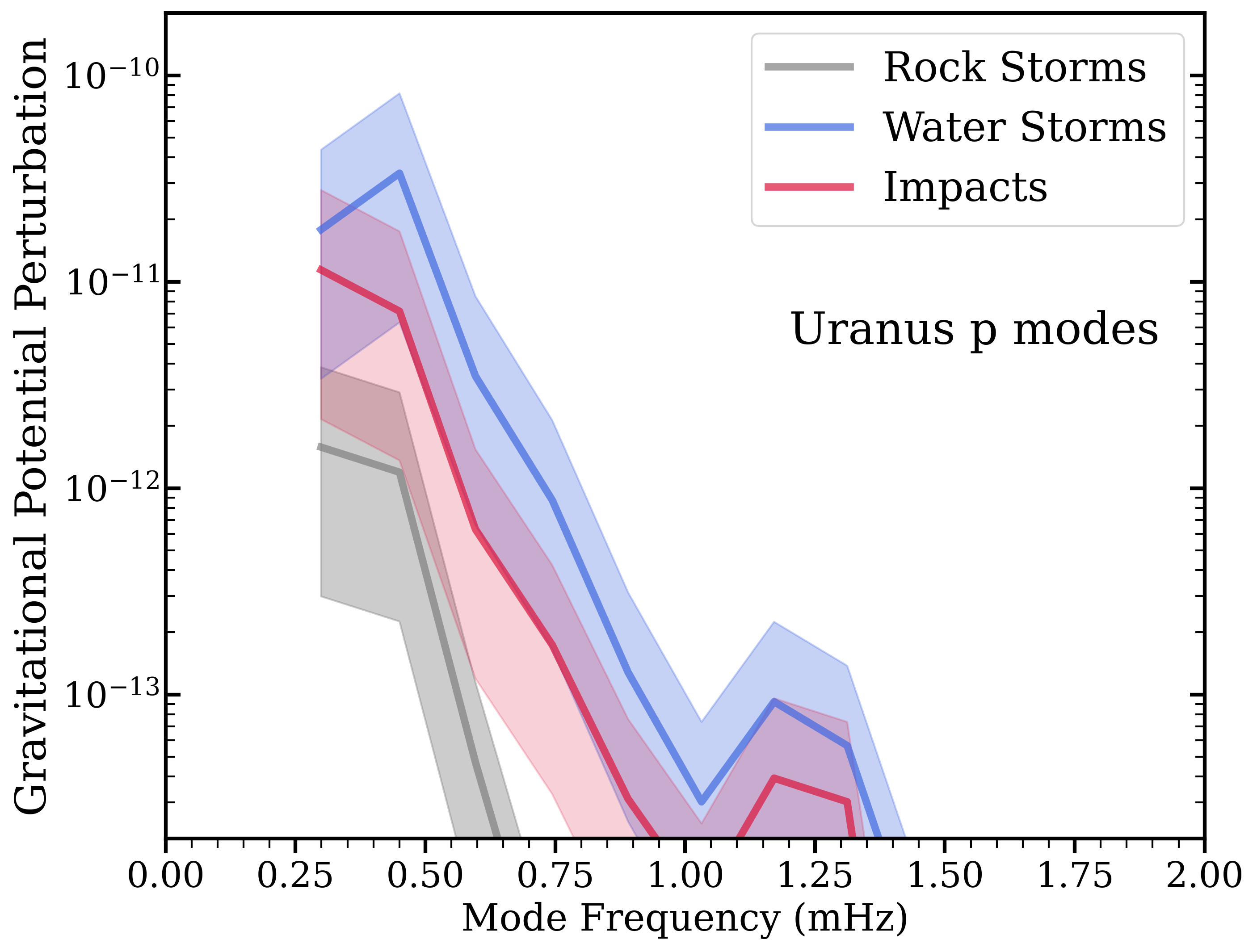}
\caption{\label{fig:phimodesUranus} Similar to Figure \ref{fig:phimodes}, but now for the $\ell=m$ f~modes of Uranus (top) and $\ell=2$ p~modes (bottom).}
\end{figure}

\begin{figure}
\includegraphics[scale=0.36]{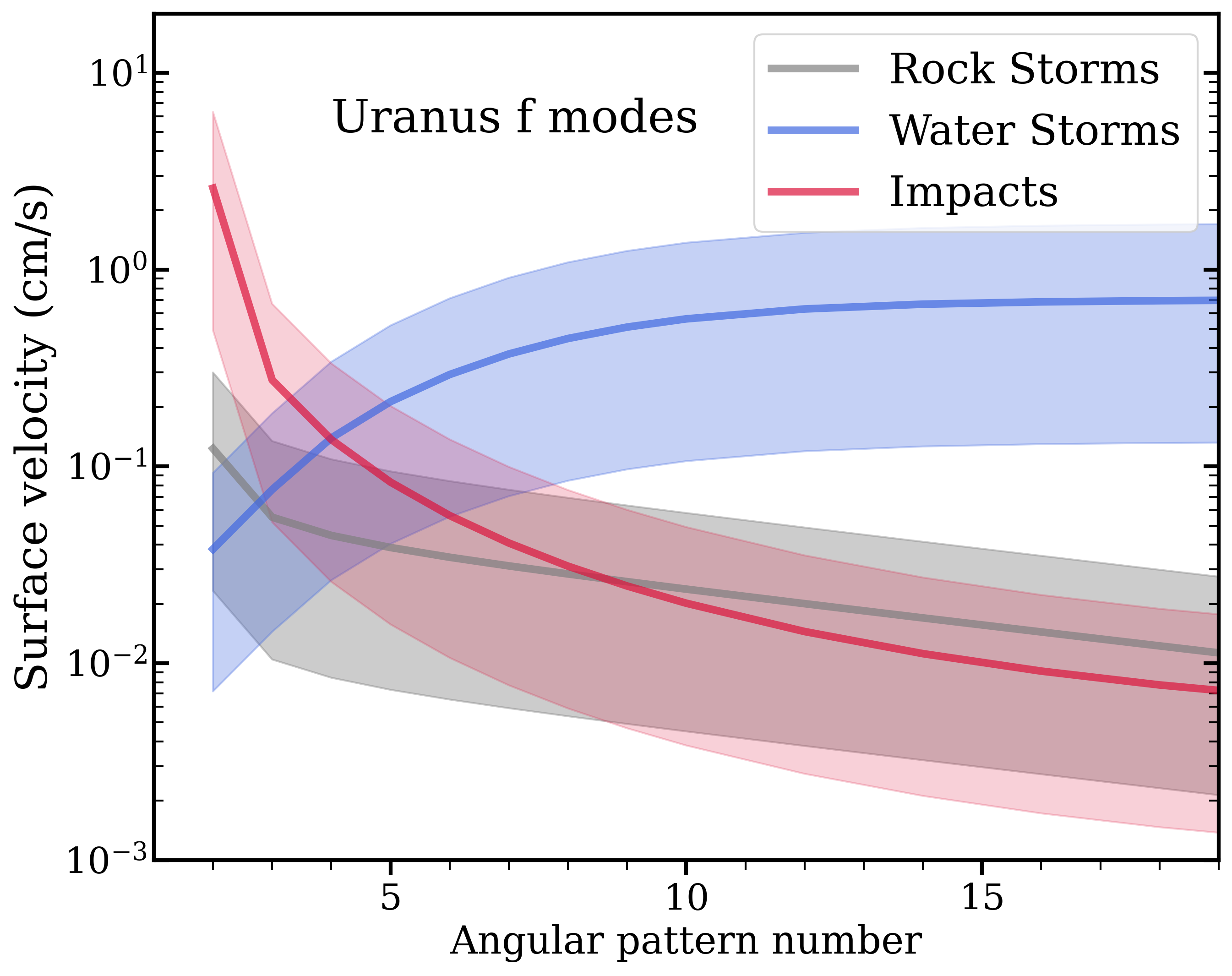}
\includegraphics[scale=0.36]{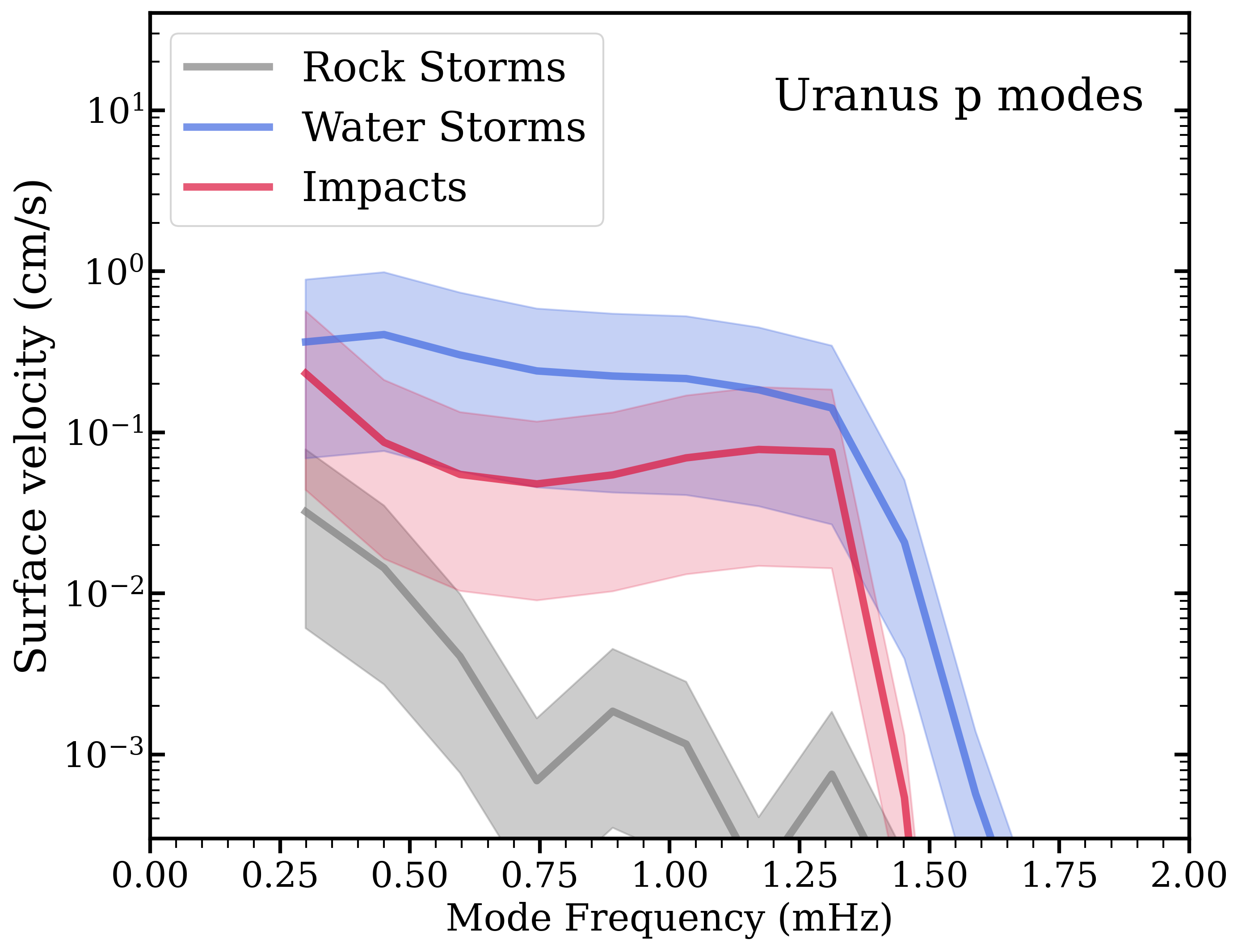}
\caption{\label{fig:vmodesUranus} Similar to Figure \ref{fig:vmodes}, but now for the $\ell=m$ f~modes of Uranus (top) and $\ell=2$ p~modes (bottom).}
\end{figure}

Figure \ref{fig:EmodesUranus} shows our predicted mode energies for Uranus. Like Jupiter, the low-$\ell$ f~modes could have large energies due to the long damping times. However, the high-$\ell$ f~modes are predicted to have smaller energies due to the larger damping from convective viscosity in our Uranus model. Water storms are an important excitation source in Uranus, because the weather layer is deeper due to its cooler surface temperature. In our models, the water weather layer occurs at a temperature of $T = 400 \, {\rm K}$ and pressure of $P \sim 100$ bar, with a typical storm energy $E \sim 7 \times 10^{28} \, {\rm erg}$, more than 100 times larger than in Saturn. The storm turnover frequency $\omega_{\rm storm}/(2 \pi) \sim 0.23 \, {\rm mHz}$ is high enough to excite f~modes and low-frequency p~modes.

Figure \ref{fig:phimodesUranus} shows predicted gravitational perturbations for Uranus' modes. As in Jupiter, the low-$\ell$ modes are predicted to have the largest $\delta \Phi$. Impacts could excite low-$\ell$ f~modes to potentially detectable amplitudes $\delta \Phi/(GM/R) \sim 10^{-7}$. The higher-$\ell$ f~modes are predicted to have lower amplitudes, though still comparable to those of Saturn with $\delta \Phi/(GM/R) \sim 10^{-10}$. The f~modes may therefore be able to excite detectable density waves in Uranus's rings, enabling Uranus ring seismology \citep{ahearn:22,mankovich:25}. The p~modes are not predicted to produce large gravity perturbations in any of our models.

Figure \ref{fig:vmodesUranus} shows the predicted surface velocities produced by Uranus's modes. Similar to Jupiter, the f~modes are expected to produce surface velocities smaller than 1 cm/s, apart from the $\ell=2$ f~mode which could produce a velocity of a few cm/s. Unlike Jupiter and Saturn, the p~modes of Uranus are also predicted to have low amplitudes of less than $\sim$1 cm/s. This results from the shorter damping times discussed above, which results in small energies from the impact model that scale as $t_{\rm damp}^3$. Water storms may contribute to excitation but are expected to produce amplitudes smaller than 1 cm/s because our estimated storm turnover frequencies are lower than the p~mode frequencies.

\section{Discussion}
\label{sec:discussion}

\subsection{Detecting Modes with Doppler Tracking}

The mode amplitudes we predict could likely be detected via Doppler tracking of an orbiting satellite at sufficiently close separation to the planet \citep{parisi:25}. Indeed, both the Cassini and Juno satellites have likely detected oscillations of Saturn/Jupiter \citep{markham:20,durante:22} using this technique. However, the eccentric orbits of those missions make it difficult to measure the angular pattern numbers or frequencies of the modes responsible. 

In the case of Saturn, \cite{markham:20} found that p~modes with frequencies of $f \sim 0.6 \, {\rm mHz}$ could explain the Cassini data. However, the required p~mode amplitudes corresponded to surface displacements of $\xi_r \sim$1 km (surface velocities of $v \sim 400 \, {\rm cm/s}$), whereas our impact model predicts lower much lower surface amplitudes of $\xi_r \sim 10 \, {\rm m}$ ($v \sim 4 \, {\rm cm/s}$). If the conclusions of \cite{markham:20} are correct, the p~modes should be readily detectable via radial velocity observations.

In the case of Jupiter, \cite{durante:22} found that \textit{Juno} data could be explained by low-order p~modes with frequencies $f \sim 1 \, {\rm mHz}$ and surface amplitudes of $v \sim 10-100 \, {\rm cm/s}$. These amplitudes are a factor of $\sim$10 larger than our predictions in Figure \ref{fig:vmodesJupiter}, and in mild tension with detections/upper limits from ground-based radial velocity measurements \citep{Gaulme2011,shaw:22}. \cite{durante:22} also found that f~modes should have amplitudes $v \lesssim 1 \, {\rm cm/s}$, in slight tension with our predictions for the lowest-$\ell$ f~modes. 

For both Jupiter and saturn, our models predict that the low-$\ell$ f~modes should produce larger gravitational perturbations than the p~modes, by a factor of $\sim \! 10^2-10^3$. Therefore, we expect the f~modes (whose frequencies are lower than the p~modes by a factor of $\sim$10) should dominate the Doppler tracking signal, in conflict with \cite{markham:18} and \cite{durante:22}. Note that Figure \ref{fig:phimodes} is a little misleading in this regard: the amplitudes of the $\ell \neq m$ f~modes of Saturn are predicted to be much higher than $\ell=m$ modes, due to the absence of ring damping for the former. Our models predict $\ell \neq m$ f~modes have $\delta \Phi/\Phi \sim 10^{-8}$, a factor of $\sim$100 larger than the $\ell=m$ f~modes.

The simplest explanation for this discrepancy is that our models overpredict f~mode amplitudes relative to p~modes, for an unknown reason. Given the large uncertainties inherent to our predictions, this is certainly possible.

Another possibility is that the Doppler tracking signal does not accurately reflect the true signal from planetary modes. The Doppler residuals represent a time-integrated quantity (with the actual measurement being the phase of the radio signal), and therefore do not necessarily capture the instantaneous character of the normal mode signal. It may be possible for mode perturbations to be partially or totally absorbed into estimates of other elements of the orbital solution (e.g. gravity field moments, tidal parameters), which could lead to biases in the data interpretations. We have seen a similar effect in injection-recovery exercises, where the residual signal after subtracting out a best-fit orbital model (including the static gravity field but not including modes) is different from the expected signal from the injected mode \citep{parisi:25}. It is possible that additional Juno data will help resolve this issue, and we encourage further investigation.


\subsection{Radial Velocity Monitoring}

Perhaps the most likely chance for seismic analyses is to detect radial velocity (RV) variations produced by planetary oscillation modes. Multiple ground-based RV campaigns are already underway, such as JOVIAL/JIVE \citep{schmider:13} and PMODE \citep{shaw:22}.

As discussed in Section \ref{sec:amplitudes}, our models predict RV variations of order $\sim$5 cm/s for the p~modes of Jupiter and Saturn, and less than 1 cm/s for Uranus. These predictions are only a few times lower than current detections or detection limits \citep{Gaulme2011,shaw:22}, suggesting that further RV monitoring may be able to detect oscillations if the measurement precision can be slightly improved. However, we also emphasize that our predictions are uncertain by at least an order of magnitude (see Section \ref{sec:uncertainties}), and it is certainly possible that true mode amplitudes too small to be detected with existing instruments. 

As noted above, \cite{durante:22} concluded that the orbit of \textit{Juno} is perturbed by p~modes with amplitudes of $\sim$10-100 cm/s, in apparent conflict with limits from RV measurements \citep{Gaulme2011,shaw:22}. \cite{markham:20} concluded the orbit of \textit{Cassini} is perturbed by p~modes with amplitudes of a few m/s, which could be detected by RV instruments. Our predicted amplitudes are 1-2 orders of magnitude smaller than these estimates, but they are very uncertain, and p~mode amplitudes of a few m/s are feasible.

If p~modes can be detected, it could allow for improved planetary seismic inferences. However, measuring key asteroseismic quantities such as the large frequency spacing, $\Delta \nu$, requires detecting multiple radial overtones for each $\ell$. Based on our results, this would require measurement precisions several times higher than that needed to detect the highest amplitude mode.

While our plots show predictions for $\ell=2$ p~modes, we find similar results for p~modes of other $\ell$ values. Measuring p~modes of multiple $\ell$-values would allow for measurements of the small frequency spacing $\delta \nu_{02}$, forward modeling techniques, and may allow for seismic inversions for the planetary structure. 

Unlike solar-like oscillations of main sequence stars that have lifetimes of $\sim$days, our predicted mode lifetimes of planetary oscillations are $t_{\rm damp} \gtrsim 10^4$ years. This means that stochastic fluctuations of mode amplitudes would not be observable on human time scales, which would make it difficult to detect modes whose amplitudes happen to be low at the current epoch. Nonetheless, the detection of a single p~mode would indicate many other modes could be detected with further monitoring, allowing for asteroseismic analyses.

\subsection{Saturn's Low-Amplitude Modes}

An intriguing feature of Saturn's observed f~modes is that the $\ell=m=5-8$ modes have smaller energies, by a factor of $\sim$100, than modes of both lower and higher $\ell$. Whether these lower amplitudes is a physical effect or a statistical abnormality is unclear, but it could indicate that those modes are more strongly damped (or less strongly driven) for some unknown reason. We could not identify any compelling mechanism. Most conceivable mechanisms would scale with $\ell$ or $\omega$, so it would be difficult for them to affect the $\ell=5-8$ modes without suppressing either lower or higher $\ell$ modes.

One possibility is that f~modes in this frequency range have an avoided crossing with a g~mode in Saturn's core, which somehow causes extra damping. However, an avoided crossing tends to produce two mixed f~mode/g~modes for each $m$, which are bumped in frequency relative to a pure f~mode. But only one f~mode for each value of $\ell=m=5-8$ is observed, and all of the observed frequencies are very close to those of pure f~modes in the model of \cite{Dewberry2021}. Another possibility is that the avoided crossing is with a Rosette mode \citep{takata:13,saio:14,takata:14}, whose short wavelength structure could increase the damping rate. Investigating this possibility would require non-perturbative treatments of rotation when computing oscillation modes, but could be examined in future work. 

\subsection{Neptune}
\label{sec:neptune}

We have not explicitly applied our models to Neptune. For the most part, we expect similar results for Neptune and Uranus. The planets have similar masses, radii, surface temperatures, differential rotation profiles, etc. The cometary impact rate is uncertain but estimates appear similar to that of Uranus \citep{zahnle:03}. Mode detection will likely be more difficult on Neptune due to its faintness. In addition, unlike an anticipated mission to Uranus \citep{NAP26522}, it appears unlikely an orbiter will be sent to Neptune in the near future.

\subsection{Comparison to Prior Work}
\label{sec:comparison}

Our work builds upon recent investigations of mode damping/driving processes in giant planets such as \cite{dederick:18}, \cite{markham:18}, \cite{wu:19}, and \cite{zanazzi:25}.
A new contribution from this work is the suggestion of enhanced convective viscosity due to differential rotation (Section \ref{sec:diffrot}). According to our models, this is the dominant source of damping for f~modes and low-order p~modes in Jupiter, Saturn, and Uranus (apart from Saturn's low-$\ell=m$ f~modes which are damped by the rings). We thus believe convective damping is more important than estimated in prior work.

Some of our results are similar to those discussed above, but we highlight a few important differences. \cite{dederick:18} investigated mode excitation by water storms in Jupiter. Their detailed storm modeling predicted velocities of $v_{\rm storm} \sim $50 m/s compared to our simple estimate of $v_{\rm storm} \sim $100 m/s. The storm models of \cite{hueso:01} predict maximum velocities ranging from $\sim30-200$ cm/s, so both estimates appear reasonable. \cite{dederick:18} did not rigorously calculate the transfer of storm energy to modes via Reynold's stresses, and our work suggests the coupling is not strong enough to excite modes to detectable amplitudes. 

\cite{markham:18} also investigated various damping and driving processes in Jupiter. We agree with the arguments of \cite{wu:19} that \cite{markham:18} overestimated damping rates via radiation losses. In terms of driving, we agree that convection and water storms are ineffective in Jupiter, while rock storms could play an important role.
However, in our model, rock storms can only effectively drive f~modes and not p~modes, because the storm turnover frequency of $f_{\rm st}$ is smaller than p~mode frequencies. Low-frequency suppression was not included in the calculation of \cite{markham:18}, even though they also find very low rock storm turnover frequencies. Unlike \cite{markham:18}, we find that cometary impacts can be important for exciting p~modes because our predicted mode damping times are larger than theirs. 

Comparing our work to \cite{wu:19} and \cite{zanazzi:25}, we reach many of the same conclusions, with a few differences. One is that they have an incorrect factor of $\sqrt{2 \ell + 1}$ in their mode excitation as discussed in Section \ref{sec:impacts}.
When computing excitation via storms, we believe \cite{wu:19} incorrectly computed the source term (see Section \ref{sec:storms}), which should arise from Reynold's stresses rather than an external momentum source.

Because \cite{wu:19} estimate weaker convective and radiative damping, they predict larger amplitude modes. The rock storm and impact models shown in their Figure 4 predict much larger energies for high-$\ell$ f~modes because of the very weak ring damping used in their calculations. Those predictions are incompatible with Saturn's relatively flat dependence of mode energy on $\ell$ for $\ell \gtrsim 10$ (Figure \ref{fig:Emodes}). This indicates a source of damping other than the rings for $\ell \gtrsim 10$ modes, possibly supporting our sheared convective viscosity model. Similarly, the impact model shown in their Figure 6 overpredicts Jupiter's p~mode amplitudes, predicting radial velocity variations of $\sim 10^3$ cm/s that can be ruled out \citep{Gaulme2011,shaw:22}. The larger damping rates we estimate are important for tempering mode amplitudes to levels within observational constraints.

\cite{mosser:95} estimate radiative mode damping rates of Jupiter. They estimate damping times of only $\sim \! 300$ years ($Q \sim 10^7$) at frequencies of 1 mHz, and damping times of only $\sim 10^{-2}$ years ($Q \sim 1000$) at frequencies of 3 mHz. Their calculation relies on radiative damping in the optically thin upper atmosphere where $P\sim 10^{-6}$ bar, but the details are not provided. Since the upper atmosphere is very optically thin, radiative losses are proportional to the Planck opacity $\kappa_p$ at low pressures. Since opacities typically fall off proportional to pressure/density (see \citealt{markham:18}) which we verify in our MESA models, the radiative emissivity falls exponentially with height in the upper atmosphere. We therefore believe that \cite{mosser:95} overestimate radiative damping rates. 

\cite{friedson:23} examined low-frequency r-modes in Saturn, which may excite some density waves in Saturn's rings (those discovered in \citealt{hedman:22}, not those shown in Figure \ref{fig:phimodes}). Intriguingly, they found that some of these modes could be driven unstable by Saturn's differential rotation (see more discussion below). However, this instability is only possible for modes whose pattern frequencies are less than the amount of differential rotation in Saturn, $f_\alpha/m \lesssim \Delta \Omega \sim 6 \times 10^{-3} \, {\rm mHz}$. Hence, low-$\ell$ f~modes and p~modes cannot be excited via this mechanism, but it could be important for low-frequency gravito-inertial modes, or high-$\ell$ f~modes ($\ell \gtrsim 100$).

\subsection{Known Unknowns}
\label{sec:uncertainties}

The physics of planetary mode excitation and damping is extremely uncertain because it relies on several physical processes that are not well understood. The shaded regions of our figures in Section \ref{sec:amplitudes} only indicate statistical fluctuations due to stochasticity and not systematic uncertainties in the underlying physics. 

We have argued that differential rotation can greatly enhance convective viscosity because it stretches convective eddies, giving them larger correlation lengths (or higher turnover frequencies) to produce viscous stresses. However, this is a new idea that needs to be confirmed/calibrated with numerical experiments.
While we believe equation \ref{eq:nuconef2} is the right order of magnitude for shear-modified convective viscosity, our calculations do not determine whether this viscosity is positive or negative. Indeed, \cite{Duguidetal2020a,Duguidetal2020b} found both positive and negative convective viscosity in hydrodynamic simulations. A negative viscosity would excite oscillation modes via linear instability. While this is possible in principle, we believe it is unlikely to occur in Saturn because the observed oscillation modes have very low amplitudes, but this possibility should be explored in future work.

In our models, mode damping occurs primarily in a transition layer between the differentially rotating envelope and rigidly rotating interior, and the viscosity scales with the transition layer width $\Delta R$ as $\nu \propto \Delta R^{-2}$. After integrating over this layer, the mode damping rate scales approximately as $\Delta R^{-1}$.
The differential rotation profiles of planets are not perfectly measured, especially for Uranus, where Mankovich et al. (in prep) find the differential rotation could extend significantly deeper than our estimate of $r/R =0.97$. We find that decreasing this to $r/R =0.94$ can slightly increase the damping times and amplitudes of Uranus's high-$\ell$ f~modes and low-order p~modes. 

The spectrum of convective eddy sizes and turnover frequencies is important for mode driving/damping, but it is not totally understood. We have used rotating mixing length theory \citep{stevenson:79,barker:14} to estimate the sizes and speeds of the energy-bearing eddies. However, the spectrum of sub-scale eddies with higher turnover frequencies is not well understood, and our results are sensitive to the high-frequency tail of the convective frequency spectrum. Some works such as \cite{mininni:10a} find a nearly Kolmogorov spectrum in terms of wavenumber. Simulations of convection exhibit a power spectrum that drops at frequencies $\omega \gtrsim \omega_{\rm con}$, but there is some debate about the slope of the frequency spectrum (\citealt{couston:18,horst:20,edelmann:19,anders:23,herwig:23,lecoanet:23}). Simulations that develop differential rotation exhibit a similar power spectrum at $\omega \gtrsim \omega_{\rm con}$ \citep{fuentes:25}. We find that the power spectrum would need to be nearly flat at $\omega > \omega_{\rm con}$ in order for convection to be able to excite Saturn's modes to their observed amplitudes, and this seems highly unlikely.

Mode excitation due to storms (equation \ref{eq:stormpower}) scales strongly with storm turbulent velocity $\dot{E}_\alpha \propto v_{\rm storm}^4$, and $v_{\rm storm}$ is likely uncertain by a factor of a few. Similarly, the storm sizes are very uncertain, and the excitation rate scales as $\dot{E}_\alpha \propto H_{\rm storm}^3$. If some unknown process increases/decreases the correlation lengths of turbulent motions in storms, the mode excitation rates would be greatly affected. Finally, the storm recurrence time is uncertain. Our method for computing storm energetics and recurrence times requires $\sim$5\% of each planet's luminosity to be channeled into rock and water storms. A less efficient conversion of internal energy into storm power would increase storm recurrence times and decrease mode excitation rates.

Exciting modes via impacts is also quite uncertain. There is roughly $\sim$1 order of magnitude uncertainty in impact rates based on differences between published estimates \citep{zahnle:03,nesvorny:23,brasser:25}. More importantly, the largest and least frequent impacts dominate mode excitation (see Section \ref{sec:impacts}), so the expected mode amplitudes become very sensitive to mode damping times, which are highly uncertain as discussed above.

\subsection{Unknown unknowns}

We have discussed many of the uncertainties that affect our estimated mode damping/excitation rates. But there are many other atmospheric and internal processes that could potentially contribute to mode damping/excitation.
We have not accounted for diurnal thermal forcing produced by the Sun's radiation, which certainly affects atmospheric dynamics. Magnetic fields could play an important role in some unknown way, though both \cite{markham:18} and \cite{wu:19} predict they are unimportant for mode damping. \cite{Dandoyetal2023} show that mode interaction with vortices can cause damping, but it is not clear how to calculate the energy dissipation rate in solar system planets. In any case, new ideas should be pursued, especially if measured mode amplitudes cannot be understood via the processes discussed here.

One mechanism that could play a role is fluid instabilities caused by zonal flows, which contain much more kinetic energy than storms or convective flows in giant planets. Zonal flows are observed to cause shear and baroclinic instabilities in simulations (e.g., \citealt{baruteau:13,bekki:22,blume:24,fuentes:25}).  However, the modes directly excited by shear instability typically have frequencies comparable to the shear itself, which is much smaller than f~mode and p~mode frequencies of gaseous planets.
Additionally, Saturn's observed f~modes have energies that increase with $\ell$ (Figure \ref{fig:Emodes}). This suggests an excitation mechanism near the surface of the planet by high-frequency motion, since f~modes become more confined to surface layers and obtain higher frequencies as $\ell$ increases. We have explored adhoc models with mode driving in deeper layers, and they nearly always predict larger energies for low-$\ell$ f~modes, suggesting the modes are excited by near-surface mechanisms such as storms or impacts.

Another possibility is driving/damping due to condensation and evaporation processes. Temperature changes induced by a mode could drive solutes like water to evaporate/condense, creating an entropy perturbation. Naively, this would damp modes because the thermal energy contained in compressed regions would be used to evaporate the condensate, sapping energy from the mode. This process should be examined in more detail. Along similar lines, \cite{bercovici:87} argued that modes should be driven due to energy release by ortho-para hydrogen conversion. They estimated a very short growth time of only $\sim$10 years for Jupiter. We believe this is unlikely based on the upper limits of $\sim$20 cm/s for Jupiter's oscillation mode amplitudes, but it is a topic that warrants more investigation.

\section{Conclusion}
\label{sec:conclusion}

Seismology is an exciting prospect for measuring the internal structures of gaseous planets in the solar system, but it requires oscillations of giant planets to be detectable. Predicting oscillation mode amplitudes is very difficult because it requires some understanding of excitation and damping processes, both of which are very uncertain in gaseous planets. We have explored several different damping and excitation mechanisms, making predictions for the amplitudes of fundamental modes (f~modes) and pressure modes (p~modes) of Jupiter, Saturn, and Uranus. These theories can be tested with observed f~modes of Saturn from ring seismology, allowing them to be extrapolated (albeit with great uncertainty) to predict the amplitudes of different types of modes in different planets.

We find that convective viscosity can be greatly enhanced by differential rotation, which stretches convective eddies such that they obtain larger correlation lengths. We estimate that this decreases mode damping times by a factor of $\sim$100 in solar system planets relative to a standard mixing length estimate, producing f~mode damping times ranging from $\sim 10^4-10^7$ years. In our models, this convective viscosity sets the damping rates of f~modes and low-order p~modes, while radiative diffusion sets the damping rates of higher frequency p~modes.

In agreement with prior work, we find that modes cannot be excited to detectable levels by convective turbulence as they are in the Sun. Instead, the most promising excitation mechanisms are storms and cometary impacts. Storms enhance mode excitation because they allow thermal energy to be stored as latent heat in evaporated water or silicates, and then to be suddenly converted into turbulent kinetic energy during the outbreak of a storm. A storm can thus have much larger kinetic energy and higher convective turnover frequency than an ordinary convective eddy, greatly enhancing mode excitation.

In our models, rock storms could excite Jupiter's f~modes to detectable amplitudes, while water storms could do the same for Uranus's f~modes. Both types of storms may be at play in Saturn's observed f~modes. These f~modes could likely be detected via Doppler tracking of an orbiting satellite, given a close enough orbit. Storms are unlikely to excite p~modes to detectable amplitudes in any of the planets, because water storms typically have too little energy, while rock storms have too low convective turnover frequencies. 

We find cometary impacts to be the most promising method for exciting p~modes to detectable amplitudes. We predict maximum radial velocity amplitudes of $v \sim 5 \, {\rm cm/s}$ for low-order p~modes in Jupiter and Saturn, with frequencies of $\sim$1 mHz. Impacts can also potentially excite Saturn's f~modes to their observed amplitudes, and excite Jovian and Uranian f~modes to detectable amplitudes via satellite Doppler tracking.

Unfortunately, all of our predictions are extremely uncertain. Since our computed damping/excitation rates result in mode amplitudes within an order of magnitude of Saturn's observed f~mode amplitudes, we consider these mechanisms to be good candidates for mode damping/excitation in giant planets, but they require more careful examination in future work. In particular, the shear-enhanced convective viscosity (and whether its sign is positive or negative) should be investigated with numerical simulations incorporating shear flows. The sizes, flow speeds, and recurrence time of storms in the planets are all uncertain, translating to very large uncertainties in mode excitation rates. Cometary impact rates are also uncertain, and resulting mode amplitudes from this channel are especially sensitive to uncertain mode damping rates.

Nonetheless, our calculations demonstrate that f~modes and p~modes of giant planets can plausibly be excited to detectable levels for existing or upcoming measurements. The best way to test our models is to go out and look. If we are fortunate, planetary oscillation modes will be detectable, allowing for a new era of solar system seismology.

\begin{acknowledgments}

This research was carried out at the Jet Propulsion Laboratory and the California Institute of Technology under a contract with the National Aeronautics and Space Administration (80NM0018D0004) and funded through the President's and Director's Research \& Development Fund Program. This work benefited from ideas from the participants of the 2023 September workshop ``Determining the Interior Structure of Uranus” organized by the W.M. Keck Institute for Space Studies. JF is grateful for the hospitality of the University of Tokyo and the Max Planck Institute for Astrophysics, where part of this work was carried out. He thanks Chris Mankovich, Masao Takata, Huazhi Ge, Mark Hofstadter, and Yanqin Wu for useful discussion, and Simon Mueller and Ravit Helled for help constructing MESA models.

\end{acknowledgments}

\bibliography{bib,CoreRotBib,SatBib}

\appendix

\section{Convective Viscosity with Shear}
\label{sec:shearvisc}

Here we present a more careful discussion of convection in the presence of shear, and implications for mode damping.

\subsection{Analytics}

We begin by considering the velocity perturbation imparted to a mode by convection, as discussed in Section \ref{sec:convdamp}. Considering only the inertial terms, a mode's velocity perturbation $\vec{v}_w$ changes with time as
\begin{equation}
    \frac{\partial}{\partial t} \vec{v}_w \approx \left( \vec{v}_{\rm con} \cdot \vec{\nabla} \right) \vec{v}_w + \left( \vec{v}_{\rm w} \cdot \vec{\nabla} \right) \vec{v}_{\rm con} \, .
\end{equation}
The first term is the one usually associated with convective viscosity. The second one has been argued to be important by \cite{Terquem2021,Terquem2023}, but shown to integrate to zero by \cite{Barker2021}, so we will not consider it further.
Over the lifetime of the interaction $\Delta t \sim {\rm min} (\omega_{\rm con}^{-1},\omega)$, the mode's velocity is changed by
\begin{equation}
\label{eq:deltavw}
    \Delta \vec{v}_w \sim \Delta t \left( \vec{v}_{\rm con} \cdot \vec{k} \right) \vec{v}_w \,
\end{equation}
where $\vec{k}$ is the mode's wavevector.

Next we consider how differential rotation affects the structure of convective eddies. For simplicity, we consider only the effect of inertial terms, and assume that a convective eddy would be approximately isotropic in the absence of shear. As above, the convective velocity is changed by
\begin{equation}
    \frac{\partial}{\partial t} \vec{v}_{\rm con} \approx \left( \vec{v}_{\rm rot} \cdot \vec{\nabla} \right) \vec{v}_{\rm con} + \left( \vec{v}_{\rm con} \cdot \vec{\nabla} \right) \vec{v}_{\rm rot} \, .
\end{equation}
The first term represents advection of the convective motion by the mean flow, while the second represents advection of the mean flow by the convection.

Since the rotation is only in the $\phi$-direction, we have
\begin{equation}
    \frac{\partial}{\partial t} \vec{v}_{\rm con} \approx \left( v_{\rm rot} h_{{\rm con},\phi}^{-1} \right) \vec{v}_{\rm con} + \left( \vec{v}_{\rm con} \cdot \vec{\nabla} \right) v_{\rm rot} \hat{\phi} \, ,
\end{equation}
where $h_{{\rm con},\phi}$ is the length of the original convective eddy in the $\phi$-direction. Over the lifetime of the convective eddy, the differential rotation causes a velocity change 
\begin{equation}
\label{eq:deltavcon}
    \Delta \vec{v}_{\rm con} \sim \omega_{\rm con}^{-1} \Big[ \left( v_{\rm rot} h_{{\rm con},\phi}^{-1} \right) \vec{v}_{\rm con} + \left( \vec{v}_{\rm con} \cdot \vec{\nabla} \right) v_{\rm rot} \hat{\phi} \Big] \, .
\end{equation}

We can choose a frame centered on a convective eddy that is co-rotating with the local mean flow, such that $v_{\rm rot}$ is zero at the center of the eddy. At a coordinate $\vec{h}_{\rm con}$ relative to the center of the eddy, the rotational velocity is $v_{\rm rot} \sim (\vec{h}_{\rm con} \cdot \vec{\nabla}) v_{\rm rot}$. Defining $\omega_{\rm shear} = | \vec{\nabla} v_{\rm rot}|$, to order of magnitude we have 
\begin{equation}
\label{eq:deltavcon}
    \Delta \vec{v}_{\rm con} \sim \omega_{\rm con}^{-1} \Big[ \omega_{\rm shear} \vec{v}_{\rm con} + \omega_{\rm shear} v_{\rm con} \hat{\phi} \Big] \, .
\end{equation}
We can see both terms have the same order of magnitude but are in different directions.

Plugging equation \ref{eq:deltavcon} in equation \ref{eq:deltavw}, the effect of shear on the mode-convection interaction is
\begin{equation}
\label{eq:deltavwshear}
    \Delta \vec{v}_w \sim \Delta t \frac{\omega_{\rm shear}}{\omega_{\rm con}} \Big[ \left( \vec{v}_{\rm con} + v_{\rm con} \hat{\phi} \right) \cdot \vec{k} \Big] \vec{v}_w \,
\end{equation}
Assuming the original convective eddies have similar velocities in all directions, we have 
\begin{equation}
\label{eq:deltavwshear2}
    \Delta \vec{v}_w \sim \Delta t \frac{\omega_{\rm shear}}{\omega_{\rm con}} v_{\rm con} k (1 + k_\phi/k) \vec{v}_w \, .
\end{equation}
The factor of $(1 + k_\phi/k)$ is of order unity so we will neglect it from here forward.

As described in the text, the mode loses energy per volume at a rate $\dot{\varepsilon}_w \sim \omega_{\rm con} \rho |\Delta v_w|^2$, giving
\begin{equation}
    \label{eq:edotw}
    \dot{\varepsilon}_w \sim \rho  \left( \Delta t \, \omega_{\rm shear} \, v_{\rm con} \, k \, v_w \right)^2 \omega_{\rm con}^{-1}  \, .
\end{equation}
The corresponding damping rate is $\gamma = \dot{\varepsilon}_w/(\rho v_w^2)$, with an effective viscosity $\nu_{\rm con,shear} = \gamma/k^2$ yielding
\begin{equation}
    \label{eq:nucon_ap}
     \nu_{\rm con,shear} \sim \left( \Delta t \, \omega_{\rm shear} \, v_{\rm con} \right)^2 \omega_{\rm con}^{-1} \, .
\end{equation}
Using $v_{\rm con} \sim h_{\rm con} \omega_{\rm con}$, 
\begin{equation}
    \label{eq:nucon_ap2}
     \nu_{\rm con,shear} \sim \left( \Delta t \, \omega_{\rm shear} \right)^2 v_{\rm con} h_{\rm con} \, .
\end{equation}
When the mode frequency $\omega$ is larger than the convective turnover frequency, we expect $\Delta t \sim \omega^{-1}$ and we obtain
\begin{equation}
    \label{eq:nuconshear}
     \nu_{\rm con,shear} \sim \left( \frac{\omega_{\rm shear}}{\omega} \right)^2 v_{\rm con} h_{\rm con} \, ,
\end{equation}
the same result obtained through the heuristic calculation of Section \ref{sec:diffrot}.

\subsection{Second point of view}

While the expression above yields the correct scaling for convective energy dissipation, it is useful to derive a similar expression from a first-principles calculation. Following \citep{Barker2021}, the rate at which energy is transferred from a mode into convective motions is
\begin{equation}
\label{eq:edot}
    \dot{E} = - \int dV \, \rho \, \vec{v}_{\rm con} \cdot \left(\vec{v}_{\rm con} \cdot \vec{\nabla} \right) \vec{v}_w \, .
\end{equation}
Here we have replaced their $v_e$ (corresponding to tidal distortion) with our $v_w$ corresponding to an oscillation mode. We agree with \citep{Barker2021} that other terms arising from convection-tide interaction integrate to nearly zero and will henceforth be ignored.

Equation \ref{eq:edot} vanishes upon integration over time or space, unless we account for the fact that convective flows are altered by the mode. The tidal shearing and convective motions perturb the convective velocity to $\Delta \vec{v}_{\rm con}$ as
\begin{align}
\label{eq:dconvdt}
    \frac{\partial}{\partial t} \Delta \vec{v}_{\rm con} \approx - \left( \vec{v}_{\rm con} \cdot \vec{\nabla} \right) \vec{v}_{\rm w} - \left( \Delta \vec{v}_{\rm con} \cdot \vec{\nabla} \right) \vec{v}_{\rm con} - \left( \vec{v}_{\rm con} \cdot \vec{\nabla} \right) \Delta \vec{v}_{\rm con} \nonumber \\ + \textrm{Buoyancy \& Pressure Terms} \, .
\end{align}
We are ignoring the term $-\left( \vec{v}_w \cdot \vec{\nabla} \right) \vec{v}_{\rm con}$ which describes the advection of the convection and will vanish upon integrating over the distorted volume.

We cannot easily solve this due to the complicated and time-dependent terms on the right side. But the first term will induce variation in $\Delta v_{\rm con}$ at a frequency $\sim \omega + \omega_{\rm con}$, while the others will induce variation at frequency $\sim \! \omega_{\rm con}$, which is stochastically varying. The convective velocities are changing in sign and magnitude and so $\omega_{\rm con} = \omega_{\rm con,r} + i \omega_{\rm con,i}$ is complex. To order of magnitude, this equation has a solution
\begin{align}
\label{eq:dconvdt2}
    \Delta \vec{v}_{\rm con} &\sim - \frac{ \left( \vec{v}_{\rm con} \cdot \vec{\nabla} \right) \vec{v}_{\rm w} }{ (i \omega + \omega_{\rm con}) } \nonumber \\ 
    & \sim - \frac{ -i \omega - i \omega_{\rm con,i} + \omega_{\rm con,r} }{ (\omega+\omega_{\rm con,i})^2 + \omega_{\rm con,r}^2 } \left( \vec{v}_{\rm con} \cdot \vec{\nabla} \right) \vec{v}_{\rm w}
\end{align}

Replacing $\vec{v}_{\rm con}$ in equation \ref{eq:edot} with $\vec{v}_{\rm con} + \Delta \vec{v}_{\rm con}$ yields several terms. We ignore any terms that are first or third order in $v_w$ because they integrate to zero over time or space. Similarly, terms proportional to $i \omega v_w^2$ also integrate to nearly zero over time (they amount to integrating $\int dt \sin(\omega t) \cos(\omega t) \approx 0$). The remaining terms yield an energy transfer rate
\begin{align}
\label{eq:edot2}
    \dot{E} &\sim \int dV \, \rho \, \frac{\omega_{\rm con,r} }{ (\omega+\omega_{\rm con,i})^2 + \omega_{\rm con,r}^2 } \, \vec{v}_{\rm con} \cdot \left( \left[ (\vec{v}_{\rm con} \cdot \vec{\nabla}) \vec{v}_w \right] \cdot \vec{\nabla} \right) \vec{v}_w \nonumber \\
    & + \int dV \, \rho \,  \frac{\omega_{\rm con,r} }{ (\omega+\omega_{\rm con,i})^2 + \omega_{\rm con,r}^2 } \, | (\vec{v}_{\rm con} \cdot \vec{\nabla} ) \vec{v}_w  |^2 \, .
\end{align}

These two terms have the same order of magnitude, but their net sign can be positive or negative, depending on correlations between convective velocities. The order of magnitude in energy dissipation per unit volume is 
\begin{equation}
    \dot{\varepsilon} \sim \frac{\omega_{\rm con}}{\omega^2 + \omega_{\rm con}^2} \rho v_{\rm con}^2 k^2 v_w^2 \, .
\end{equation}
The corresponding convective viscosity is
\begin{equation}
    \nu_{\rm con} = \frac{\dot{\varepsilon}}{\rho k^2 v_w^2} \sim \frac{\omega_{\rm con}}{\omega^2 + \omega_{\rm con}^2} v_{\rm con}^2\, .
\end{equation}
In the case of fast convection with $\omega_{\rm con} \gg \omega$ we obtain 
\begin{equation}
    \nu_{\rm con} \sim \omega_{\rm con}^{-1} v_{\rm con}^2 \sim v_{\rm con} h_{\rm con} \, .
\end{equation}
This is the ordinary expectation for convective viscosity. In the case of slow convection with $\omega \gg \omega_{\rm con}$ we obtain 
\begin{equation}
    \nu_{\rm con} \sim (\omega_{\rm con}/\omega^2) v_{\rm con}^2 \sim (\omega_{\rm con}/\omega)^2 v_{\rm con} h_{\rm con} \, .
\end{equation}
We thus obtain the quadratic scaling predicted by \cite{goldreich:77} and \cite{goodman:97}. However, this scaling does not depend on considering smaller eddies in a Kolmogorov cascade with $\omega_{\rm con} \sim \omega$. This is consistent with numerical simulations \cite{Duguidetal2020a,Duguidetal2020b} who find this scaling, but who also show the dissipation is not produced by sub-scale eddies.

Proceeding with equation \ref{eq:edot2}, we now try to include the effects of differential rotation. When $\omega_{\rm shear} \gg \omega_{\rm con}$ the convective velocity is roughly that of equation \ref{eq:deltavcon}, 
\begin{equation}
    \delta \vec{v}_{\rm con} \approx \omega_{\rm con}^{-1} \Big[ \left( v_{\rm rot} h_{{\rm con},\phi}^{-1} \right) \vec{v}_{\rm con} + (\vec{v}_{\rm con} \cdot \vec{\nabla} v_{\rm rot}) \, \hat{\phi} \Big] \, .
\end{equation}
Plugging into equation \ref{eq:edot2} and keeping only the second term for simplicity, the energy transfer rate is
\begin{align}
\label{eq:edot3}
    \dot{E} &\sim  \int dV \, \rho \, \frac{\omega_{\rm con,r} }{ (\omega+\omega_{\rm con,i})^2 + \omega_{\rm con,r}^2 } \, \nonumber \\ 
    &\bigg| \left( \omega_{\rm con}^{-1} \Big[ \left( v_{\rm rot} h_{{\rm con},\phi}^{-1} \right) \vec{v}_{\rm con} + (\vec{v}_{\rm con} \cdot \vec{\nabla} v_{\rm rot}) \, \hat{\phi} \Big] \cdot \vec{\nabla} \right) \vec{v}_w  \bigg|^2 \, .
\end{align}
For slow convection with $\omega_{\rm con} < \omega_{\rm shear} < \omega$, this expression has order of magnitude
\begin{align}
\dot{\varepsilon} &\sim \omega^{-2} \omega_{\rm con}^{-1} \omega_{\rm shear}^2 \rho v_{\rm con}^2 k^2 v_w^2 \,
\end{align}
with corresponding viscosity
\begin{equation}
    \nu_{\rm con,shear} = \frac{\dot{\varepsilon}}{\rho k^2 v_w^2} \sim \left( \frac{\omega_{\rm shear}}{\omega} \right)^2 v_{\rm con} h_{\rm con} \, .
\end{equation}
We thus obtain the same result as the preceding section.

\subsubsection{Extension to molecular viscosity}

One might be tempted to argue that the energy dissipation rate from equation \ref{eq:edot2} averages to nearly zero when integrating over different convective elements, because the sign of $\omega_{\rm con,r}$ can be positive or negative. Here we show this is misleading by extending the previous result to the limit of ordinary molecular viscosity, obtaining a positive-definite energy dissipation rate.

Ordinary viscosity corresponds to replacing $\omega_{\rm con,r}$ in equation \ref{eq:edot2} with the particle collision frequency $t_{\rm col}^{-1}$, and replacing the fluid velocity $\vec{v}_{\rm con}$ with the particle velocity $\vec{v}_{\rm p}$. The particle collision frequency $t_{\rm col}^{-1}$ is much larger than the wave frequency $\omega$ for any fluid wave, so $t_{\rm col}^{-1} \gg \omega,\omega_{\rm con,i}$. Assuming isotropic and uncorrelated particle motions, the average energy dissipation rate of equation \ref{eq:edot2} simplifies to
\begin{align}
\label{eq:edotvisc2}
    \dot{E} = 2 \int dV \, \rho \, t_{\rm col} v_{\rm p}^2  | \vec{\nabla} \vec{v}_w |^2 \, ,
\end{align}
which is positive-definite. Recognizing that the viscosity is $\nu = t_{\rm col} v_{\rm p}^2$, we have
\begin{align}
\label{eq:edotvisc3}
    \dot{E} = 2 \int dV \, \rho \, \nu | \vec{\nabla} \vec{v}_w |^2 \, ,
\end{align}
which is the correct expression for the viscous dissipation rate of an incompressible flow.

\subsection{Simulations}

\begin{figure}
\includegraphics[scale=0.36]{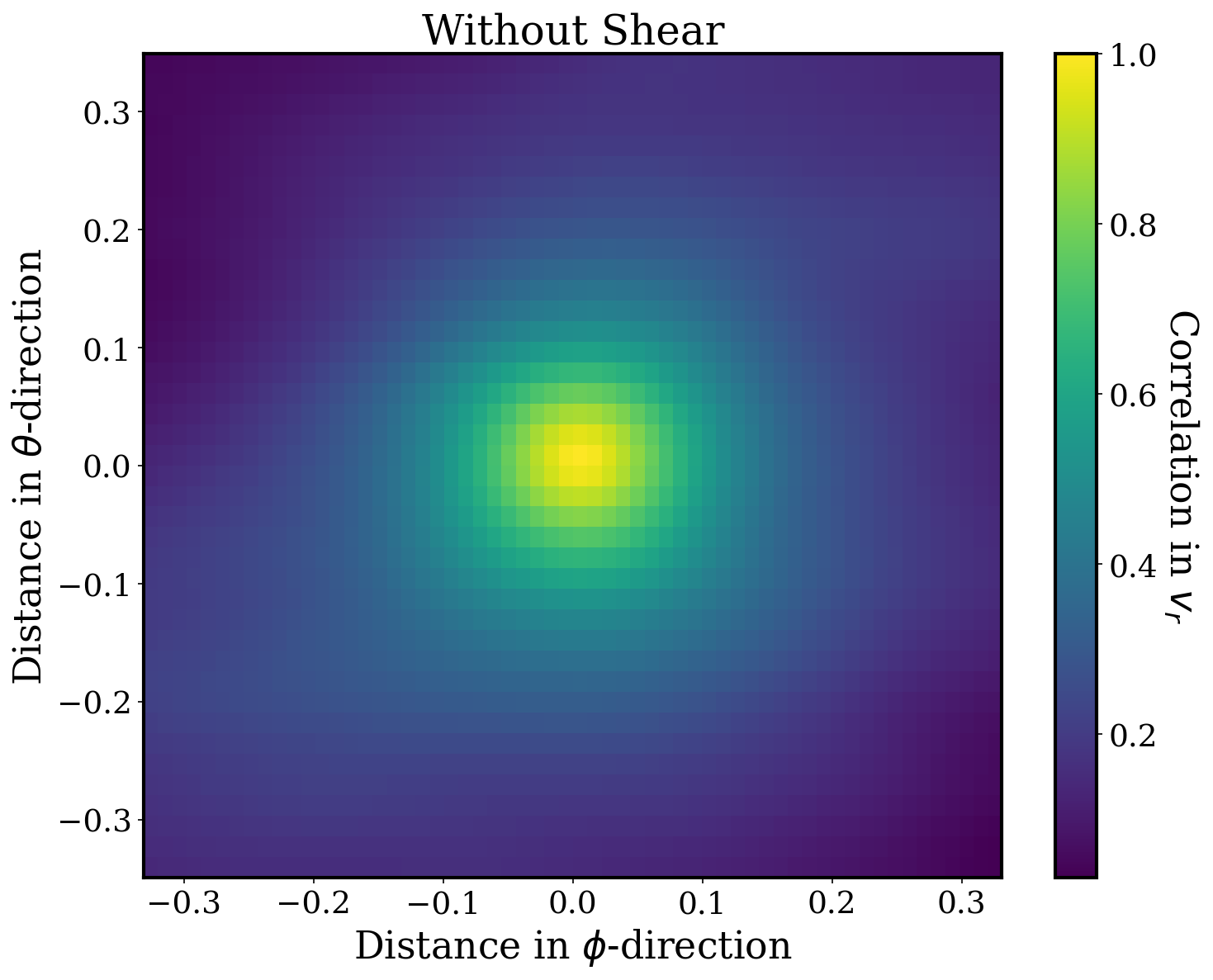}
\includegraphics[scale=0.36]{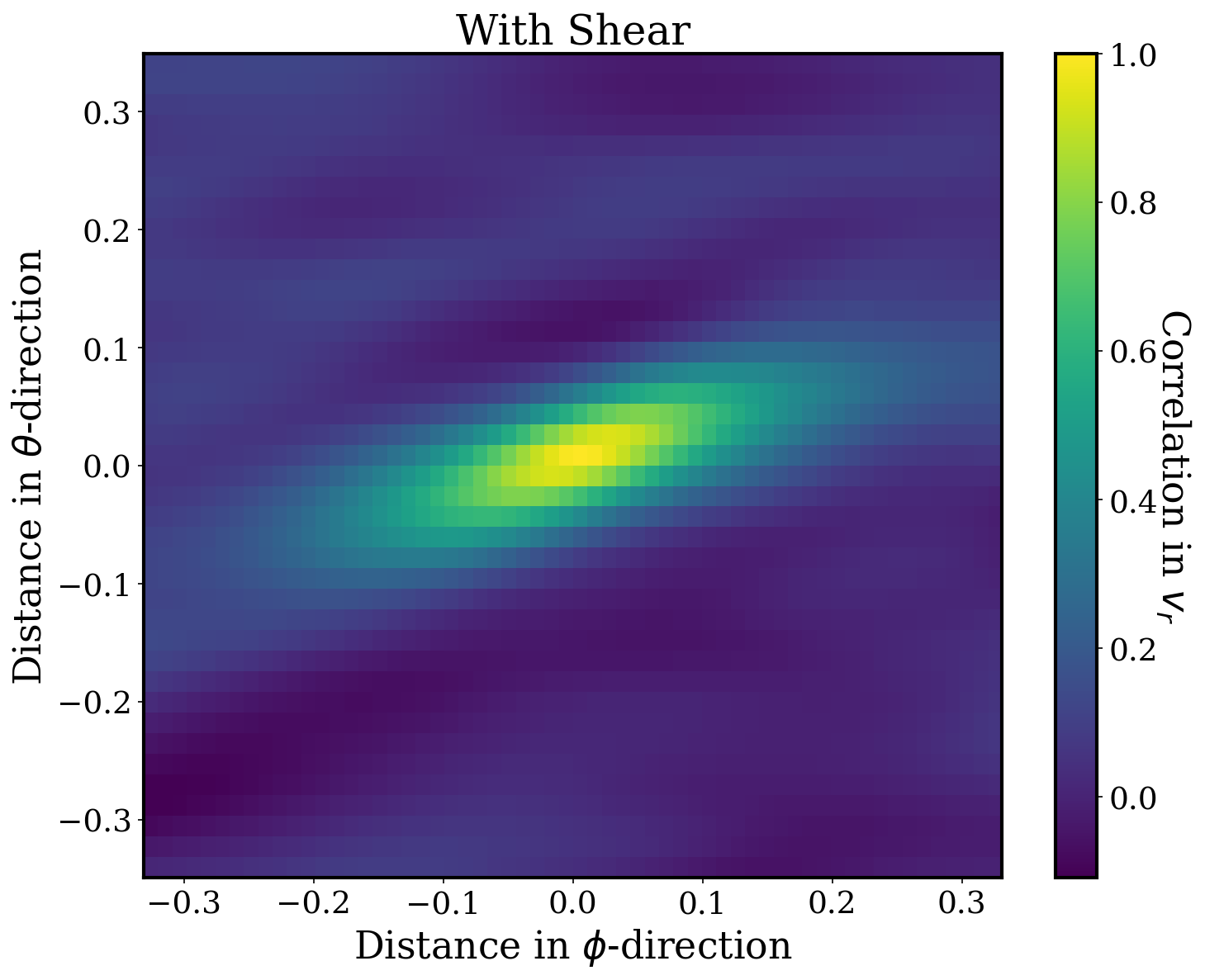}
\caption{\label{fig:correlations} Correlation $C_r$ of radial velocity variations in a simulation of convection in a spherical shell. $C_r$  is plotted as a function of distance in the $\phi$ and $\theta$ directions, for both a non-rotating case (left) and a differentially rotating case (right). The differential rotation clearly stretches convective eddies in the $\phi$-direction.}
\end{figure}

We are not aware of any simulations including large scale shear interacting with modes that can be used to test our predictions for mode damping. However, there are several existing simulations of convection in the presence of shear \citep{lee:18,blass:21,zhong:23}, which demonstrate that convective eddies are greatly stretched in the direction of the shear (the $\phi$-direction in the case of differential rotation).

We can quantify the sizes of convective eddies by defining a radial velocity correlation function
\begin{equation}
    C_r(\Delta \phi, \Delta \theta) = \frac{ \langle v_r(r,\phi,\theta) v_r(r, \phi + \Delta \phi,\theta + \Delta \theta) \rangle }{\langle v_r^2(r,\phi,\theta) \rangle } \, .
\end{equation}
Here, $\langle ... \rangle$ indicates a summation over different grid cells and/or times. Without rotation or shear, we expect the correlation function to be axisymmetric in the horizontal direction, dropping off on a length scale $\sim \! h_{\rm con}$. With shear, we expect the correlation pattern to be stretched in the $\phi$-direction, forming an elongated ellipse tilted at angle $\sim \omega_{\rm con}/\omega_{\rm shear}$ relative to the $\phi$-direction.

We calculate $C_r$ from the simulations of rotating and sheared convection in \cite{fuentes:25}, for a case with no rotation or shear, and a case with rotation and latitudinal shear $\omega_{\rm shear} \sim \omega_{\rm con}$ (the ${\rm Ro}_c \approx 0.7$ case shown in their Figures 1 and 2). We perform the calculation at fixed radius $r=0.85$ and latitude $\theta \simeq 0.7$ and perform the $\langle ... \rangle$ sum over all $\phi$ coordinates and 125 separate output times. Similar patterns emerge when considering different radial coordinates, latitudes, or different components of the convective velocities.

Figure \ref{fig:correlations} shows $C_r$ in the cases with and without shear. It is clear that the eddies are stretched in the $\phi$-direction by the shear. They have similar area to the non-rotating case, but smaller correlation length in the $\theta$-direction and larger correlation length in the $\phi$-direction. These results are consistent with correlation lengths enhanced by a factor of $\sim \omega_{\rm shear}/\omega_{\rm con}$, as calculated in the previous section.

\section{Mesa Models}
\label{sec:mesa}

We generate planetary models using MESA \citep{paxton:11,paxton:13,paxton:15,paxton:18,paxton:19,jermyn:23}. Our models are adapted from those of \cite{muller:20,muller:21,helled:25}, which use a custom equation of state allowing for higher metallicity than MESA's default grids.\footnote{Inlists and run files for our models can be accessed at \href{https://zenodo.org/records/18512364}{https://zenodo.org/records/18512364}.} Our Jupiter and Saturn models are coreless, with bulk metallicities $Z=0.06$ and $Z=0.20$, and equilibrium irradiation temperatures of $T_{\rm eq} = 120 \, {\rm K}$ and $T_{\rm eq} = 100 \, {\rm K}$ respectively. Our Uranus model includes an inert core of 10 Earth masses and an envelope of $Z=0.20$. In order to prevent numerical problems, we irradiate it at a level slightly higher than reality ($T_{\rm eq} = 130 \, {\rm K}$) and inject a luminosity of $10^{22} \, {\rm erg/s}$ (about 3x Uranus's observed surface flux) at the outer boundary of its core. We evolve the models to ages of 4.5 Gyr. The radii and luminosities of our Jupiter and Saturn models are similar to those observed, but our Uranus model is too large and luminous due to its extra heating, with a radius of $\sim$5 Earth radii.


We then compute the non-adiabatic oscillation modes of our planetary models with GYRE. We find that the modes are always damped, with most of the damping occurring in the near-surface layers where the pressure is less than $\sim$10 bars and the optical depth is less than $\sim$100. To map the damping rates from these modes to those used for the rest of our calculations, we interpolate based on the dimensionless frequency $\omega_\alpha/\omega_{\rm dyn}$. For Uranus, the model's larger luminosity may artificially enhance the radiative damping. Better estimates should be explored with more accurate planetary models.

\end{document}